\documentclass[sigplan,twocolumn,nonacm]{acmart}

\acmSubmissionID{129}
\renewcommand\footnotetextcopyrightpermission[1]{}
\setcopyright{acmlicensed}
\copyrightyear{2026}
\acmYear{2026}
\acmDOI{XXXXXXX.XXXXXXX}
\acmConference[EUROSYS '26]{European Conference on Computer Systems (EuroSys)}{April 27-30, 2026}{Edinburgh, Scotland}

\usepackage{amsmath,amsfonts}
\usepackage{booktabs}
\usepackage{algorithmic}
\usepackage{graphicx}
\usepackage{textcomp}
\usepackage{xcolor}
\usepackage{algorithmic}
\usepackage{textcomp}
\usepackage{xcolor}
\usepackage{fancyhdr}
\usepackage{microtype}[final]
\usepackage[normalem]{ulem}
\usepackage{tikz}
\usepackage{tcolorbox}
\usepackage[linesnumbered,ruled,vlined]{algorithm2e}
\usepackage{multirow}
\usepackage{colortbl}
\usepackage[bottom]{footmisc}
\usepackage{stfloats}
\usepackage{soulutf8}
\usepackage{enumitem}
\usepackage{multicol}
\usepackage{qcircuit}
\usepackage{fvextra}
\usepackage{forest}
\usepackage{booktabs}
\usepackage{listings}
\usepackage[labelformat=simple]{subcaption}

\renewcommand{\theenumii}{\theenumi.\arabic{enumii}.}

\newcommand{\subheading}[1]{\noindent \textbf{#1:}}

\newcommand{\rev}[1]{\textcolor{black}{#1}}

\definecolor{aliceblue}{rgb}{0.97, 0.96, 1.0}

\newtcolorbox{boxK}{
    sharpish corners, 
    boxrule = 0pt,
    toprule = 1pt, 
    enhanced,
    fuzzy shadow = {0pt}{-2pt}{-0.5pt}{0.5pt}{black!35} 
}

\newtcolorbox{boxH}{
    sharpish corners,
    boxrule = 0pt, 
    leftrule = 2pt 
}

\begin{document}

\title{A Case for Elastic Quantum Error Correction Decoders}

\author{Satvik Maurya}
\affiliation{
  \institution{University of Wisconsin-Madison}
  \city{Madison, WI}
  \country{USA}
}
\author{Abtin Molavi}
\affiliation{
  \institution{University of Wisconsin-Madison}
  \city{Madison, WI}
  \country{USA}
}
\author{Aws Albarghouthi}
\affiliation{
  \institution{University of Wisconsin-Madison}
  \city{Madison, WI}
  \country{USA}
}
\author{Swamit Tannu}
\affiliation{
  \institution{University of Wisconsin-Madison}
  \city{Madison, WI}
  \country{USA}
}

\begin{abstract}
\sloppy

Large-scale quantum computers promise transformative speedups, but their viability hinges on fast and reliable quantum error correction (QEC). At the center of QEC are \emph{decoders}—classical algorithms running on hardware such as FPGAs, GPUs, or CPUs that process error syndromes to detect errors every microsecond to preserve fault-tolerance. Quantum processors, therefore, operate not in isolation, but as accelerators tightly coupled with powerful classical digital hardware. A key challenge is that decoder demand fluctuates unpredictably: bursts of activity can require orders of magnitude more decodes than idle periods. Provisioning hardware for the worst case wastes resources, while provisioning for the average case risks catastrophic slowdowns. We show that this mismatch is a systems problem of capacity planning and scheduling, and propose a two-level framework that treats decoders as shared accelerators managed by the quantum operating system. Our approach reduces decoder requirements by 10--40\% across fault-tolerant benchmarks, demonstrating that efficient decoder scheduling is essential to making FTQC practical.

\end{abstract} 

\maketitle

\section{Introduction}


Quantum computers are not just faster machines; they are transformative domain-specific accelerators designed to solve problems that are fundamentally intractable for classical systems, enabling breakthroughs in cryptography, search, chemistry, and materials discovery~\cite{Shor1997, Grover1996, Lloyd1996, AspuruGuzik2005}.
However, their viability depends on continuous, effective quantum error correction (QEC). QEC protects information by repeatedly computing and tracking parities between qubits. QEC is necessary since quantum systems are intrinsically noisy: every gate, idle step, or measurement risks error, and without correction, noise will overwhelm computation. QEC is thus not just a hardware feature but the foundation of the software stack for fault-tolerant quantum computing (FTQC), where user programs run alongside a perpetual loop of error detection and correction. The need for FTQC systems has prompted comprehensive industry roadmaps to build warehouse-scale quantum computers by the end of this decade~\cite{GoogleQuantumRoadmap, IBMQuantumRoadmap, QuantinuumQuantumRoadmap}.

At the core of QEC are \emph{decoders}—classical algorithms that process measurement outcomes called \emph{syndromes} (bit-strings indicating which parity checks detected an error) to identify and correct errors. The preparation and measurement of one syndrome constitutes a QEC cycle. Decoding is generally NP-hard~\cite{iyer2015hardness} yet must finish within microsecond-level QEC cycles. Software alone cannot meet these demands on leading platforms such as superconducting qubits, making \emph{hardware decoders} essential~\cite{riverlaneLCD2024, Wu2025, Alavisamani2024, Vittal2023, muller2025improved, Das2022afs, Das2022lilliput, Ravi2023}. Industry efforts include NVIDIA’s CUDA-QX GPU-accelerated tensor-network decoders~\cite{nvidiaCUDAQX2025} and Riverlane’s “Deltaflow” FPGA/ASIC stack for real-time decoding across hundreds of qubits~\cite{riverlaneDeltaflow2025}. These decoders are costly, shared resources whose performance directly affects system throughput and reliability. Their role can be summarized by the “\textbf{four C’s}”:
\begin{itemize}[topsep=0pt, leftmargin=*]
\item \textit{Classical:} serve quantum systems but run on classical hardware (CPUs/FPGAs/ASICs/GPUs).
\item \textit{Configurable:} adapt to diverse codes and noise channels.
\item \textit{Critical:} real-time in lockstep with quantum hardware.
\item \textit{Catastrophic:} any delay or failure breaks fault-tolerance.
\end{itemize}

A natural way to accelerate decoding is through parallelism via \emph{windowed decoding}, where a decoding task is divided into smaller regions (“windows”) processed simultaneously~\cite{Skoric2023, FuhuiLin2025, Bombin2023}. We refer to highly parallel variants as \emph{parallel windowed decoding (PWD)}. While this reduces latency, it multiplies hardware demand, since each window needs its own decoder. Even one decoder per logical qubit is costly, but parallelism makes the requirement prohibitive. For example, a modest 32-bit quantum adder uses about 300 logical qubits; parallel decoding inflates demand by $3$–$5\times$, pushing the total beyond a thousand decoders even for a small quantum program. The burden is compounded because decoding is required not only for data qubits but also for numerous auxiliary qubits that support computation and movement. In many cases, auxiliaries even outnumber data qubits, further amplifying decoder demand. One major driver of this variability is \emph{lattice surgery}, a technique that implements logical operations by dynamically merging and splitting encoded logical qubits.

\begin{figure}[t]
    \centering
    \includegraphics[width=0.9\linewidth]{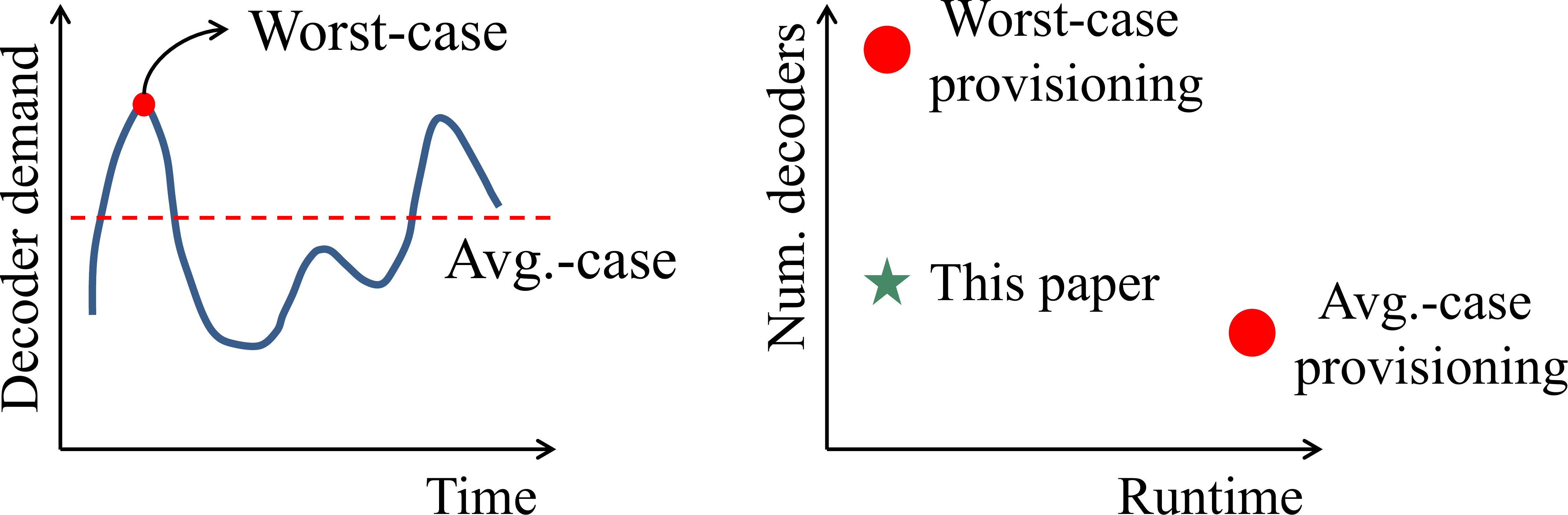}
    \caption{Decoder provisioning highlights the tension between worst-case (wasteful) and average-case provisioning (risky), motivating elastic decoder scheduling.}
    \Description[some figure]{}
    \label{fig:fig1}
    \vspace{-0.15in}
\end{figure}

All decoding tasks are not created equal -- some are more urgent than others. We define a \emph{critical decode} as one whose result is immediately required for program progress. Qubits actively involved in computation (and thus requiring critical decodes) must be decoded immediately to continue program computation, while idle memory qubits can tolerate some delay. This asymmetry creates an opportunity: we can time-multiplex decoders by prioritizing active qubits on the critical path while buffering and deferring decoding of idle qubits. By sharing a smaller pool of decoders across both active and idle qubits in this manner, we can reduce hardware overhead while meeting the strict real-time deadlines of FTQC.

From a systems perspective, this is a classic utilization problem analogous to accelerator scheduling in classical platforms such as GPUs, where finite compute units must be multiplexed across dynamic workloads. Figure~\ref{fig:fig1} shows that over-provisioning guarantees performance but wastes resources, while average-case provisioning risks severe stalls under bursts. We therefore introduce the notion of \emph{elastic decoders}: a scheduling abstraction in which a fixed physical pool of decoder hardware is dynamically allocated to logical qubits based on demand. This differs from the traditional systems notion of \emph{elastic resources}, which typically assumes that additional hardware can be provisioned on demand. In contrast, decoder elasticity refers strictly to time-multiplexing a fixed hardware budget. Our study shows that decoder demand is highly variable, driven by bursts of lattice surgery operations and sporadic non-Clifford gates. Static allocation fails under these conditions. Instead, we treat decoders as shared accelerators managed by the quantum operating system (QOS)~\cite{giortamis2025qos, CorriganGibbs2017}, and introduce scheduling policies that balance throughput, latency, and fairness.

This paper is the first to address the \emph{capacity planning} problem for FTQC: given a workload, how many hardware decoders should be provisioned to ensure timely error correction without over-provisioning costly resources? We argue that decoder allocation and scheduling are core responsibilities of the QOS. The decoding infrastructure is the backbone of any FTQC system, and is thus an integral part of the QOS.

\noindent \textbf{This paper makes the following contributions:}
\begin{itemize}[topsep=0pt, leftmargin=*]
\sloppy
    \item \textbf{Workload characterization.} We quantify decoder demand across a range of benchmarks (QFT, Shor, adders, Ising), showing that decoder demand comes in bursts and dominated by a small set of critical operations. This variability makes static allocation ineffective.
    
    \item \textbf{Capacity planning.} We present the first systematic study of how many hardware decoders are required to sustain FTQC workloads. We show that provisioning for the worst case wastes up to $40\%$ of decoders, while provisioning for the average case risks catastrophic slowdowns.
    
    \item \textbf{Scheduling framework.} We introduce a two-level scheduling design for the quantum operating system (QOS): coarse-grained scheduling to prioritize critical-path and spatially-coupled decodes, and fine-grained scheduling to fairly allocate remaining decoders across qubits.
    
    \item \textbf{Scheduling policies.} We evaluate several fine-grained policies—Most Frequently Decoded (MFD), Round-Robin (RR), and Minimize Longest Undecoded Sequence (MLS). MLS consistently achieves the best tradeoffs, reducing decoder needs by up to $19\%$ compared to RR.
    
    \item \textbf{Open-source workflow.} We release an open-source workflow~\cite{vader} that compiles programs into lattice surgery IR, tracks decoder demand, and enables reproducible evaluation of scheduling policies. Using this, we demonstrate that careful scheduling yields $10$--$40\%$ decoder savings even with long decoder tail latencies.
\end{itemize}

\section{Quantum Error Correction and Decoding}
\label{sec:background}




In this section, we cover high-level details of Quantum Error Correction (QEC) and the role of decoders.

\begin{figure}[t]
    \centering
    \begin{subfigure}[b]{0.29\linewidth}
        \begin{subfigure}[t]{\linewidth}
            \centering
            \includegraphics[width=0.8\linewidth]{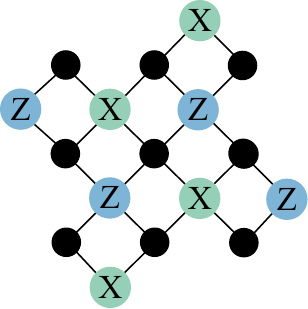}
            \caption{}
            \label{subfig:patch}
        \end{subfigure}
        \\
        \begin{subfigure}[t]{\linewidth}
            \centering
            \includegraphics[width=\linewidth]{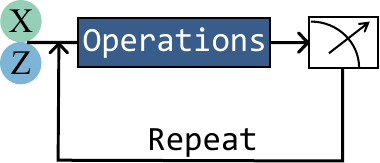}
            \caption{}
            \label{subfig:ops}
        \end{subfigure}
    \end{subfigure}
    \hfill
    \begin{subfigure}[b]{0.69\linewidth}
        \begin{subfigure}[t]{\linewidth}
            \centering
            \includegraphics[width=\linewidth]{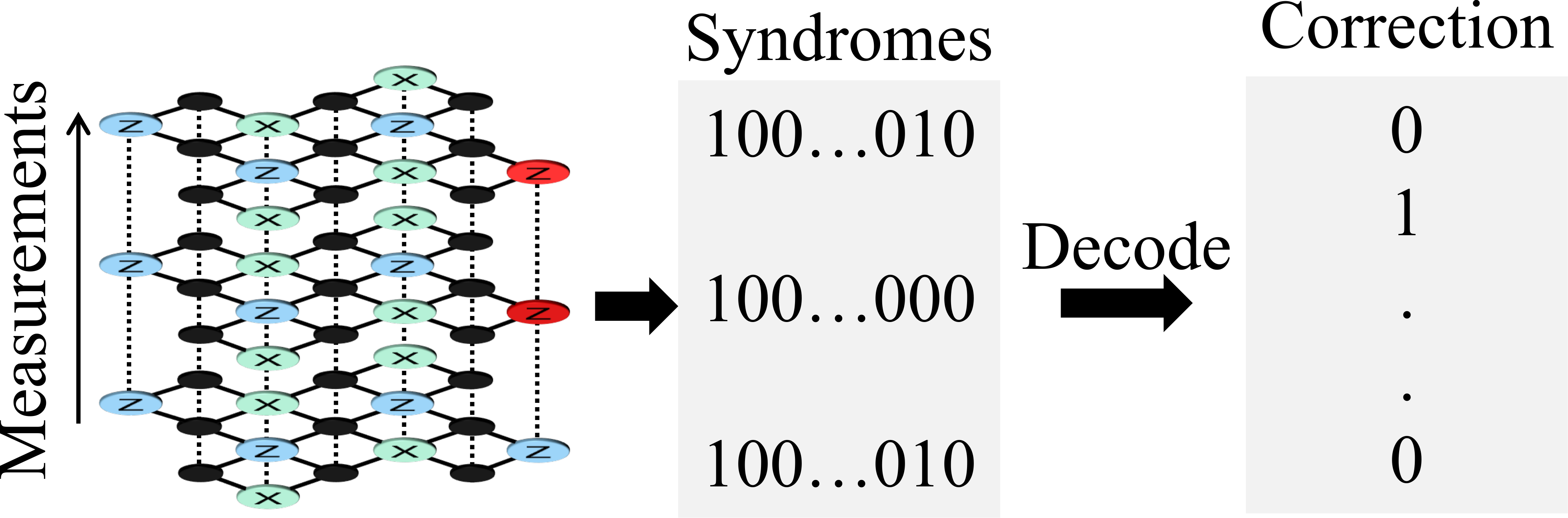}
            \caption{}
            \label{subfig:decode2}
        \end{subfigure}
        \\
        \begin{subfigure}[t]{\linewidth}
            \centering
            \includegraphics[width=\linewidth]{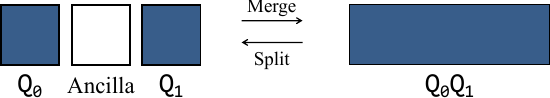}
            \caption{}
            \label{subfig:ls}
        \end{subfigure}
    \end{subfigure}
    
    \caption{
    \rev{
    (a) A surface code logical qubit (patch);
    (b) QEC works by repeatedly performing operations followed by measurements to generate syndromes;
    (c) Decoding with measurement errors -- multiple rounds of measurements are decoded collectively;
    (d) The fundamental split and merge operations of lattice surgery used for logical computation: each square is a logical qubit shown in (a).
    }
    }
    \Description[some figure]{}
    \label{fig:background}
    \vspace{-0.2in}
\end{figure}

\begin{figure*}[htbp]
    \centering
    \begin{subfigure}[b]{0.30\linewidth}
        \centering
        \includegraphics[width=\linewidth]{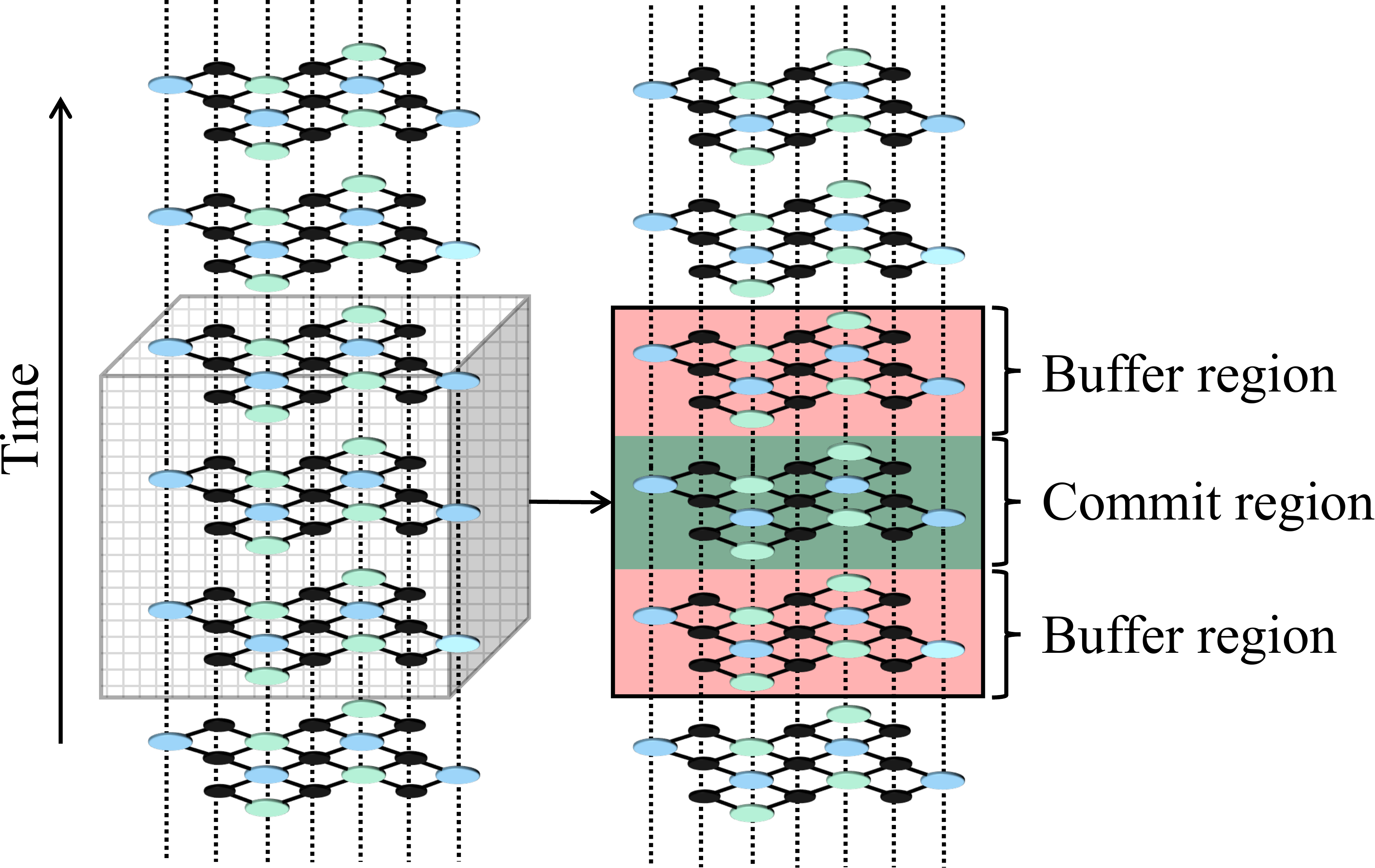}
        \caption{}
        \label{subfig:blocks_proc}
    \end{subfigure}
    \hfill
    \begin{subfigure}[b]{0.19\linewidth}
        \centering
        \includegraphics[width=\linewidth]{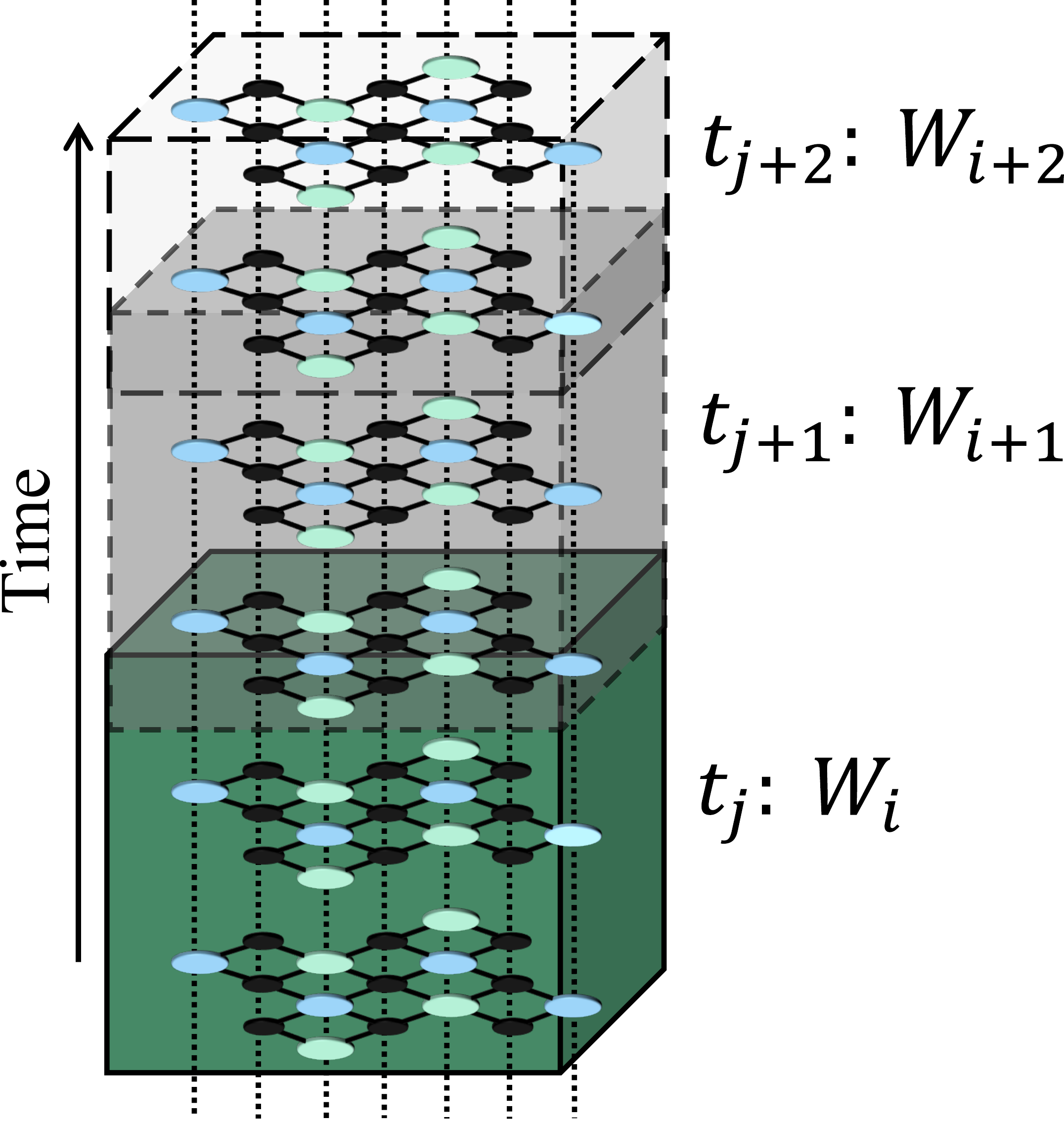}
        \caption{}
        \label{subfig:swd_proc}
    \end{subfigure}
    \hfill
    \begin{subfigure}[b]{0.19\linewidth}
        \centering
        \includegraphics[width=\linewidth]{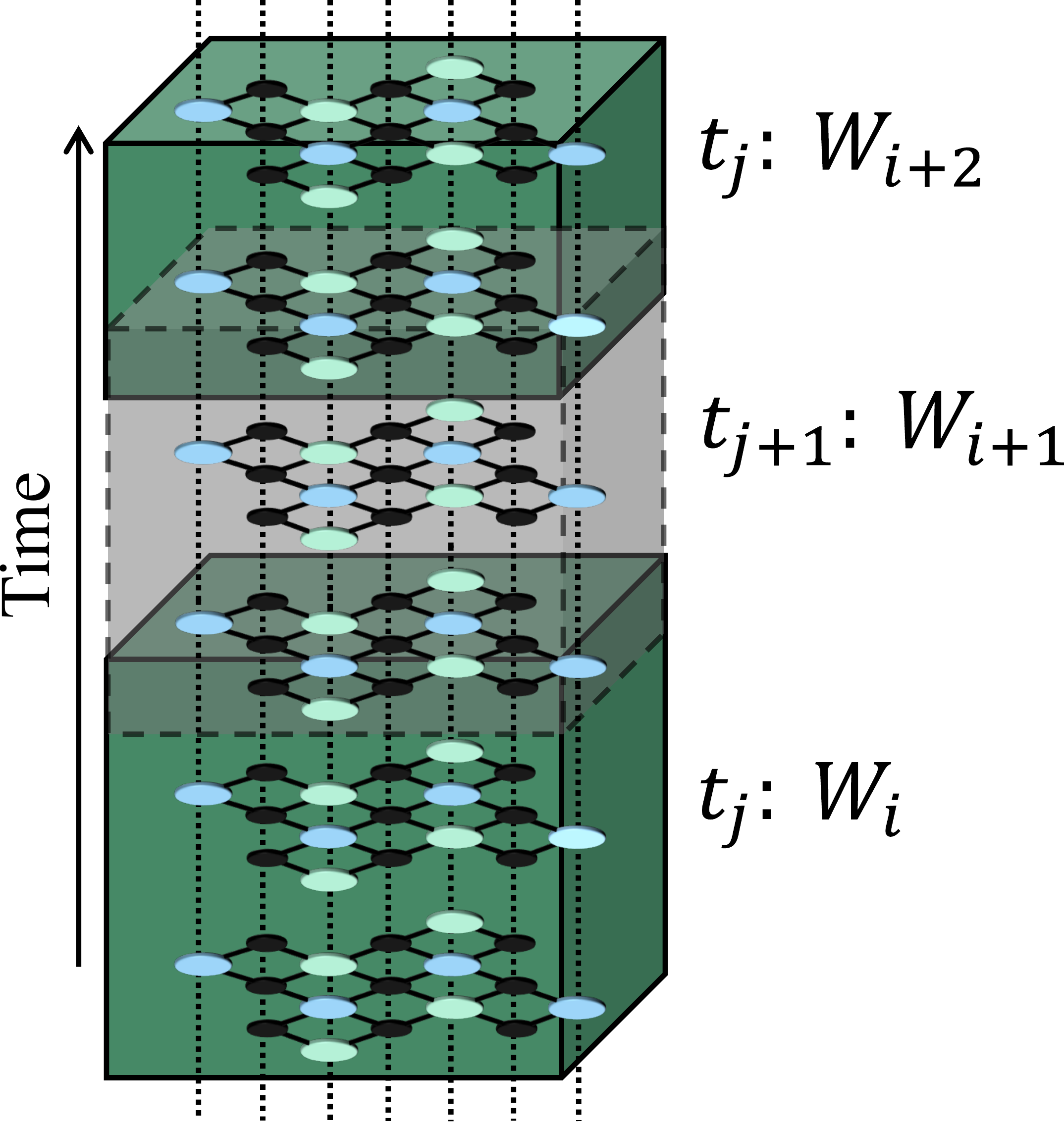}
        \caption{}
        \label{subfig:pwd_proc}
    \end{subfigure}    
    \hfill
    \begin{subfigure}[b]{0.30\linewidth}
        \centering
        \includegraphics[width=\linewidth]{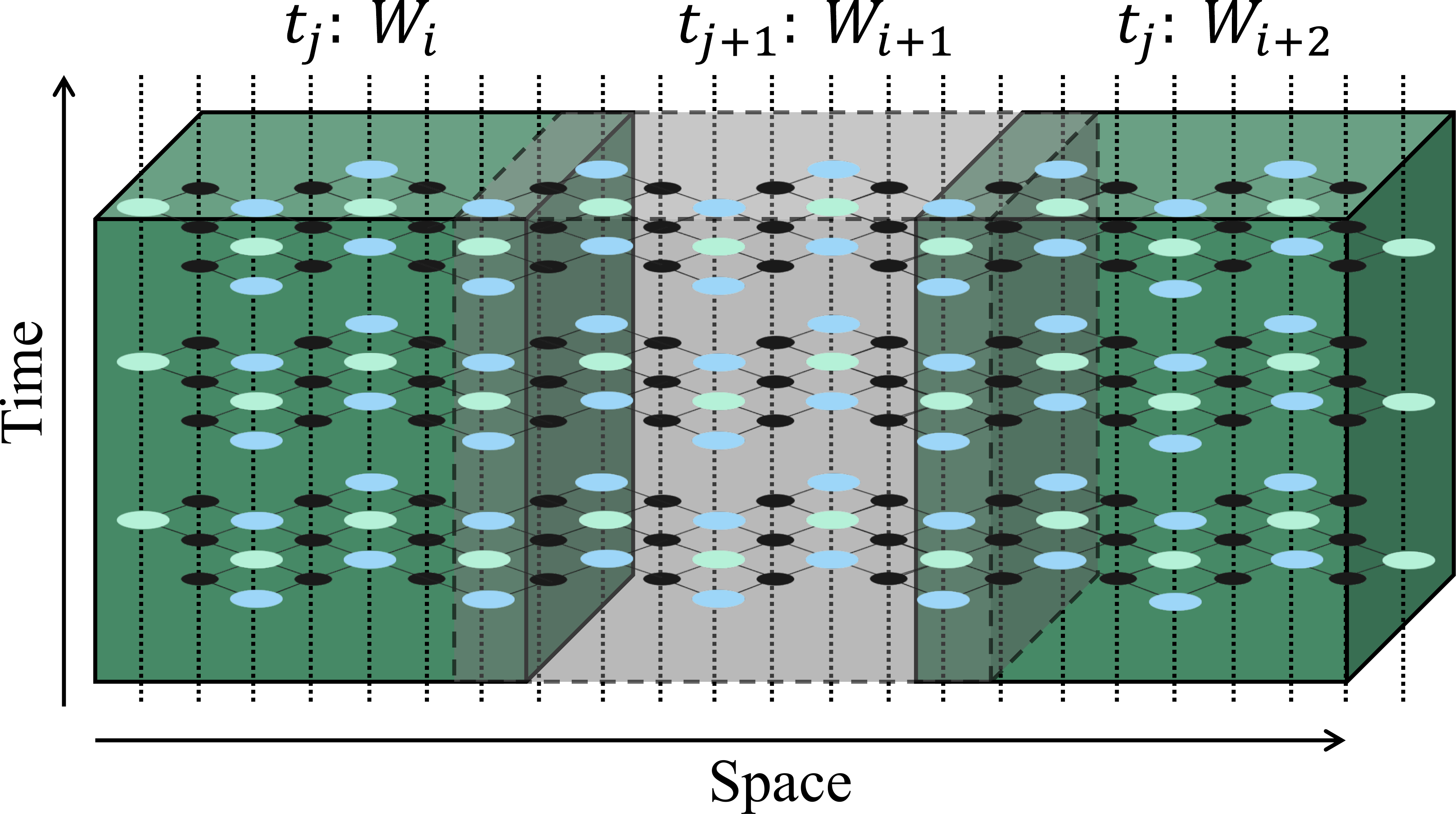}
        \caption{}
        \label{subfig:pwd_proc_spatial}
    \end{subfigure}
    
    \caption{
    (a) Streams of syndromes are processed in windows represented by \emph{blocks} which are fixed decoding volumes: each block consists of a region in which corrections are committed, while buffer regions are used for merging corrections with overlapping windows;
    (b) Sliding windowed decoding (SWD) is inherently sequential: window $W_{i+1}$ can only be processed after $W_i$ has been processed;
    Parallel windowed decoding (PWD) can be (c) temporal and/or (d) spatial: in either case, $W_{i}, W_{i+2}$ are treated as independent tasks and processed independently in parallel at time step $t_j$. $W_{i+1}$, which includes overlaps with $W_{i}, W_{i+2}$, is processed in the next time step.
     }
    \Description[some figure]{}
    \label{fig:processing}
    \vspace{-0.1in}
\end{figure*}

\def\lc{\left\lceil}   
\def\rc{\right\rceil}

\subsection{Quantum error correction}
Qubits are fragile: a variety of processes can introduce an error in a qubit's state. Such errors occur far more often than in classical storage (e.g., SRAM, DRAM, RAID)~\cite{Resch2021}, preventing reliable execution on today’s devices.
Quantum Error Correction (QEC) mitigates this by encoding a few logical qubits into many physical ones—similar in spirit to ECC in DRAM but able to correct both bit- and phase-flip errors.
Among candidate codes, the surface code~\cite{Fowler2012} is especially compelling for its compatibility with current hardware connectivity and is the focus of this work, though our ideas generalize to other codes.

QEC codes are defined by their code distance $d$, which determines how many errors can be reliably detected and corrected.
A surface code logical qubit of distance $d$, for example, can correct error chains up to length $\tfrac{d-1}{2}$.
Figure~\ref{subfig:patch} illustrates a $d=3$ surface code logical qubit comprising many physical qubits: data qubits (black) and check qubits ($X$ and $Z$, which detect phase- and bit-flips, respectively).
Each QEC cycle applies a sequence of gates (Figure~\ref{subfig:ops}) and measures all check qubits, producing syndromes that reveal the presence and type of errors for correction.
Every cycle takes $T_{cycle}$ time units.

\subsection{Decoding}

A QEC code is defined by its parity-check matrix $H$, which specifies the relationship between check and data qubits in the encoded logical qubit(s). Decoding can be expressed as solving a linear system for $F$: $$HF=\sigma$$ 
Where $\sigma$ is the measured syndrome and $F$ is a correction that restores the logical state. Since $H$ typically has more columns than rows, this system is \emph{under-determined}, yielding many possible corrections---a challenge unique to quantum codes, known as \emph{quantum degeneracy}. QEC decoders thus rely on maximum likelihood decoding (MLD), selecting the correction most probable for a given set of errors that can affect the system, making it an NP-Hard problem. 
Since every operation involving qubits is susceptible to errors, measurements of the check qubits can be erroneous too. For this reason, syndromes are collected over multiple measurement rounds and decoded collectively, as shown in Figure~\ref{subfig:decode2}.

\subsection{Logical operations}
\label{sec:logical_ops}
\sloppy

Once logical qubits are encoded, computation proceeds through logical operations between them. In the surface code, the leading method is \emph{lattice surgery} (LS)~\cite{Horsman2012, Fowler2019, Litinski2019}, which consists of two primitives (Figure~\ref{subfig:ls}): \emph{merge}, which combines multiple logical qubit patches into one, and \emph{split}, which divides a patch into two. Sequencing these primitives enables a wide range of logical operations.

\begin{equation*}
\Qcircuit @C=1em @R=1em {
\lstick{|Q_0\rangle} & \multigate{1}{Merge~\&~Split} & \qw & \qw & \gate{Clifford} & \qw & \rstick{|\psi\rangle} \\
\lstick{|Q_1\rangle} & \ghost{Merge~\&~Split} & \meter & \cw & \control \cw \cwx[-1]
}
\end{equation*}

In the circuit above, $Q_1$ is measured after lattice surgery and its outcome drives a corrective \emph{feed-forward} operation on $Q_0$. Since $Q_1$ is a logical qubit, its measurement outcome cannot be read directly --- it must first be inferred by a classical \emph{decoder} processing noisy syndrome data. This places the decoder on the critical path of execution: computation cannot proceed until decoding completes. We call such decodes \textbf{critical decodes}. This imposes a hard real-time requirement: the average decoding latency $T_{\mathrm{decode}}$ must satisfy $T_{\mathrm{decode}} < T_{\mathrm{cycle}}$, where $T_{\mathrm{cycle}}$ is the QEC cycle time. If violated, unprocessed syndromes accumulate and computation suffers an exponential slowdown --- the \textbf{backlog problem}~\cite{Terhal2015}.

\subsection{Processing streams of syndromes}
\label{sec:paradigms}
Logical qubits continuously generate syndromes that must be decoded.  
Instead of decoding all syndromes at once, decoding is performed on \emph{windows} of syndromes.  
Each window of size $n_W$ forms a 3-D decoding volume, visualized as a block in Figure~\ref{subfig:blocks_proc}.  
A block is divided into buffer and commit regions.  
Buffer regions are required to handle corrections spanning multiple overlapping blocks.  
Corrections fully contained in a commit region are finalized, while buffer-region corrections are deferred.  
With this block-based structure, syndrome streams can be decoded using the following paradigms.

\subsubsection{Sliding windowed decoding (SWD)}
SWD is inherently sequential, as shown in Figure~\ref{subfig:swd_proc}.  
At time step $t_j$, the decoder processes window $W_i$.  
Once complete, the next window is processed at the following step.  
The overlap between $W_i$ and $W_{i+1}$ corresponds to their buffer regions.

\subsubsection{Parallel windowed decoding (PWD)}
PWD~\cite{Skoric2023} extends SWD by exploiting temporal (Figure~\ref{subfig:pwd_proc}) and spatial (Figure~\ref{subfig:pwd_proc_spatial}) parallelism.  
In Figure~\ref{subfig:pwd_proc}, windows $W_i$ and $W_{i+2}$ are independent and can be decoded simultaneously with two decoders.  
Window $W_{i+1}$ is processed in the next step, with buffers overlapping those of $W_i$ and $W_{i+2}$.  
Similarly, in Figure~\ref{subfig:pwd_proc_spatial}, a large patch formed by lattice surgery can be decoded using spatial parallelism~\cite{FuhuiLin2025}.  
PWD increases decoding throughput but requires more decoders.  
As shown in Figures~\ref{subfig:pwd_proc},~\ref{subfig:pwd_proc_spatial}, compared to SWD, at least one extra decoder is needed.  
In general, decoder requirements grow linearly with the size of the decoding task in both temporal and spatial dimensions.

\begin{figure}
    \begin{subfigure}[b]{0.33\linewidth}
        \centering
        \includegraphics[width=\linewidth]{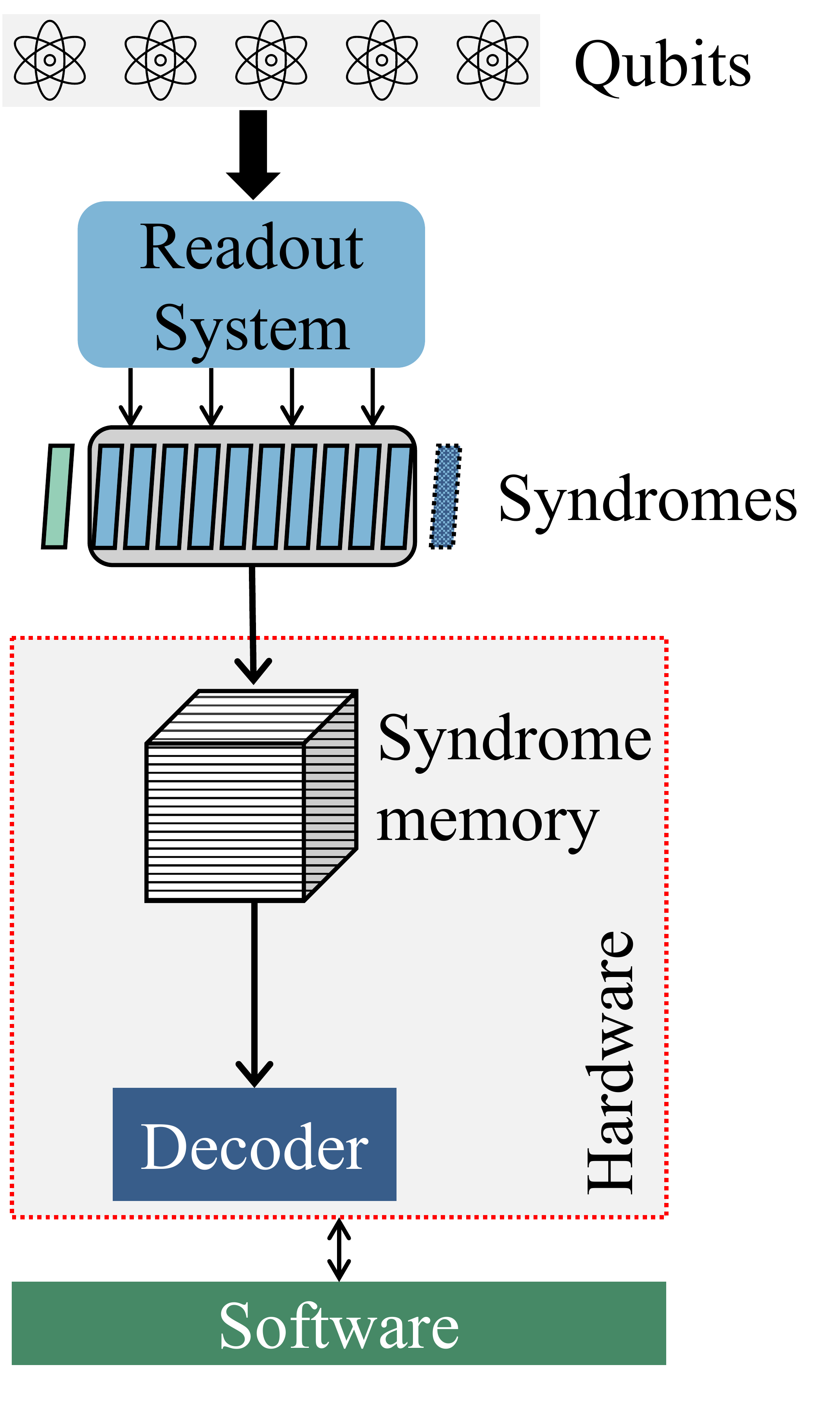}
        \caption{}
        \label{subfig:pipeline}
    \end{subfigure}
    \hfill
    \begin{subfigure}[b]{0.65\linewidth}
        \centering
        \includegraphics[width=\linewidth]{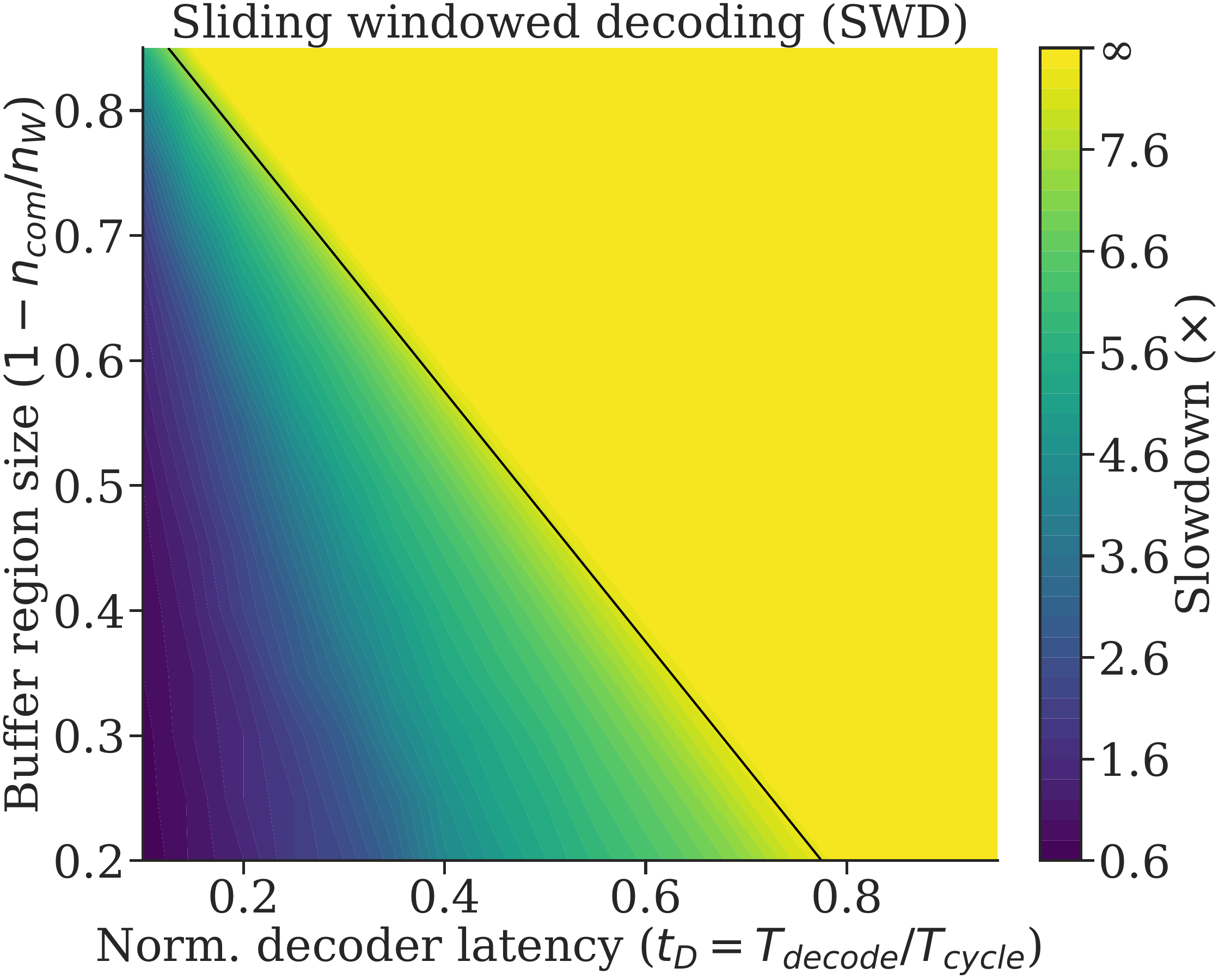}
        \caption{}
        \label{subfig:swd}
    \end{subfigure}
    
    \caption{
    (a) A full system using hardware (FPGA/GPU/ASIC) decoders: qubits are measured using the readout system, which produces syndromes that are buffered until a window has been collected. Then, the decoder uses these collected syndromes to determine a correction, which is communicated to software;
    (b) Effect of the normalized decoder latency $t_D$ on the slowdown in processing $5d$ rounds of syndromes using SWD. The buffer region size is the normalized size of the buffer regions required ($n_W=3d, n_{com}=d$ would yield an overlap of 0.66). }
    \Description[some figure]{}
    \label{fig:motivation}
    \vspace{-0.2in}
\end{figure}

\begin{figure*}[b]
    \vspace{-0.1in}
    \centering
    \begin{subfigure}[b]{0.32\linewidth}
        \centering
        \includegraphics[width=\linewidth]{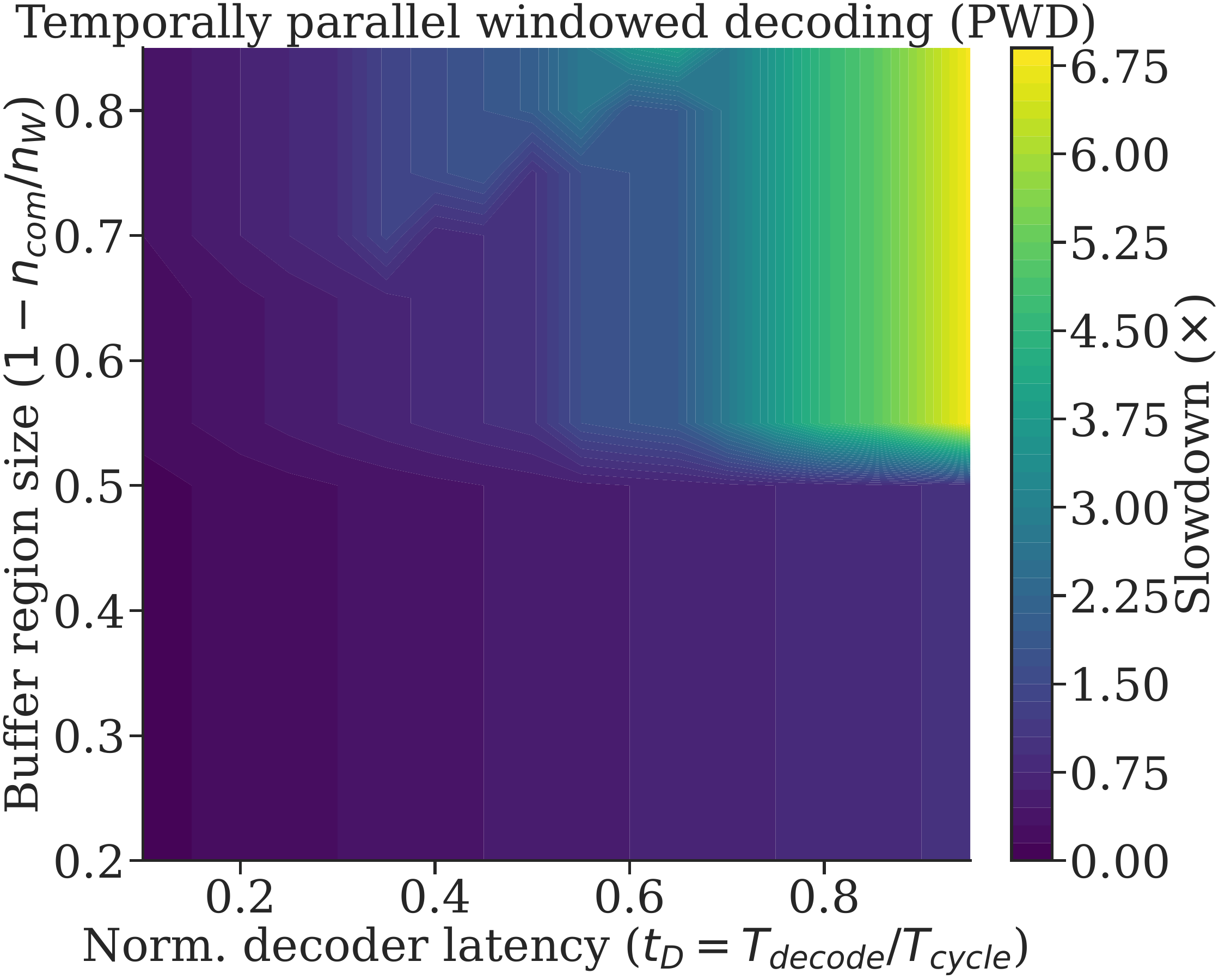}
        \caption{}
        \label{subfig:tpwd}
    \end{subfigure}
    \hfill
    \begin{subfigure}[b]{0.25\linewidth}
        \centering
        \includegraphics[width=\linewidth]{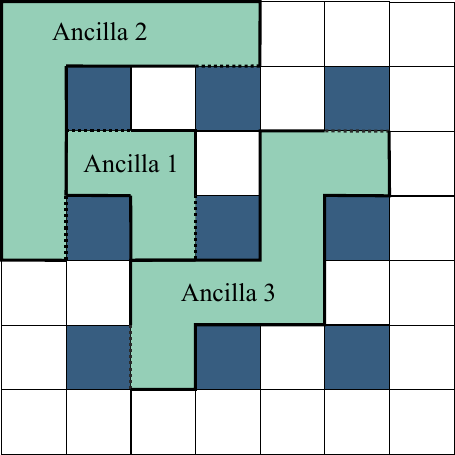}
        \caption{}
        \label{subfig:routing}
    \end{subfigure}
    \hfill
    \begin{subfigure}[b]{0.32\linewidth}
        \centering
        \includegraphics[width=\linewidth]{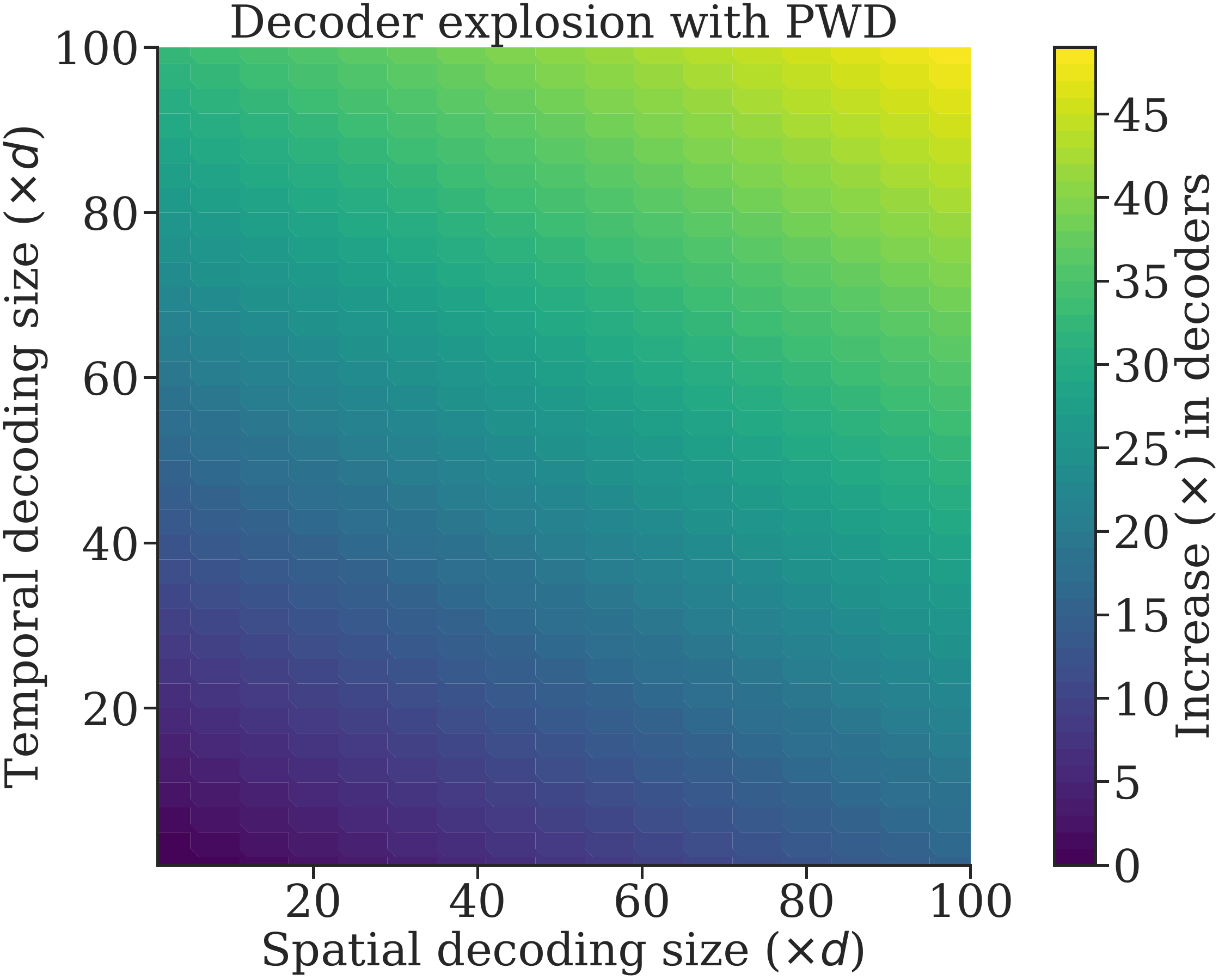}
        \caption{}
        \label{subfig:pwd_explodes}
    \end{subfigure}    
    
    \caption{
    (a) Using simulations, effect of the normalized decoder latency $t_D$ on the slowdown in processing $5d$ rounds of syndromes using Temporally parallel windowed decoding (PWD);
    (b) Logical qubits arranged in the EDPC layout with ancilla bridges of varying sizes to facilitate LS operations;
    (c) PWD is applicable in both space \emph{and} time -- this leads to an a variable and sudden increase in the number of decoders required for implementing PWD (assuming $t_W=3d$).
     }
    \Description[some figure]{}
    \label{fig:heats}
\end{figure*}

\section{Decoder Allocation is Hard}
\label{sec:classical_req}
In this section we will discuss the need for accelerated, hardware decoders, and how it is non-trivial to allocate these decoders for a system in a resource-efficient manner.

\subsection{Hardware decoders}
The requirement for syndromes to be processed faster than they are generated has motivated research to build fast and accurate hardware decoders for the surface code~\cite{Smith2023, Ravi2023, Riverlane2023, Vittal2023, Alavisamani2024, Wu2025, riverlaneLCD2024}, especially for systems using superconducting qubit architectures due to their fast gate times. GPU-accelerated decoding has also been demonstrated for neutral atom systems~\cite{nvidia_quera}. Figure~\ref{subfig:pipeline} shows the general structure of a real-time decoding system that uses accelerated hardware decoders. Syndromes are generated after readout, and these syndromes are buffered until a decoder can operate on them. Such hardware systems are necessary for ensuring that the backlog problem is avoided~\cite{Terhal2015}. Figure~\ref{subfig:pipeline} also shows that $T_{decode}$ is dependent on not just the actual decoder latency but also the communication/memory access latencies required for storing and providing the syndromes to the decoder. 

\subsection{SWD is not scalable}
As discussed in Section~\ref{sec:paradigms}, decoding in the SWD paradigm is inherently sequential.  
This makes SWD too slow and fundamentally unscalable for large systems.  
Figure~\ref{subfig:swd} illustrates this issue for $5d$ rounds of pending syndromes from a single logical qubit.  
The figure shows slowdown as a function of buffer region size and normalized decoder latency $t_D = T_{decode}/T_{cycle}$. Slowdown accounts for both the initial $5d$ rounds of pending syndromes and the additional syndromes generated during their processing; we measure the time required for the decoder to catch up to the live syndrome stream. Buffer regions are expressed in multiples of code distance $d$, so a buffer of size $d$ corresponds to $d$ rounds of overlap between adjacent windows.
Larger buffer regions improve accuracy in windowed decoding~\cite{Skoric2023, FuhuiLin2025}, but also increase slowdown.  
The results are clear: SWD avoids the backlog problem only for unrealistically small buffer sizes and very optimistic $t_D$.  
For instance, $t_D=0.2$ in a superconducting system with $T_{cycle}\simeq 1\,\mu s$ requires an average decoder latency of $200$\,ns — far below what any proposed hardware decoder has achieved.  
When $t_D>0.8$, the slowdown diverges, meaning the decoder can never catch up with the incoming syndrome stream.

\subsection{PWD introduces non-determinism}

Figure~\ref{subfig:tpwd} shows the same initial problem used for the SWD paradigm in Figure~\ref{subfig:swd}: even for values of $t_D$ close to one, the slowdown is still finite. This example clearly highlights the utility and scalability of the PWD paradigm -- Figure~\ref{subfig:tpwd} would be the same for any initial size of pending syndromes.

This instance of PWD exploits \emph{temporal} parallelism: windows in time are executed in parallel to increase decoder throughput. However, as discussed in Section~\ref{sec:paradigms}, PWD can also have a spatial interpretation. We now discuss how the spatial interpretation of PWD is non-deterministic and its interplay with the temporally PWD.

\subsubsection{The impact of gate routing}
Logical operations in the surface code are performed via lattice surgery, where qubits are merged through ancilla regions that form routing paths. 
The size of these ancilla bridges varies, leading to unpredictable decoding volumes. 
Figure~\ref{subfig:routing} illustrates this in the EDPC layout~\cite{Beverland2022edpc}, where blue squares are logical qubits and white squares are ancillas. 
Ancilla~1 connects nearby qubits with a short bridge of length~3. 
Ancilla~2 connects similarly adjacent qubits but requires a longer path of length~7 to avoid Ancilla~1. 
Ancilla~3, unaffected by others, also has length~7 since it spans distant qubits. 
Such routing conflicts and varying patch sizes create non-deterministic decoder demands, as the number of required decoders scales linearly with patch size (Section~\ref{sec:paradigms}).

\subsubsection{Interplay of temporal and spatial requirements}
Consider the logical program shown in Section~\ref{sec:logical_ops}. 
This lattice surgery operation could be between qubits that have either a large temporal decoding task (pending syndromes), a large spatial decoding task (due to the physical distance and the subsequent routing between them), or both.
We attempt to capture this interplay between the spatial and temporal decoder requirements in Figure~\ref{subfig:pwd_explodes}. 
In the simplest case, both qubits do not have any pending syndromes, and they are next to each other on the device, which means that they need at most a decoder each to decode all generated syndromes. 
However, if there is a spatial and/or temporal element to the decoding task, the number of decoders required can explode to up to 50$\times$, just for two logical qubits!

\begin{figure}[t]
    \centering
    \includegraphics[width=\linewidth]{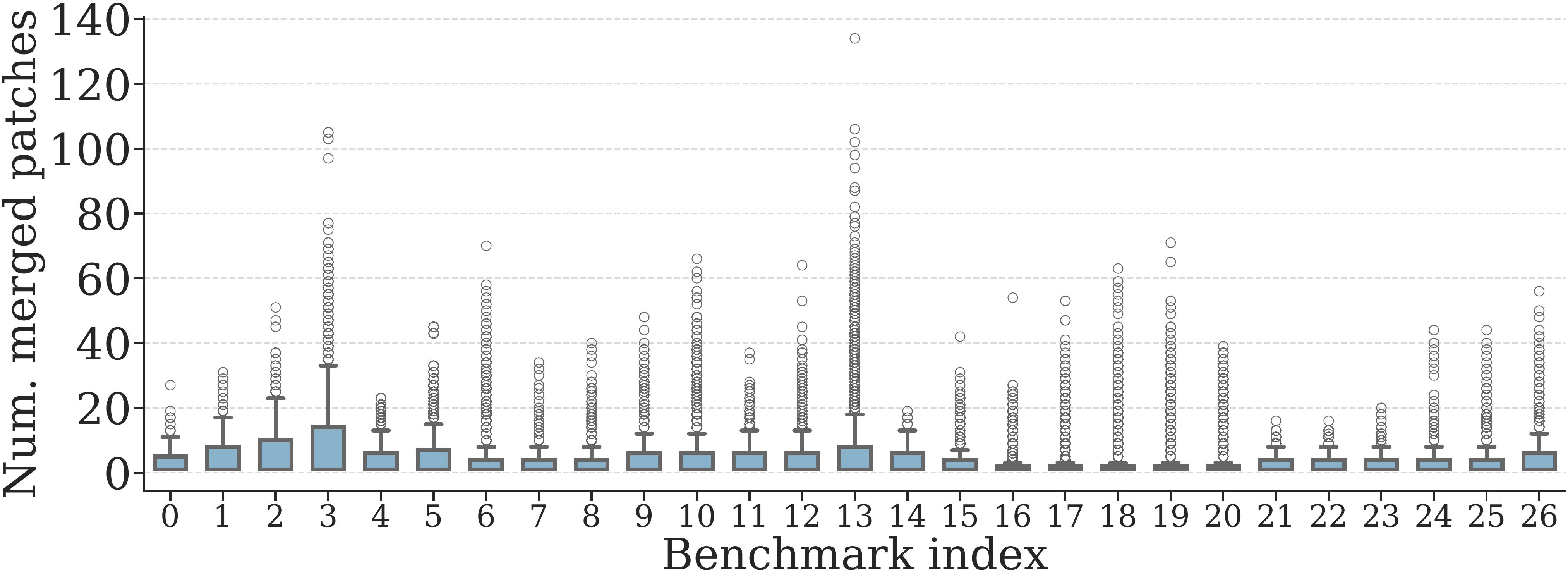}
    \caption{Distribution of the number of patches merged to perform lattice surgery-based logical measurements for different workloads -- most workloads require short-range interactions on average, but some large outliers are possible due to routing constraints.}
    \label{fig:distances}
    \Description[some figure]{}
    \vspace{-0.1in}
\end{figure}

\begin{figure}[t]
    \centering
    \includegraphics[width=\linewidth]{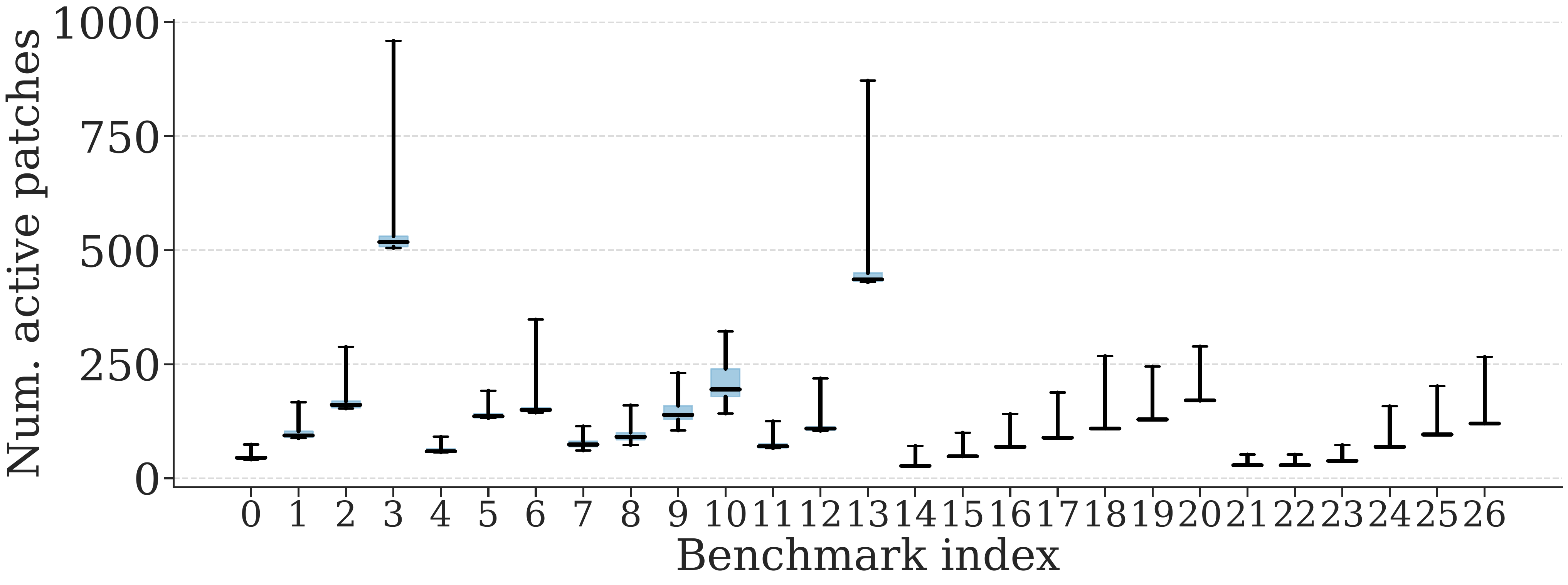}
    \caption{Distribution of the decoder demand over time for different benchmarks -- these error bars capture the fluctuations in active logical patches that require decoding during the computation.}
    \label{fig:demand}
    \Description[some figure]{}
    \vspace{-0.1in}
\end{figure}

\subsection{Need for decoder scheduling}
In the sections above, we showed that the number of decoders required under the PWD paradigm can be highly variable and, in some cases, untenable when both spatial and temporal decoding demands grow.  
Figure~\ref{fig:distances} highlights this variability in the spatial dimension: different benchmarks (see Section~\ref{sec:framework} for more information on benchmarks) exhibit a wide range in the number of merged patches, with significant outliers.  
This directly affects the number of logical qubits active at any given time, as illustrated in Figure~\ref{fig:demand}, which shows the distribution of active qubits across benchmarks.  
Since every logical qubit continuously produces syndromes that must be decoded, the key challenge becomes:  

\begin{boxH}
How do we allocate decoders for benchmarks with such diverse and dynamic demands?
\end{boxH}

The simplest allocation strategies are to provision (a) for the worst-case demand (i.e., the maximum number of active qubits from Figure~\ref{fig:demand}), or (b) for the average-case demand.  
Option (a) is prohibitively resource-intensive, since each decoder may require nearly all the resources of a single FPGA~\cite{Wu2025, riverlaneLCD2024}.  
Option (b) risks substantial slowdowns: spatial decoding tasks of large size would be delayed, reducing throughput and degrading program performance.  

Thus, static allocation is insufficient.  
The problem of decoder allocation is better viewed as a \emph{scheduling problem}, where a finite pool of decoders must be dynamically assigned to decoding tasks over time.  
Like CPU scheduling in classical computers, decoder scheduling must balance throughput, latency, and fairness, while avoiding bottlenecks from workload spikes.  
In this paper, our goal is to determine how many decoders are needed and how they should be scheduled to ensure that benchmarks execute without slowdowns.

\begin{figure}[t]
    \centering
    \includegraphics[width=\linewidth]{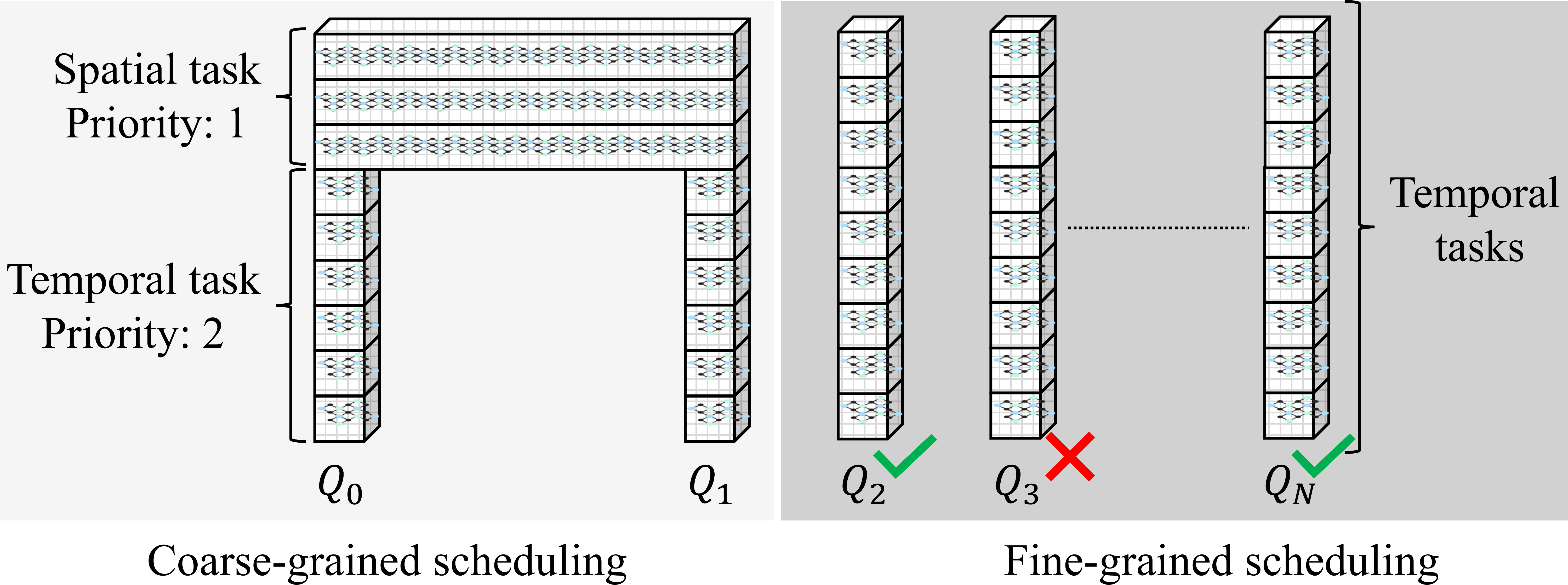}
    \caption{Distinction between coarse- and fine-grained scheduling. (Left) Coarse-grained decoder scheduling handles the prioritization between spatial and temporal decoding tasks. This includes handling critical decodes, which will always have some spatial component to them due to lattice surgery. (Right) Any decoders left over after handling critical decoding tasks are assigned to the temporal decoding tasks associated with logical qubits not involved in any operation in that time step. Scheduling decoders among this subset of qubits is what we refer to as fine-grained scheduling.}
    \label{fig:coarse_fine}
    \Description[some figure]{}
    \vspace{-0.1in}
\end{figure}

\begin{figure*}[b]
    \vspace{-0.1in}
    \centering
    \begin{subfigure}[b]{0.08\linewidth}
        \centering
        \includegraphics[width=\linewidth]{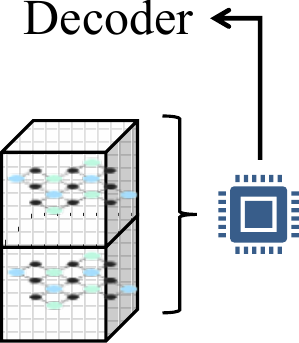}
        \caption{}
        \label{subfig:volume_1}
    \end{subfigure}
    \hfill
    \begin{subfigure}[b]{0.08\linewidth}
        \centering
        \includegraphics[width=\linewidth]{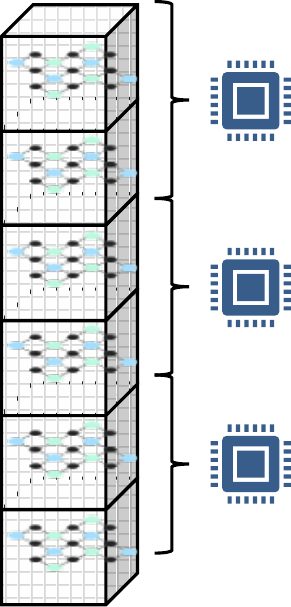}
        \caption{}
        \label{subfig:volume_1_time}
    \end{subfigure}
    \hfill
    \begin{subfigure}[b]{0.36\linewidth}
        \centering
        \includegraphics[width=\linewidth]{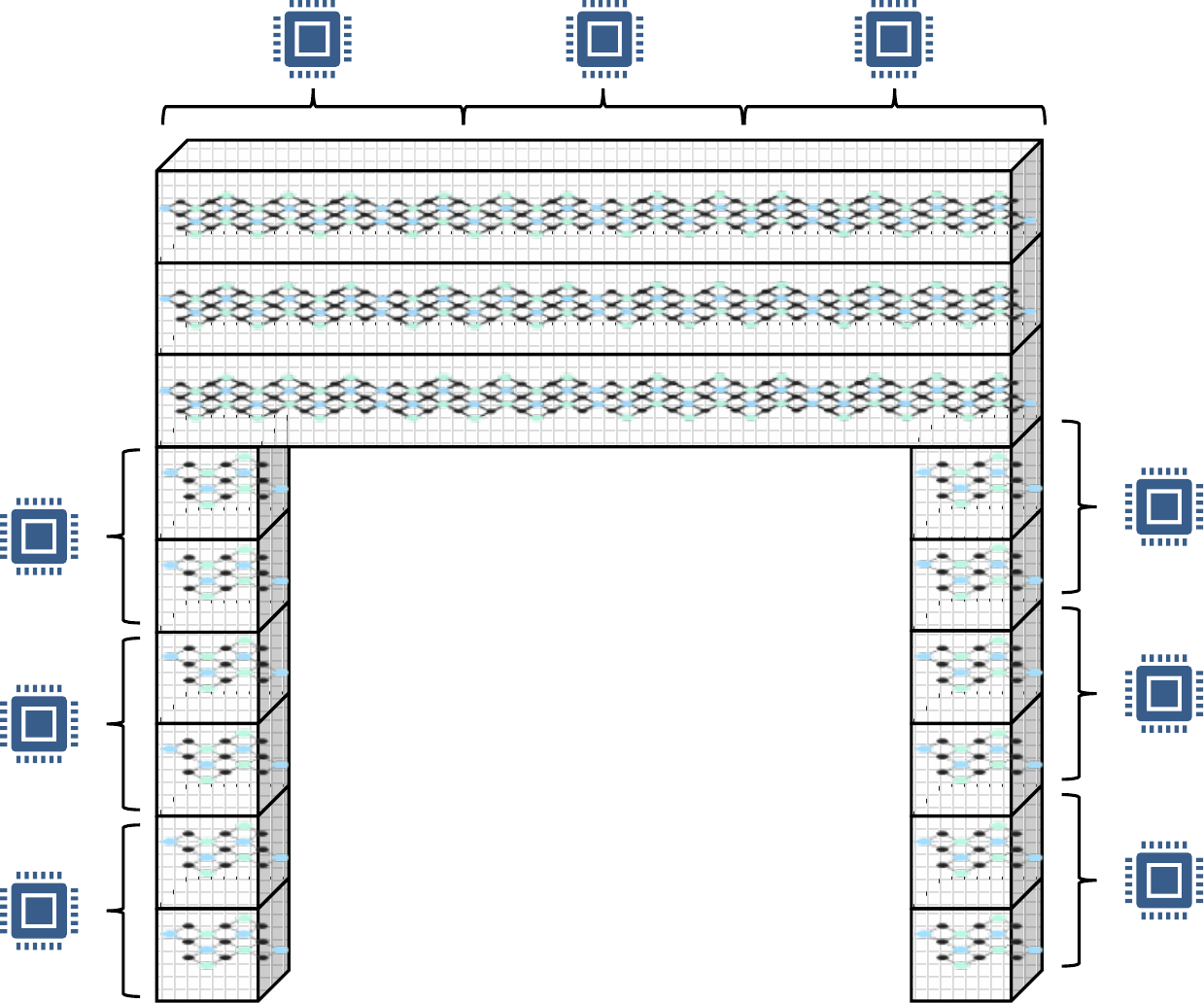}
        \caption{}
        \label{subfig:volume_space_time}
    \end{subfigure}
    \hfill
    \begin{subfigure}[b]{0.46\linewidth}
        \centering
        \includegraphics[width=\linewidth]{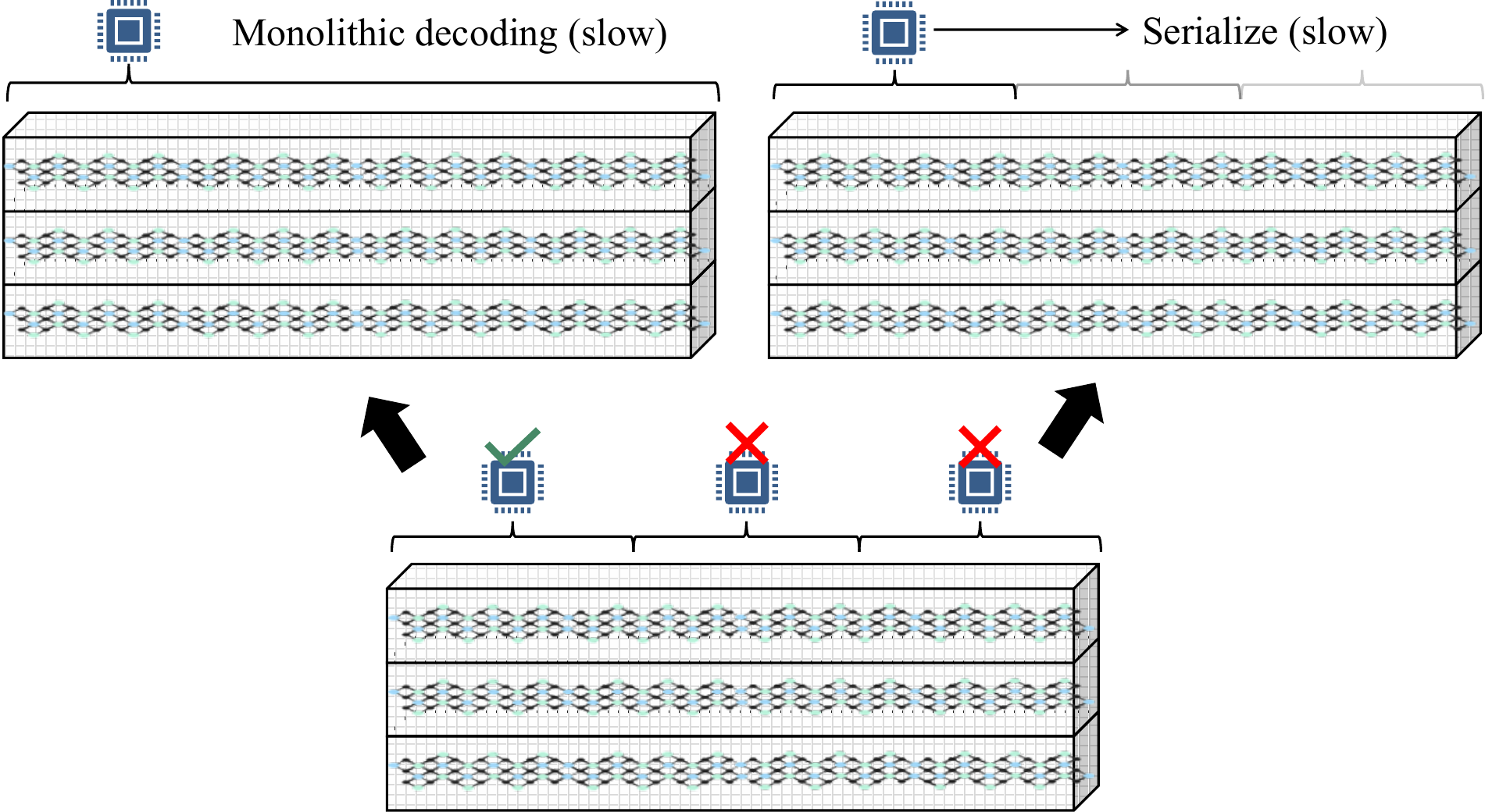}
        \caption{}
        \label{subfig:spatial_delay}
    \end{subfigure}
    
    \caption{
        (a) Decoding volume of a single logical patch idling for a few rounds—only one decoder is needed; 
        (b) Idling for multiple rounds creates a linear growth in windows, requiring temporally parallel windowed decoding (PWD); 
        (c) A lattice surgery merge between distant patches temporarily increases decoder demand due to spatial parallelism, which may coincide with temporal parallelism (overlapping windows omitted for clarity); 
        (d) Spatial PWD is not robust to decoder shortages—if two of three required decoders are unavailable, one must either decode the merged patch monolithically or serialize the protocol, both slow and undesirable. Deferring the task is also problematic, as the next slice may contain a critical decode that must wait for the deferred task, further slowing execution.
    }

    \label{fig:decoding_vols}
    \Description[some figure]{}
\end{figure*}




\section{Coarse-Grained Scheduling}
\label{sec:coarse-grained}

We define decoder scheduling at two levels—coarse-grained and fine-grained—as illustrated in Figure~\ref{fig:coarse_fine}.
Coarse-grained scheduling prioritizes critical decodes and resolves conflicts between spatial and temporal tasks on the same qubits.
For example, in Figure~\ref{fig:coarse_fine}, the spatial decode on qubits $Q_0, Q_1$ from a lattice surgery operation takes precedence over the temporal decode.
After coarse-grained allocation, remaining decoders are assigned to other qubits.
In the figure, $Q_2$ and $Q_N$ receive decoders while $Q_3$ does not, demonstrating fine-grained scheduling, which selects among tasks from individual qubits.
Because critical and spatial decodes impose rigid constraints, we adopt a single coarse-grained policy to maximize utilization, while Section~\ref{sec:scheduling} presents multiple fine-grained policies that operate in tandem.



\subsection{Decoding space-time volumes}
Figures~\ref{subfig:volume_1},~\ref{subfig:volume_1_time}, and~\ref{subfig:volume_space_time}, show space-time decoding volumes for three distinct scenarios under the PWD paradigm. Decoding volumes could be simple in both space and time, or require multiple decoders for a large volume in time, or have a large volume in \emph{both} space and time. Each case shows how the decoding volume can be different in the same program, which requires decoders to be \textit{elastic} such that an increase or decrease in the demand can be dealt with without any detrimental effects on the performance of a program.

\begin{figure}[t]
    \centering
    \begin{subfigure}[b]{\linewidth}
        \centering
        \includegraphics[width=\linewidth]{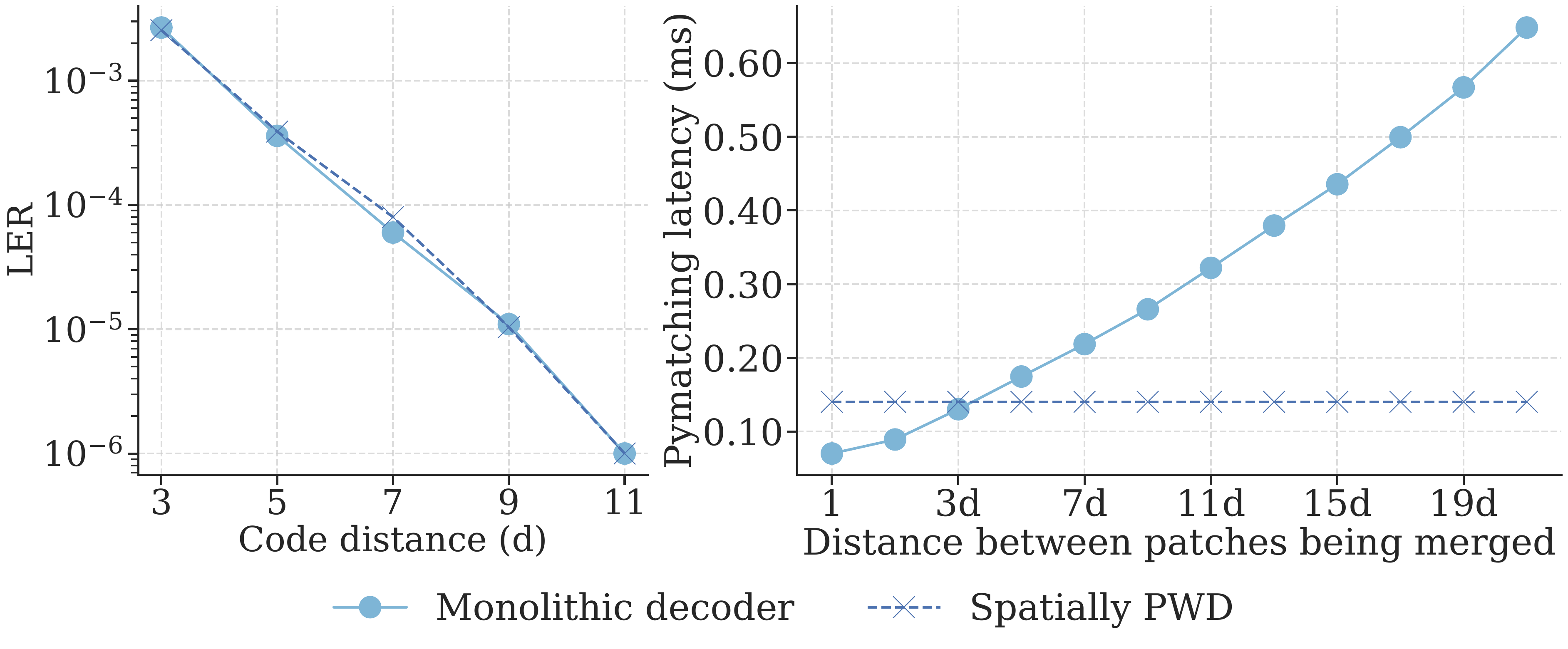}
        \caption{}
        \label{fig:monolithic-spwd}
    \end{subfigure}
    \\
    \begin{subfigure}[b]{\linewidth}
        \centering
        \includegraphics[width=\linewidth]{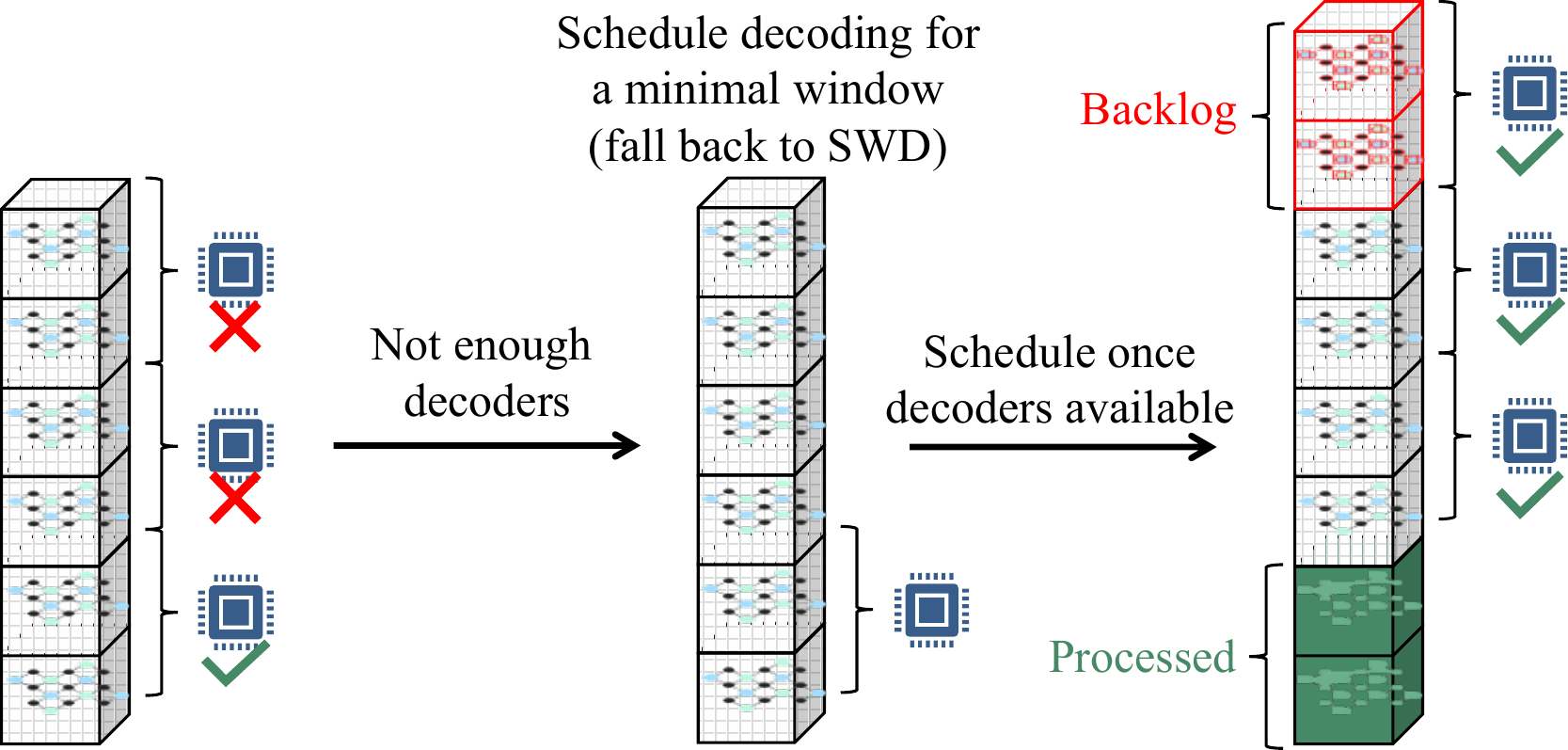}
        \caption{}
        \label{subfig:temporal_delay}
    \end{subfigure}
    \caption{
    (a) (Left) Logical error rates (LER) for various lattice surgery configurations with and without spatially parallel window decoding (PWD) at circuit-level noise $p=0.1\%$. (Right) Spatial parallelism does not affect LER but reduces decoder latency when merged patches are separated by $\geq 3d$ data qubits, assuming sufficient decoders for all distances (for $d=11$, window size $=3d$).  
    (b) Temporal PWD remains robust under decoder resource constraints: with only one of three decoders available, the decoding task can be processed incrementally by the single decoder. This creates a qubit backlog, which can be rapidly cleared once additional decoders become available via temporal parallelism.
    }

    \label{fig:pwd_results}
    \Description[some figure]{}
    \vspace{-0.1in}
\end{figure}

\subsection{Prioritizing between spatial and temporal parallelism}
As shown in Figure~\ref{subfig:volume_space_time}, lattice surgery can require decoders for both spatial and temporal parallelism, raising a key scheduling question: \textbf{Which type of parallelism should take priority when both are needed simultaneously?}  

Consider first the effect of decoder scarcity on spatial PWD. Suppose a large patch requires three decoders: the first layer uses all three for non-overlapping windows, while the second layer reuses two to handle overlaps and boundary errors~\cite{Bombin2023, FuhuiLin2025}. If only one decoder is available, the protocol cannot proceed as designed, forcing one of three fallbacks: (i) switch to a monolithic decoder, (ii) serialize window decoding, or (iii) defer decoding to the next slice. Option~(iii) risks slowdown if followed by a critical non-Clifford operation. As Figure~\ref{fig:pwd_results} shows, spatial PWD and monolithic decoding yield similar logical error rates (LERs), but monolithic decoding incurs far higher latency for large patches\footnote{Spatial PWD is only beneficial beyond $3d$ data qubits (or $\sim 3$ patches); smaller patches are treated as monolithic in our evaluations.}. While demonstrated using PyMatching~\cite{Higott2023}, the same trends hold for hardware decoders. Thus, fallback options (i) and (ii) degrade performance, and spatial PWD loses its latency advantage without concurrent decoder availability.  

In contrast, temporally PWD is more resilient to decoder constraints. As a parallelized form of sliding windowed decoding (SWD), it can always fall back to SWD when decoders are scarce. Available decoders are simply assigned to the oldest undecoded windows, ensuring correctness. Once more decoders free up, temporally PWD clears accumulated backlogs in \emph{constant time}, provided sufficient parallelism is available. This adaptability is its main strength: \textbf{as long as delays do not exceed the system’s recovery capacity, efficient scheduling remains feasible despite temporary decoder shortages.}



\subsection{Distinguishing between spatial decoding tasks}
Lattice surgery operations that require spatially PWD can either be due to critical decodes arising due to non-Clifford gates, or due to other operations between $Y$-states~\cite{watkins2023high}. 
Critical decodes must be serviced, which is why those spatial decoding tasks are given the highest priority, followed by any temporal decoding tasks needed for the critical decode. If there are non-critical spatial decoding tasks required at the same time step, they are given a priority lower than critical decodes but higher than that of decodes needed by logical qubits not involved in any operation in that time step. 
This allows spatial decoding tasks, critical or not, to be prioritized over all other temporal decoding tasks.

\section{Fine-Grained Scheduling Policies}
\label{sec:scheduling}


In this section, we define fine-grained decoder scheduling policies. These policies act on decoders that are available after all other higher priority spatial decoding tasks have been allocated decoders. These fine-grained policies are similar to process scheduling policies used by modern operating systems where decoding tasks of individual qubits are analogous to processes and decoders are analogous to CPU cores.

\subsection{Connection with real-time scheduling systems research}

Decoder scheduling maps naturally to a \textbf{mixed-criticality, soft real-time model}~\cite{Baruah2011, Vestal2007}: decoding windows are sporadic tasks with bursty arrivals; critical decodes (from non-Clifford gates) impose firm deadlines on dependent computation; non-critical decodes tolerate bounded delay; and the decoder pool represents a set of identical parallel processors. Our two-level scheduler mirrors hierarchical scheduling architectures~\cite{Davis2011}: coarse-grained scheduling handles criticality levels (critical > spatial > temporal), analogous to priority-based admission control; fine-grained policies allocate residual capacity among admitted tasks for backlog control.

While conventional real-time systems research often provides hard schedulability guarantees (e.g., rate-monotonic analysis~\cite{liu1973scheduling}), this approach is unattainable for QEC decoders due to fundamental differences:

\begin{enumerate}
    \item \textbf{Non-stationary arrivals:} Task generation depends on dynamic program structure—lattice surgery operations, gate scheduling, and routing decisions—rather than fixed periodic or sporadic arrival models with known minimum inter-arrival times.
    
    \item \textbf{Data-dependent execution:} Decoding latency varies with syndrome patterns and error distributions, making tight worst-case execution time (WCET) bounds impractical. Even for fixed code distance, execution time can vary 2-3× based on error chain complexity.
    
    \item \textbf{Elastic resource requirements:} Spatial decoding tasks require anywhere from 1 to 50+ decoders depending on patch size and routing distance—a form of moldable parallelism~\cite{Masko2006} uncommon in traditional real-time systems.
    
    \item \textbf{Temporal dependencies:} Overlapping decoding windows create precedence constraints between consecutive tasks on the same logical qubit, violating the task independence assumption central to most schedulability analyses.
\end{enumerate}

Despite these constraints, certain scheduling principles from queuing theory and soft real-time systems remain applicable. Work-conserving policies that account for task backlog and waiting time can prevent starvation and control queue growth under bursty, non-stationary arrivals~\cite{tassiulas1990stability}. Similarly, policies that prioritize tasks based on accumulated service deficits can provide fairness guarantees without requiring detailed arrival models~\cite{shreedhar1996efficient}. These insights inform our fine-grained policy design, which we describe next.

\subsection{Scheduling for individual qubits}

\begin{figure}[t]
    \centering
    \begin{subfigure}[b]{0.46\linewidth}
        \centering
        \includegraphics[width=\linewidth]{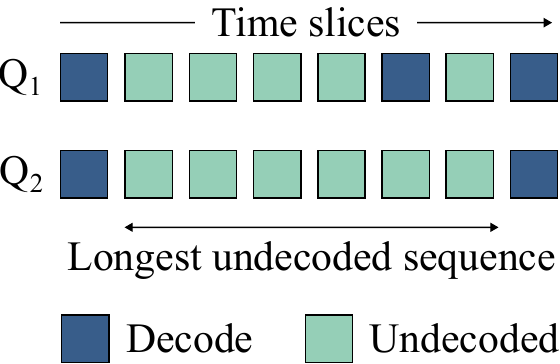}
        \caption{}
        \label{subfig:longest_undecoded}
    \end{subfigure}
    \hfill
    \begin{subfigure}[b]{0.46\linewidth}
        \centering
        \includegraphics[width=\linewidth]{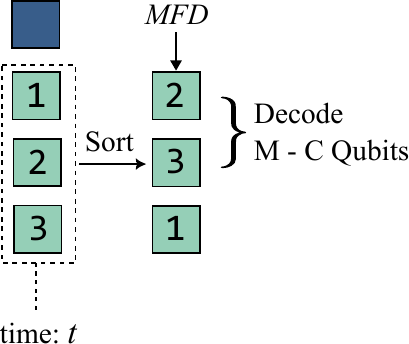}
        \caption{}
        \label{subfig:mfd}
    \end{subfigure}
    \\
    \begin{subfigure}[b]{0.46\linewidth}
        \centering
        \includegraphics[width=\linewidth]{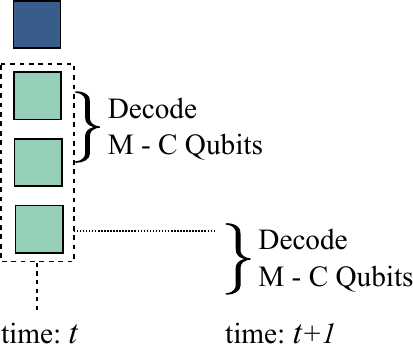}
        \caption{}
        \label{subfig:rr}
    \end{subfigure}
    \hfill
    \begin{subfigure}[b]{0.46\linewidth}
        \centering
        \includegraphics[width=\linewidth]{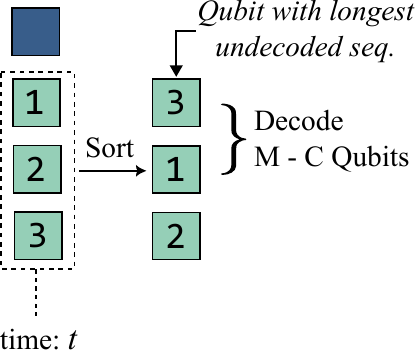}
        \caption{}
        \label{subfig:mls}
    \end{subfigure}
    
    \caption{
    (a) Illustration of the longest undecoded sequence -- $Q_1$ has the longest undecoded sequence before the last decode (each square is a decoding block/window); 
    (b) \texttt{MFD} policy -- undecoded qubits are sorted according to the number of critical decodes they are involved in after time slice $t$ and $M - C$ qubits are selected from this sorted list; 
    (c) \texttt{RR} policy -- the $M - C$ qubits decoded in time slice $t$ are not decoded in time slice $t+1$; 
    (d) \texttt{MLS} policy -- undecoded qubits are sorted according to their undecoded sequence lengths and $M - C$ qubits are selected from this sorted list.}
    \Description[some figure]{}
    \label{fig:policies}
    \vspace{-0.1in}
\end{figure}

Fine-grained scheduling determines when each logical qubit's syndromes (current and pending) are provided access to a decoder. If a policy starves qubits of decoding, its measured syndromes will accumulate, leading to longer decoding latencies (since the total size of the decoding problem increases with time). This thus necessitates fine-grained policies to ensure fair access to decoders for all qubits that are not involved in lattice surgery operations.


\subheading{Longest undecoded sequence}
To quantify the fairness of a fine-grained scheduling policy, we use a metric `\emph{Longest Undecoded Sequence}', which measures how well the decoders are servicing all logical qubits. A large undecoded sequence length implies that a qubit has been left undecoded for a long time. A qubit left undecoded for a long time will also require more processing from the decoder for it to be up to date with the latest syndrome. Figure~\ref{subfig:longest_undecoded} shows an example of determining the longest undecoded sequence length. 


Consider an arbitrary time slice $t$ in the execution of a quantum program. There are $N$ logical qubits and $M$ hardware decoders ($N > M$). All decoding scheduling policies will have two components: The first will assign the decoders necessary for all critical decodes $C$ in the time slice $t$. The second will assign all the remaining $M - C$ hardware decoders to the $N - C$ qubits based on the scheduling policy used. We now discuss three decoder scheduling policies (all policies are illustrated in Figure~\ref{subfig:mfd} -- Figure~\ref{subfig:mls}):

\subsubsection{Most frequently critically decoded (\texttt{MFD})} 
A logical qubit that consumes a significant number of $T$-states during the execution of a program would have a frequent requirement of critical decodes -- leaving such a logical qubit undecoded for more than a few slices would make subsequent critical decodes take longer, thus slowing down computation. This motivates the \texttt{MFD} scheduling policy that prioritizes decoding of logical qubits that have numerous critical decodes in the future at any given time (Figure~\ref{subfig:mfd}). The \texttt{MFD} policy will ensure that future critical decodes have a minimized number of undecoded syndromes for qubits with frequent critical decodes. The \texttt{MFD} policy is predicated on quantum programs being static in nature. While logical operations to perform $S$ corrections during magic state consumption represent `dynamic' instructions, they do not add any additional non-Clifford gates to the program, thus ensuring that any statically compiled program using the \texttt{MFD} policy will always work.

\subsubsection{Round-robin (\texttt{RR})}
Derived from CPU scheduling policies used in operating systems, the \texttt{RR} policy does not prioritize any specific logical qubits -- rather, it chooses $M - C$ qubits in a round-robin manner (Figure~\ref{subfig:rr}) in every time slice to ensure fairness for all qubits in the system. This policy was also alluded to in prior works~\cite{Das2022afs, Byun2022} for scheduling decoders for different logical qubits.

\subsubsection{Minimize longest undecoded sequence (\texttt{MLS})}
The longest undecoded sequence length at any time slice is an indicator of how well the fine-grained scheduling policy is servicing all qubits in the system. We use this as the base for the \texttt{MLS} policy, which greedily minimizes the longest undecoded sequence at every time slice (Figure~\ref{subfig:mls}). The \texttt{MLS} policy works as follows: at any time slice $t$, qubits are sorted on the basis of their current undecoded sequence lengths. Then, $M - C$ qubits with the largest undecoded sequence lengths are assigned hardware decoders.
The \texttt{MLS} policy mimics longest-queue-first (LQF) scheduling~\cite{dimakis2006sufficient}, known to prevent starvation and control backlog growth under bursty demand without strong arrival distribution assumptions.

\subsection{Magic state cultivation factories}
Magic state cultivation~\cite{Gidney2024Cultivation} is a protocol that generates high-quality non-Clifford (magic) states by using one or more smaller, more error-prone logical qubits. These cultivation protocols do not require high-performance decoders during the state generation process. Instead, they rely on \textbf{post-selection}, where attempts are \textit{discarded} as soon as a syndrome bit-flip is detected. Post-selection is well-suited for non-deterministic state preparation, providing exponential error suppression at the cost of success probability~\cite{Litinski2019magic, Gidney2024Cultivation}. While some protocols support active error correction~\cite{Chamberland2020}, we assume fully post-selected cultivation protocols for our evaluation, resulting in no decoder overhead during factory operation. However, the final output magic state is stored in a high-distance logical qubit for long-term use, and thus requires decoding to ensure reliability after successful preparation. We include these magic state storage patches in our analysis.

\subsection{Scheduling policy overheads}

While fine-grained scheduling policies provide performance benefits, they introduce computational and memory overheads that must be quantified. Table~\ref{tab:overheads} summarizes the time and space complexity of each policy.

\begin{table}[ht!]
\centering
\caption{Time and space overheads for all fine-grained scheduling policies.}
\resizebox{\linewidth}{!}{%
\begin{tabular}{@{}cccc@{}}
\toprule
\textbf{Policy} & \textbf{Per-slice time complexity} & \textbf{Metadata per qubit}  & \textbf{Total metadata storage}                                           \\ \midrule
\texttt{RR}     & O(1)                      & 1-bit flag          & $N_{decoders}$ bits                                              \\
\texttt{MFD}    & O(1)                      & 1 counter (static)  & $N_{decoders} \times log(\mathrm{Total~time~steps})$ bits          \\
\texttt{MLS}    & O(NlogN)                  & 1 counter (dynamic) & $N_{decoders} \times log(\mathrm{d \times Total~time~steps})$ bits \\ \bottomrule
\end{tabular}%
}
\label{tab:overheads}
\end{table}

\subsubsection{Time complexity}
RR and MFD both achieve $O(1)$ scheduling decisions per time slice by maintaining simple state: RR uses a round-robin counter, while MFD accesses a precomputed static priority ordering based on future critical decode counts (determined at compile time). In contrast, MLS requires sorting $N$ qubits by their undecoded sequence lengths, requiring $O(N \log N)$ comparisons. For typical workloads with $N \approx 100$--1000 active qubits, this translates to $\sim$700--10,000 comparisons per slice.

\subsubsection{Space complexity}
All policies maintain per-qubit metadata. RR requires only a single bit indicating whether each qubit was decoded in the previous slice. MFD stores a counter of remaining critical decodes for each qubit, requiring $\lceil \log_2 T \rceil$ bits to represent counts up to the total program length $T$. MLS tracks the current undecoded sequence length, which can grow up to $dT$ in the worst case (accumulating $d$ syndrome rounds per time slice for $T$ slices), requiring $\lceil \log_2 (dT) \rceil$ bits per qubit. For representative parameters ($N = 800$, $T = 10^5$, $d = 11$), total metadata ranges from 100 bytes (RR) to $\sim$2.4~KB (MLS)—negligible compared to syndrome buffer memory requirements (multiple MBs).

\subsubsection{Off-critical-path execution}
Critically, scheduling latency does not impact end-to-end execution time because it occurs \emph{off the critical path}. The decoding pipeline operates as follows:
\begin{enumerate}
    \item At time slice $t$, the quantum hardware executes operations and generates syndromes for all active qubits
    \item While syndromes are being read out and buffered (requiring $T_{\text{readout}} \sim 100$--500~ns per qubit), the classical control processor computes the scheduling decision for slice $t+1$
    \item Once $M$ decoders become available (having completed their previous windows), they are immediately assigned to qubits according to the computed schedule
    \item Decoding proceeds in parallel across all $M$ decoders
\end{enumerate}

The key insight is that syndrome readout, buffering, and decoder processing all occur in parallel with scheduling for the \emph{next} slice. As long as scheduling completes before decoders finish their current windows—i.e., $T_{\text{schedule}} < t_D \cdot T_{\text{cycle}}$—there is no additional latency overhead because of scheduling.

\subsubsection{Concrete overhead analysis.}
For MLS with $N = 800$ qubits and $t_D = 0.5$ (normalized decoder latency), each decoder processes a window in $0.5 \times 1~\mu\text{s} = 500$~ns. With $M = 400$ decoders operating in parallel, aggregate decoding throughput is 400 windows per 500~ns. Meanwhile, sorting 800 integers on a modern CPU (2~GHz, $\sim$1 cycle per comparison for integer comparisons) requires $\sim$8000 comparisons $\times$ 0.5~ns/cycle $\approx 4~\mu$s. Since this is $\ll t_D \cdot T_{\text{cycle}} = 500$~ns per individual decoder but occurs in parallel with all decoder execution, the effective overhead is negligible.

For FPGA implementations, priority queues based on heap structures reduce per-qubit updates to $O(\log N)$ operations. At typical FPGA clock rates (100--200~MHz), maintaining a heap of 1000 elements requires $\sim$10 cycles $\times$ 10~ns/cycle $= 100$~ns per insertion/update—well within the syndrome readout time budget.

\section{Methodology}
\label{sec:methodology}
We now describe the methodology used to evaluate different decoder scheduling policies for the PWD paradigm.


\subsection{Compiler}
We use the Lattice Surgery Compiler (LSC)~\cite{watkins2023high} to generate Lattice Surgery Instructions (LLI) Intermediate Representations (IR) of workloads that can be executed on an error-corrected quantum computer using the Surface code with Lattice Surgery. LSC can generate IR that denote Lattice Surgery instructions from the QASM~\cite{qasm} representation of a workload. The Lattice Surgery instructions generated by LSC are a combination of Clifford operations and multi-body measurements used for implementing Pauli-Product Rotations (PPR)~\cite{Litinski2019}. LSC handles mapping and routing based on the layout provided to the compiler. We configure LSC to use a `wave' scheduling that maximizes the number of concurrent instructions executed in every time slice. LSC uses Gridsynth~\cite{gridsynth} to synthesize arbitrary rotations. 
\subsubsection{Layout}
We perform all evaluations using the Edge-Disjoint Path Compilation layout (EDPC)~\cite{Beverland2022edpc} with a single routing lane. 

\subsubsection{Magic states}
\sloppy
LSC abstracts distillation factories away -- it is assumed that the magic state storage sites arranged along the borders of the EDPC layout get magic states within a specified time period. We assume that the magic state production throughput is high enough for generating magic states every alternate slice. 

\subsubsection{Other configurations}
$Y$-states are produced using the twist-based initialization~\cite{Gidney2024YState}.


\subsection{Simulation Framework}
\label{sec:framework}
Using the IR generated by LSC, we build a framework~\cite{vader} that can parse the IR, determine the critical decodes in every slice, generate a timeline of all operations, and assign decoders to all logical qubits depending on the scheduling policy. 


%
%
\subheading{Benchmarks}
We select benchmarks that are representative of large-scale programs that will run on FTQC systems. A majority of these will be building blocks for larger algorithms. We use the following benchmarks (benchmark indices annotated in parentheses, benchmark size in terms of number of qubits specified after the hyphen): 
\begin{Verbatim}[breaklines=true, fontsize=\footnotesize]
adder-28 (0), adder-64 (1), adder-118 (2), adder-433 (3), bwt-37 (4), bwt-97 (5), heisenberg-100 (6), ising-34 (7), ising-42 (8), ising-66 (9), ising-98 (10), multiplier-45 (11), multiplier-75 (12), multiplier-350 (13), qft-20 (14), qft-40 (15), qft-60 (16), qft-80 (17), qft-100 (18), qft-120 (19), qft-160 (20), shor-9 (21), shor-15 (22), wstate-20 (23), wstate-40 (24), wstate-60 (25), wstate-80 (26)
\end{Verbatim}


\subheading{Other Software}
Stim~\cite{Gidney2021} was used for simulating stabilizer circuits to generate syndromes and error rates. Azure QRE~\cite{Beverland2022} was used for resource estimations.


\section{Evaluations}
\label{sec:eval}

In this section, we simulate the effectiveness of different decoder scheduling policies for the parallel windowed decoding (PWD) paradigm.

\subsection{Research Questions}

The key research questions we would like to answer for the PWD paradigm is: 
\begin{enumerate}[topsep=0pt]
    \item[Q1.] What is the minimum number of decoders required to avoid slowdowns in the program runtime?
    \item[Q2.] How does this minimum number of decoders compare with the baseline system?
    \item[Q3.] How do the fine-grained scheduling policies compare with each other?
    \item[Q4.] How do decoder latencies affect decoder scheduling?
\end{enumerate}

\subsection{Benchmark statistics and baselines}

\begin{figure}[t]
    \centering
    \includegraphics[width=\linewidth]{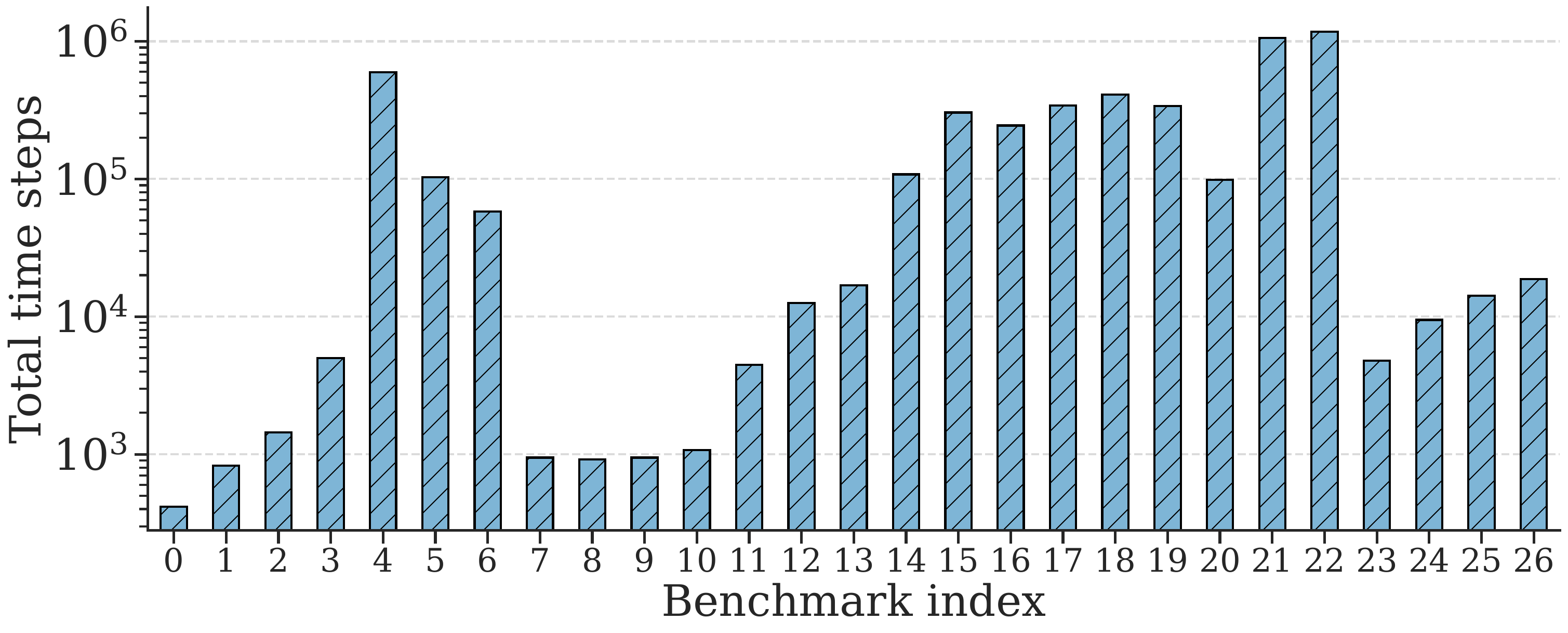}
    \caption{Total slices executed for each benchmark.}
    \Description[some figure]{}
    \label{fig:slices}
    \vspace{-0.1in}
\end{figure}
\subsubsection{Benchmark runtimes}
LSC was configured to timeout after 10 minutes to keep simulation times tractable. Figure~\ref{fig:slices} shows the number of time steps for all benchmarks. 

\begin{figure}[t]
    \centering
    \includegraphics[width=\linewidth]{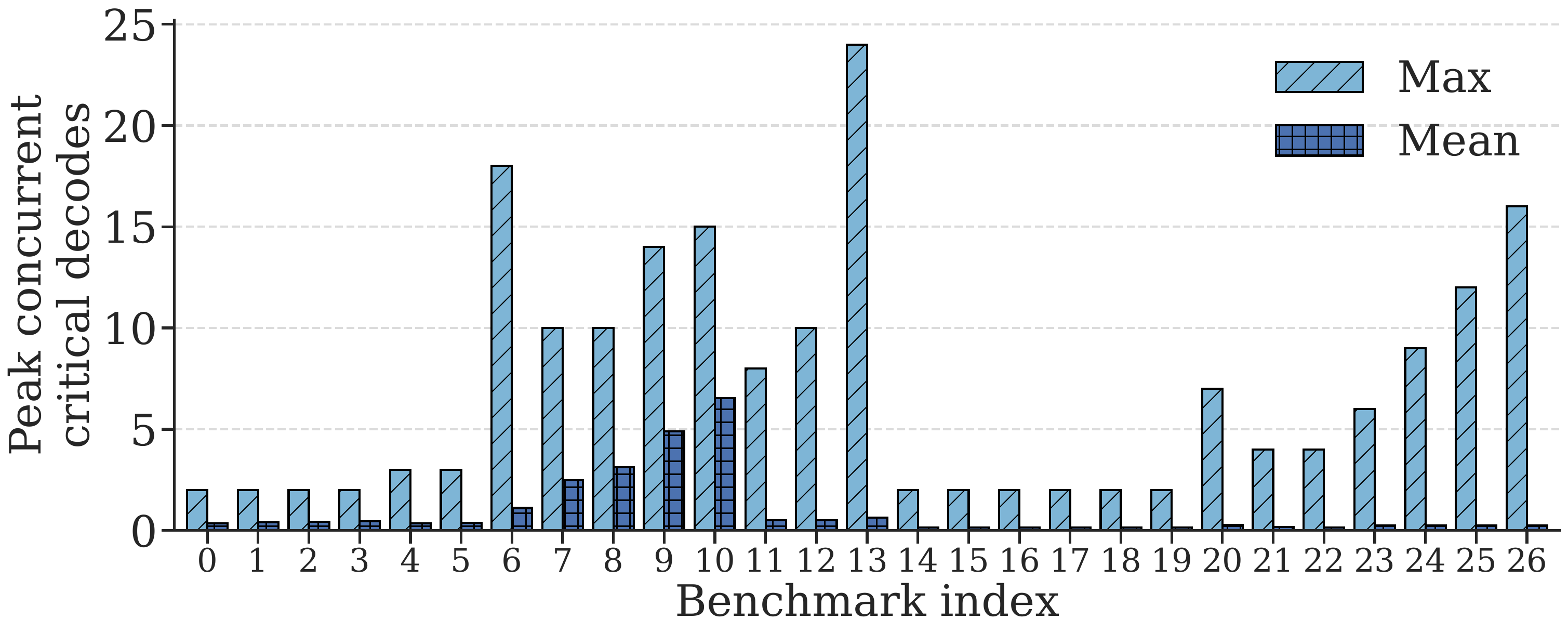}
    \caption{Peak and average concurrency for non-Clifford operations for all benchmarks -- concurrency is sporadic.}
    \Description[some figure]{}
    \label{fig:conc_crits}
    \vspace{-0.1in}
\end{figure}
\subsubsection{Concurrent critical decodes}
Since all scheduling policies prioritize decoder allocation for any critical decode in a slice, the concurrency of critical decodes plays an important role in determining the number of decoders required for a benchmark. Figure~\ref{fig:conc_crits} shows the difference between the peak and average critical decode concurrency.
This highlights the importance and need for performant decoder scheduling policies -- the limited concurrency of critical decodes in quantum programs makes efficient scheduling of decoders important for high decoder utilization. 

\begin{figure}[t]
    \centering
    \includegraphics[width=\linewidth]{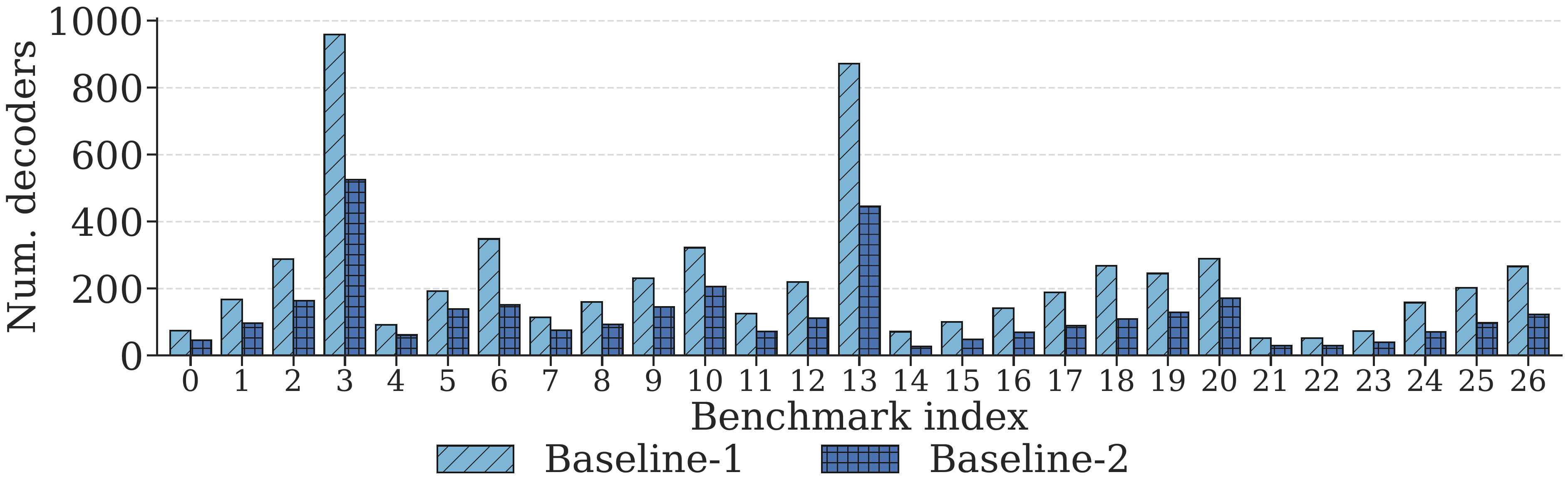}
    \caption{Number of decoders allocated for the baselines: Baselines-1 and 2 provision for the worst- and average-case decoder demands respectively.}
    \Description[some figure]{}
    \label{fig:baseline_decoders}
    \Description[some figure]{}
    \vspace{-0.1in}
\end{figure}

\subsubsection{Baselines}

We assume a baseline system to allocate decoders for every logical qubit. Since the number of active logical qubits can vary through the course of a program, we consider the following baselines: 
\begin{enumerate}[topsep=0pt, leftmargin=*]
    \item \underline{Baseline-1}: Provisions for the worst-case number of active logical qubits.
    \item \underline{Baseline-2}: Provisions for the average-case number of active logical qubits.
\end{enumerate}

Baseline-1 caters for all possible qubits that could be active for a given benchmark, and thus requires a very simple scheduling policy wherein syndromes for a logical qubit are passed to its pre-assigned decoder. On the other hand, Baseline-2 will require some scheduling policy for choosing which logical qubits are decoded at any given instant. We consider the Round-Robin (\texttt{RR}) policy as the baseline policy since its use has been suggested in prior works that allude to scheduling decoders to reduce resource overheads~\cite{Das2022afs, Byun2022}.
Figure~\ref{fig:baseline_decoders} shows the number of decoders configured for the two baselines for all benchmarks. As is to be expected, Baseline-1 requires significantly more decoders than Baseline-2.

\subsection{Scheduling for the PWD paradigm}

In our simulations of the PWD paradigm, we first assume that the normalized decoder latency is fixed at $t_D=0.5$. This is in line with existing high-performance hardware decoders~\cite{Alavisamani2024, riverlaneLCD2024}. Further research will likely reduce this latency further. We assess the impact of $t_D$ on the simulation results in subsequent sections. For all evaluations, we assume both temporal and spatial windows have the same size of $n_W=3d$. We use the coarse-grained policy presented in Section~\ref{sec:coarse-grained} in all evaluations and compare the \texttt{MFD}, \texttt{RR}, and \texttt{MLS} fine-grained scheduling policies with each other.

\begin{figure}[t]
    \centering
    \begin{subfigure}[b]{\linewidth}
        \centering
        \includegraphics[width=\linewidth]{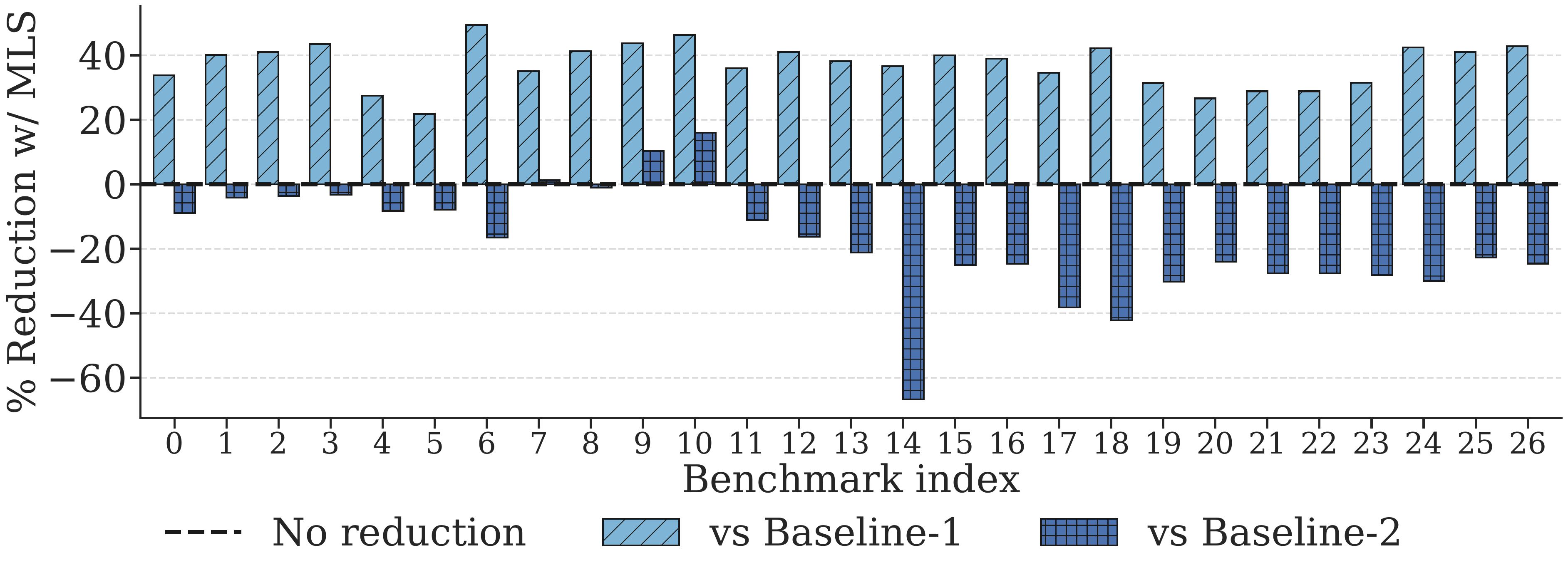}
        \caption{}
        \label{subfig:optimal_pwd}
    \end{subfigure}
    \\
    \begin{subfigure}[b]{\linewidth}
        \centering
        \includegraphics[width=\linewidth]{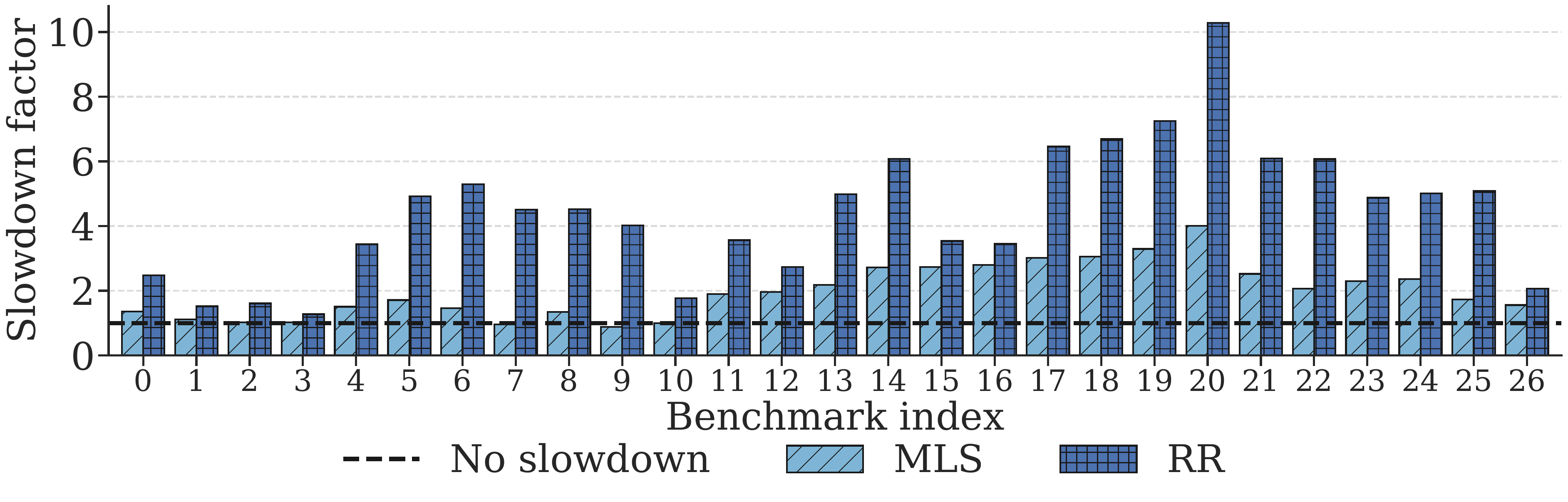}
        \caption{}
        \label{fig:baseline2_slowdown}
    \end{subfigure}
    \\
    \begin{subfigure}[b]{\linewidth}
        \centering
        \includegraphics[width=\linewidth]{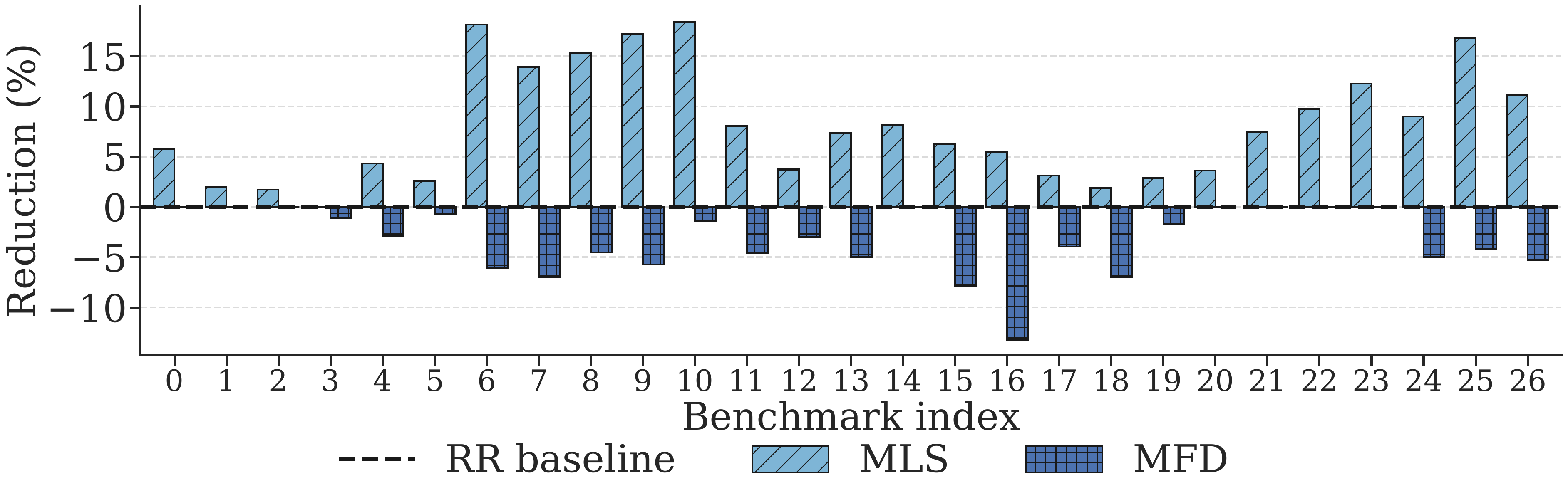}
        \caption{}
        \label{subfig:comp_opt_pwd}
    \end{subfigure}

    \caption{
    (a) The relative reduction (higher is better) in the number of decoders required by the \texttt{MLS} policy to prevent any slowdowns compared to the two baselines. Baseline-2 provisions for fewer decoders than what is required by the \texttt{MLS} policy;
    (b) Slowdown in computation when setting the number of decoders to Baseline-2 for the \texttt{MLS} and \texttt{RR} fine-grained policies;
    (c) The relative reduction in the minimum number of decoders required (higher is better) when using the \texttt{MFD} and \texttt{MLS} policies over the baseline \texttt{RR} policy. The \texttt{MFD} policy requires more decoders to prevent any slowdowns, while the \texttt{MLS} policy can reduce the number of required decoders by up to 19\%;
    }
    \Description[some figure]{}
    \label{fig:pwd_optimal_decoders}
    \vspace{-0.1in}
\end{figure}

\subsubsection{Minimum number of decoders needed to prevent any slowdowns}
We first determine the minimum number of decoders required for all fine-grained policies and benchmarks to prevent any slowdowns in the computation. We determine this optimal number by performing a binary search on the number of decoders such that the final backlog for all logical qubits is zero. 

Figure~\ref{subfig:optimal_pwd} shows the reduction in the number of decoders required when the optimal number is determined for the \texttt{MLS} policy relative to the two baselines. Compared to Baseline-1, the optimal decoder count is about \textbf{40\% lower} than the counts determined for Baseline-1 across most workloads. However, compared to Baseline-2, the optimal decoder count is greater by up to 60\% -- highlighting that the optimal number of decoders must be more than the average-case number of logical qubits active for any workload.

\subsubsection{Baseline-2 introduces significant slowdowns}
As Baseline-2 has fewer decoders allocated than the optimal number, it is expected that Baseline-2 will incur slowdowns in computation due to decoder scarcity, regardless of the scheduling policies used. Figure~\ref{fig:baseline2_slowdown} shows the slowdowns incurred by Baseline-2 across all benchmarks when the \texttt{MLS} and \texttt{RR} policies are used. For the \texttt{RR} policy, \textbf{the slowdown can be up to 10$\times$}, highlighting the sub-optimality of Baseline-2 compared to the optimal number of decoders. Unsurprisingly, the slowdowns are higher for benchmarks which have a greater difference between the optimal number of decoders and the number allocated for Baseline-2.

Slowdowns are an important factor to consider when designing fault-tolerant quantum computing systems. Despite the speed-up they offer, FTQC systems still require considerable time to finish computations. For example, it is estimated that factoring a 2048-bit number will take about a week~\cite{gidney2025factor}. Slowdowns must thus be avoided to further increase this computing time -- a 2$\times$ slowdown would increase that time to two weeks.
Figure~\ref{fig:baseline2_slowdown} also highlights that the \texttt{MLS} policy is more resilient to decoder pressure/scarcity compared to the \texttt{RR} policy. Also, Baseline-1 cannot incur any slowdowns since the number of decoders allocated to it is always more than the optimal number determined across all benchmarks.

\subsubsection{Comparing the fine-grained scheduling policies}
To further illustrate the differences between the fine-grained scheduling policies, we determine the reduction in the optimal number of decoders when the \texttt{MLS} and \texttt{RR} policies are used compared to the baseline \texttt{RR} policy, as shown in Figure~\ref{subfig:comp_opt_pwd}. The \texttt{MLS} policy achieves a relative reduction of \textbf{up to 19\%} compared to the \texttt{RR} policy, while the \texttt{MFD} policy is either as good or worse than the \texttt{RR} policy. This is likely because the \texttt{MFD} policy starves some qubits of decoding, thus requiring more decoders. Combined with the results of Figure~\ref{fig:baseline2_slowdown}, these results clearly show the benefits of the \texttt{MLS} policy over the other fine-grained policies. 

Another important observation from Figure~\ref{subfig:comp_opt_pwd} is its correlation with Figure~\ref{fig:conc_crits} -- benchmarks that exhibit higher concurrency experience greater benefits with the \texttt{MLS} policy compared to benchmarks that are more serial. 

The answers for research questions $Q1-Q3$ can be summarized from the results above as follows:

\begin{boxH}
    The optimal number of decoders is 40\% less than the worst-case decoder allocation of Baseline-1 while avoiding the slowdown of up to 10$\times$ experienced by Baseline-2. The \texttt{MLS} policy outperforms both \texttt{RR} and \texttt{MFD} policies by up to 19\%.
\end{boxH}

\begin{figure}[t]
    \centering
    \begin{subfigure}[b]{\linewidth}
        \centering
        \includegraphics[width=\linewidth]{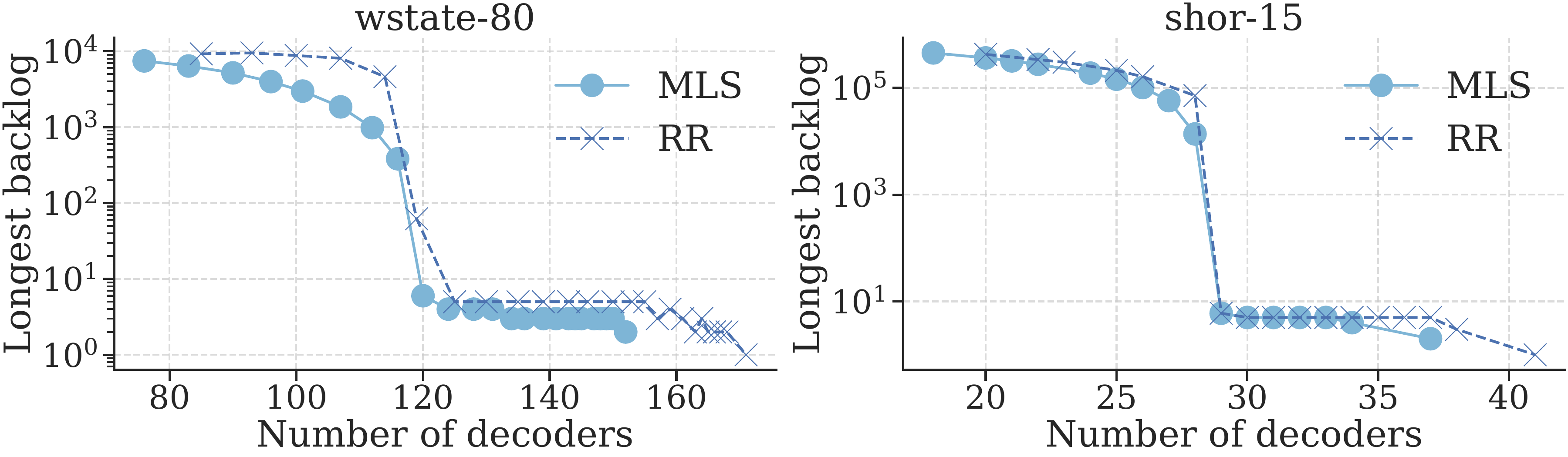}
        \caption{}
        \label{subfig:decoder_sweep}
    \end{subfigure}
    \\
    \begin{subfigure}[b]{\linewidth}
        \centering
        \includegraphics[width=\linewidth]{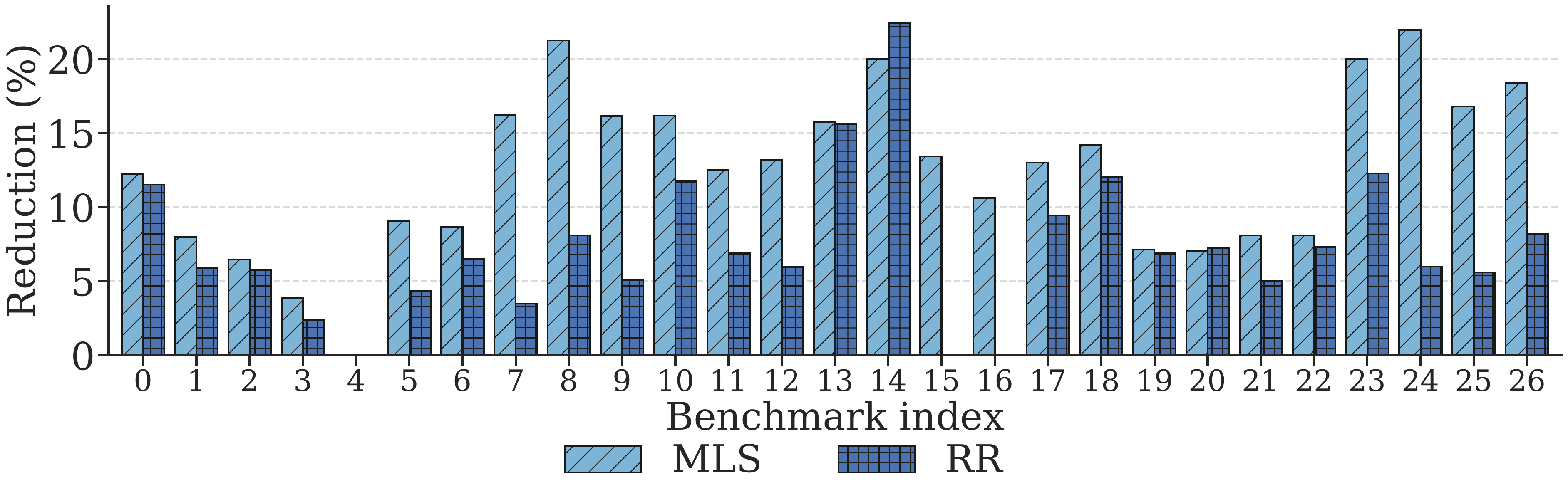}
        \caption{}
        \label{subfig:sweep_policies}
    \end{subfigure}
    
    \caption{
    (a) Effect of the number of decoders on the backlog at the end of a benchmark for workloads exhibiting high concurrency (\texttt{wstate-80}) and low concurrency (\texttt{shor-15}). When the number of decoders is less than a threshold, the backlog grows exponentially;
    (b) Reduction in the number of decoders relative to the optimal number determined in Figure~\ref{subfig:optimal_pwd} for the \texttt{MLS} and \texttt{RR} policies when a slowdown of 10\% or less is tolerable. The \texttt{MLS} policy yields a higher reduction for most workloads.
    }
    \Description[some figure]{}
    \label{fig:relaxing}
    \vspace{-0.1in}
\end{figure}

\subsubsection{Relaxing the slowdown constraint}
The optimal decoder count was determined to ensure that there would be no slowdown during the execution of the program. However, what if this slowdown constraint was relaxed? 
To understand the effect of relaxing the slowdown constraint, we first show the effect of reducing the number of decoders on the longest backlog of syndromes for two benchmarks: the first being a highly concurrent (\texttt{wstate-80}) and the second being very serial (\texttt{shor-15}). 
Figure~\ref{subfig:decoder_sweep} shows the effect of reducing the number of decoders on the longest backlog of syndromes for both benchmarks: after a certain decoder count, the backlog increases exponentially, corresponding to significant slowdowns (since these backlogs would need extra time to be processed). This exponential increase would also increase the memory requirements to store undecoded syndromes. Additionally, as was shown before, the \texttt{MLS} policy is more resilient to the scarcity of decoders compared to the \texttt{RR} policy. We omit the \texttt{MFD} policy from these evaluations for brevity, since we showed in Figure~\ref{fig:pwd_optimal_decoders} that it is inferior to both \texttt{MLS} and \texttt{RR} policies.

In Figure~\ref{subfig:sweep_policies}, we determine the reduction in the number of decoders relative to the optimal number determined in Figure~\ref{subfig:optimal_pwd} when slowdowns up to 10\% are tolerable. For all benchmarks, we see a reduction in the optimal number of decoders required, especially for \texttt{wstate}. Furthermore, the \texttt{MLS} policy achieves a higher reduction than the \texttt{RR} policy. These results show that there is a trade-off between decoder allocation and program performance: if limited slowdowns are tolerable, the classical resource requirements can be significantly reduced with good scheduling policies.

\begin{figure}[t]
    \centering
    \begin{subfigure}[b]{0.48\linewidth}
        \centering
        \includegraphics[width=\linewidth]{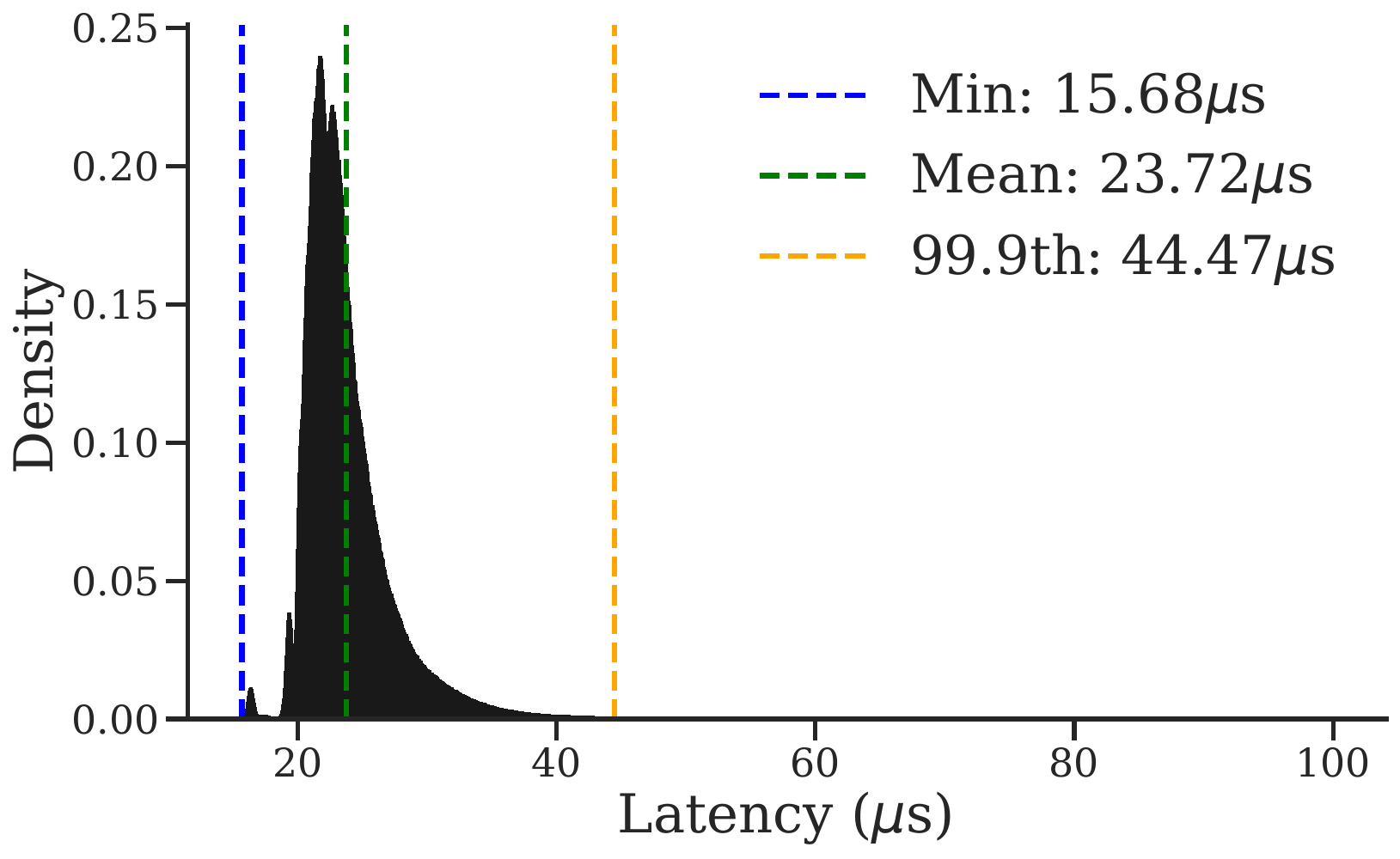}
        \caption{}
        \label{subfig:pymatching_hist}
    \end{subfigure}
    \hfill
    \begin{subfigure}[b]{0.48\linewidth}
        \centering
        \includegraphics[width=\linewidth]{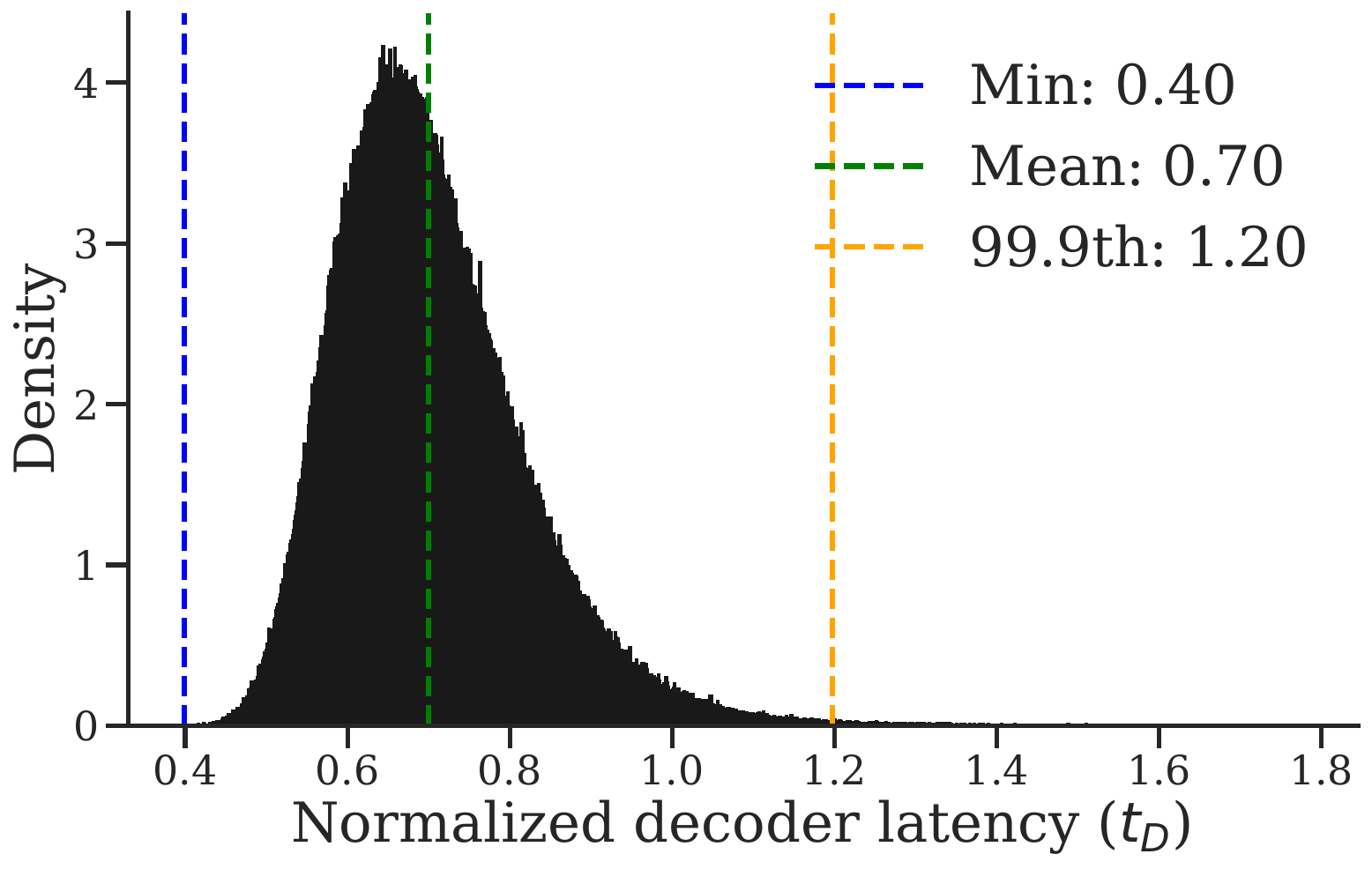}
        \caption{}
        \label{subfig:sim_hist}
    \end{subfigure}
    
    \caption{
    (a) Density histogram of latencies sampled from the PyMatching~\cite{Higott2023} decoder: the distribution can be approximated by a log-normal distribution with long tail latencies ($d=9, p=0.1\%$);
    (b) The density histogram used for our simulations with $t_D$ sampled from a log-normal distribution.
    }
    \Description[some figure]{}
    \label{fig:tails}
    \vspace{-0.1in}
\end{figure}
\subsection{Effect of decoder tail latencies}
So far, we have performed all evaluations with a fixed normalized decoder latency $t_D=0.5$ (where $t_D = \frac{T_{decode}}{T_{cycle}}$). 
However, decoders do not have a deterministic latency, and some decoding problems take longer to decode than others. In particular, some decoding tasks can have a long \emph{tail latency}. This can result in higher decoder allocations, because slower decoders are unavailable for a longer period, reducing the throughput at which decoding tasks as processed. 
Figure~\ref{subfig:pymatching_hist} shows the distribution of latencies from the state-of-the-art software decoder for the surface code, PyMatching~\cite{Higott2023} for a distance 9 ($d=9$) surface code logical qubit and a circuit-level physical error rate of 0.1\% ($p$). The distribution is centered around the mean but tail latencies are significant long. We approximate this distribution with a log-normal probability density function~\cite{Delfosse2023} of the form: 

\begingroup
\small
\setlength{\abovedisplayskip}{4pt}
\setlength{\belowdisplayskip}{4pt}
\[
PDF(x; \mu, \sigma) =
\frac{1}{x \, \sigma \sqrt{2\pi}}
\exp\!\left( -\frac{(\ln(x) - \mu)^2}{2\sigma^2} \right),
\quad x > 0
\]
\endgroup

\begin{figure}[t]
    \centering
    \includegraphics[width=\linewidth]{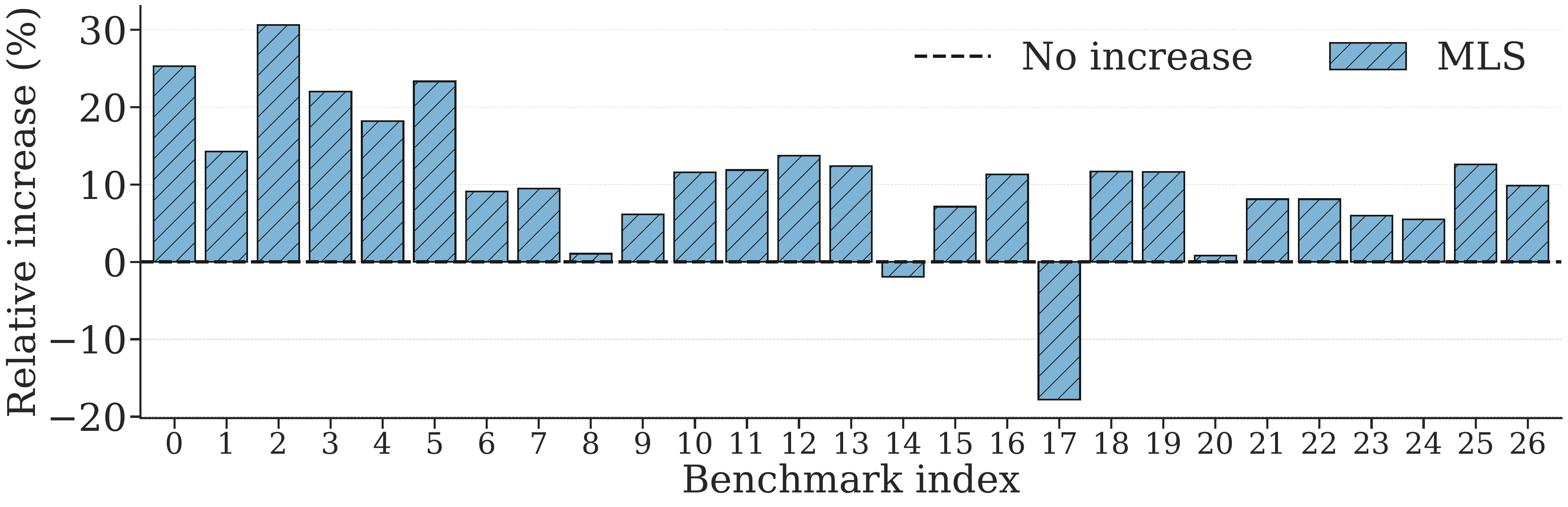}
    \caption{Relative increase in the number of optimal decoders required by the \texttt{MLS} policy to prevent slowdowns with longer non-uniform decoder latencies.}
    \Description[some figure]{}
    \label{fig:tails_decoders}
    \vspace{-0.1in}
\end{figure}


We model hardware decoder latencies with a log-normal distribution (see Figure~\ref{subfig:sim_hist}), which produces worst-case tail latencies longer than the syndrome generation cycle ($t_D > 1.0$). These long tails shift the optimal decoder count for all benchmarks compared with Figure~\ref{subfig:optimal_pwd}: most benchmarks require more decoders under the \texttt{MLS} policy, while \texttt{qft-20 (14), qft-80 (17)} require fewer because some critical decodes sampled latencies below $t_d=0.5$, reducing their decoder needs. \textbf{Overall, despite these longer, non-uniform latencies, we can still reduce decoder counts by 10–40\% across all benchmarks}.

\subsection{Limitations}

\subheading{Magic states}
We assume that magic states are produced deterministically and are always available. This represents the most resource-intensive scenario, since it maximizes the potential number of concurrent non-Clifford operations and thus critical decodes. In practice, fewer distillation factories or variable throughput would reduce concurrency and lower decoder demand. We do not evaluate these effects, so these results are an upper bound on decoder requirements.

\subheading{Fine-grained scheduling policies}
Our results show that the \texttt{MLS} policy is more effective than \texttt{RR} at reducing backlogs. However, both \texttt{MLS} and \texttt{MFD} require continuous ranking of logical qubits, which may incur non-trivial overhead in a hardware controller. While ranking can be pre-computed for statically compiled programs under deterministic magic state availability, this assumption does not hold in general. By contrast, \texttt{RR} is simple to implement and has predictable overhead. We view our evaluation as a first step, and expect future work to explore more practical, low-overhead policies that combine the efficiency of \texttt{MLS} with the simplicity of \texttt{RR}.

\subsection{Real system evaluation}
At present, no publicly available quantum computer supports the tight classical control loop required for evaluating the ideas presented in this work. This control loop requires fast transmission of syndromes from the quantum processor to the decoding hardware, an ability to route syndromes to specific decoders in the system, and fast transmission of the decoder result to the classical control software. Furthermore, the smallest workload in our study requires 20 logical qubits. Even if we make the most optimistic assumption that one logical qubit is encoded in a distance-3 ($d=3$) surface code patch, 20 logical qubits will require at least $20\times17=340$ physical qubits. Overheads due to magic state production, routing, ancillas will further inflate the qubit requirements. This is far beyond the capacity of current, cloud-accessible quantum computers. This limitation on the number of qubits also prevents offline studies where syndromes can be collected from a system and then post-processed.

\section{Related Work}

This is one of the first works to perform a workload-oriented study of the classical processing requirements and system-level scheduling policies for error-corrected quantum computers. The Round-Robin (\texttt{RR}) decoder scheduling policy was alluded to in~\cite{Das2022afs, Byun2022} for allocating decoders to different logical qubits but the efficacy of the policy and its strengths and limitations were not explored. In particular, XQ-Sim~\cite{Byun2022} proposed \emph{patch-sliding window decoding} that reduced resource and power requirements for a cryogenic control system . This patch-sliding mechanism is essentially the round-robin scheduling policy. Similarly, AFS~\cite{Das2022afs} reduced decoder resource requirements by using the conjoined decoder architecture (CDA) that shared decoder compute blocks among multiple logical qubits, with arbitration done via a round-robin policy. 

Spatial and temporal parallel window decoding are discussed in~\cite{Skoric2023, Bombin2023, FuhuiLin2025}. \cite{Skoric2023} introduced parallel windowed decoding in both space and time and showed how sliding windowed decoding is not scalable. \cite{FuhuiLin2025} further refined and discussed microarchitectural considerations for spatial windowed decoding, including scheduling different decoding windows within a decoding task. Bombín et al.~\cite{Bombin2023} introduced modular decoding, which incorporates both spatial and temporal parallelism in decoding without sacrificing accuracy.
No work focuses on scheduling decoders from a windowed decoding approach. Other works that are broadly connected to this work are summarized below.

\subheading{System-level Studies}
Network integrated FPGA-based decoding systems were introduced in~\cite{wu2024lego, liyanage2025network}. Delfosse et al.~\cite{Delfosse2023} studied the speed vs. accuracy tradeoff for decoders used in FTQC. Stein et al.~\cite{Stein2023} proposed a heterogeneous architecture for FTQC. Lin et al.~\cite{Lin2023} explored modular architectures for error-correcting codes, and scheduling for distillation factories was proposed in~\cite{Ding2018}. \cite{Kim2024} described a blueprint of a fault-tolerant quantum computer.

\subheading{Decoder Designs}
Neural network-based decoders~\cite{googleRNN, ueno2022, Overwater2022, Meinerz2022, Gicev2023, Varsamopoulos2017, Google2024QEC, Harvard2025MLDecoder}, LUT-based decoders~\cite{Das2022lilliput, Tomita2014}, decoders based on the union-find algorithm~\cite{Riverlane2023, riverlaneLCD2024}, and optimized MWPM decoders~\cite{Vittal2023, Alavisamani2024, Wu2025} have been proposed. In general, neural network decoders are slower and therefore not ideal for fast qubit technologies such as superconducting qubits. Other predecoders~\cite{Delfosse2020, Smith2023} and partial decoders~\cite{Caune2023} have also been proposed. Decoders based on superconducting logic~\cite{Ueno2021, Ravi2023} target cryogenic implementations.

\section{Conclusions}
\sloppy
Decoders form the backbone of fault-tolerant quantum computing (FTQC) systems. 
They are classical algorithms designed to continuously detect and correct errors in the quantum hardware, and must be accelerated using custom/reconfigurable hardware to meet microsecond-scale latency requirements. 
For such hardware decoders, \emph{capacity planning} and \emph{scheduling} are key to ensure optimal resource utilization and performance. Capacity planning prevents wasteful over-provisioning or risky under-provisioning, while scheduling ensures that decoders are assigned tasks such that there is no degradation to program runtime.
Capacity planning for hardware decoders is not trivial -- the use of spatial and temporal parallel windowed decoding (PWD) results in non-deterministic decoder demand throughout the course of a program's execution.
This paper is the first to address these system-level challenges for FTQC. We introduce a two-level scheduling framework that enables accurate decoder provisioning and efficient task allocation across diverse workloads. Our evaluation shows that this approach reduces decoder hardware needs by 10–40\% while sustaining fault-tolerant performance.

%

\appendix
\section{Artifact Appendix}

\subsection{Abstract}
This artifact contains program traces, benchmarks, and software to generate the main results and insights presented in the paper. 
Lattice Surgery Compiler~\cite{watkins2023high} has been used to generate program traces, and while the artifact provides the traces for ease of evaluation, the traces can be generated using instructions in the README.

This artifact can be used to reproduce Figures~\ref{fig:distances},
\ref{fig:demand},
\ref{fig:slices},
\ref{fig:conc_crits},
\ref{fig:baseline_decoders},
\ref{subfig:optimal_pwd},
\ref{fig:baseline2_slowdown},
\ref{subfig:comp_opt_pwd}, and
\ref{fig:tails_decoders} of the main text. Note that due to the inherent randomness of the sampling process used for generating decoder latencies, Figure~\ref{fig:tails_decoders} might not exactly match the generated results.

\subsection{Description \& Requirements}

\subsubsection{How to access}
We provide the code and datasets (program traces) to reproduce the results of this paper as a Zenodo repository: \url{https://doi.org/10.5281/zenodo.18555903}.

The code for generating program traces and scheduling decoders is also available on Github: \url{https://github.com/satvikmaurya/decoder-resources}. 


\subsubsection{Hardware dependencies}
$\geq$16-core workstation/server node (any Linux distro), $\geq$64 GB RAM\footnote{Our evaluations on a 32C/64T machine did not require more than 96 GB of RAM.}, and $\geq$20 GB storage.

\subsubsection{Software dependencies}
Docker.

\subsubsection{Benchmarks} 
None (all benchmarks and program traces included in artifact).

\subsection{Set-up}

This artifact utilizes Docker for a seamless installation of all underlying (C++, Python, Haskell) dependencies. 
To start off, ensure that the directory structure looks something similar to this (assuming \texttt{artifact} is the top-level directory): 

\begin{center}
\begin{forest}
for tree={font=\ttfamily\footnotesize}
[artifact/
  [decoder-resources/
    [...]
  ]
  [LLI/
  ]
]
\end{forest}
\end{center}

\noindent To build the Docker image, run: 
\begin{lstlisting}
$ cd decoder-resources/
$ docker build -t $USER/vader . 
\end{lstlisting}

\noindent We recommend using the interactive shell of the Docker container for this artifact. To prevent loss of progress, \texttt{tmux} is recommended:
\begin{lstlisting}
$ tmux new -s artifact129 
$ docker compose run app
\end{lstlisting}

\subsection{Evaluation workflow}

\subsubsection{Major Claims}

\begin{itemize}
    \item \textit{(C1): We show that the demand for decoders in fault-tolerant programs is bursty and non-deterministic.}
    \item \textit{(C2): We show that with careful capacity planning, the number of decoders can be reduced by up to 40\% compared to the worst-case provisioning, saving system costs and complexity.}
    \item \textit{(C3): We show that the MLS fine-grained scheduling policy under this bursty, non-deterministic achieves a 19\% reduction in the number of decoders compared to the baseline scheduling policy.}
\end{itemize}

\subsubsection{Experiments}

\paragraph{Experiment (E1): [End-to-end evaluation] [30 human-minutes + 24 compute-hours]:}
To simplify evaluations, we have consolidated all scripts into a single bash script that can be run from within the Docker container. We provide descriptions of some important constituent scripts below:

\begin{enumerate}
    \item \texttt{histogram.sh}: This script characterizes the patch-to-patch distances between logical qubits involved in an operation. This generates data for Figure~\ref{fig:distances}, which supports \textit{(C1)}.
    \item \texttt{spatial\_wc.py}: This script characterizes the worst-case number of concurrently active patches as a function of time for all workloads. This generates data for Figure~\ref{fig:demand}, which supports \textit{(C1)}.
    \item \texttt{baseline.sh}: This script determines the slowdown when the average-case baseline (Baseline-2) is used for provisioning decoders. This generates data for Figure~\ref{fig:baseline2_slowdown}. This script supports \textit{(C2)}.
    \item \texttt{optimize.sh}: This script runs an optimization loop to determine the optimal (minimum) number of decoders required for a workload and scheduling policy to prevent any program slowdowns. This generates data for Figures~\ref{subfig:optimal_pwd}, \ref{subfig:comp_opt_pwd}, and \ref{fig:tails_decoders}. This script supports \textit{(C2 and C3)}.
\end{enumerate}

\noindent \textit{[How to]}

\noindent To run all required scripts, simply run the \texttt{run\_all.sh} script from the \texttt{scripts/} directory: 

\begin{lstlisting}
$ cd scripts/
$ bash run_all.sh
\end{lstlisting}

\noindent The total expected runtime is at most 48 hours. Each script will generate results in a pickled format, which are then used by the plotting script for generating figures.

\noindent \textit{[Results]}

\noindent All generated figures will be available in the \texttt{artifact/decoder-resources/figs/} directory.

\subsection{Notes on Reusability}
\label{sec:reuse}
This artifact provides pre-generated program traces of various quantum computing workloads to analyze decoder provisioning and scheduling. 
However, the code provided can be used to generate program traces of other workloads too.
Furthermore, any other surface code compiler can be used for generating the program traces required by this artifact, provided the same IR is used for defining logical operations. 
A change in IR will require significant refactoring of the parsing components of this artifact, but the scheduling logic and setup should generalize. 

\subsection{Methodology}
Submission, reviewing, and badging methodology: 
\begin{itemize}
    \item \url{http://ctuning.org/ae/submission-20201122.html}
    \item \url{http://ctuning.org/ae/reviewing-20201122.html}
    \item \url{https://www.acm.org/publications/policies/artifact-review-badging}
\end{itemize}

\bibliographystyle{ACM-Reference-Format}
\bibliography{refs}

@inbook{Baruah2011,
  title = {Mixed-Criticality Scheduling of Sporadic Task Systems},
  ISBN = {9783642237195},
  ISSN = {1611-3349},
  url = {http://dx.doi.org/10.1007/978-3-642-23719-5_47},
  DOI = {10.1007/978-3-642-23719-5_47},
  booktitle = {Algorithms – ESA 2011},
  publisher = {Springer Berlin Heidelberg},
  author = {Baruah,  Sanjoy K. and Bonifaci,  Vincenzo and D’Angelo,  Gianlorenzo and Marchetti-Spaccamela,  Alberto and van der Ster,  Suzanne and Stougie,  Leen},
  year = {2011},
  pages = {555–566}
}

@inproceedings{tassiulas1990stability,
  title={Stability properties of constrained queueing systems and scheduling policies for maximum throughput in multihop radio networks},
  author={Tassiulas, Leandros and Ephremides, Anthony},
  booktitle={29th IEEE Conference on Decision and Control},
  pages={2130--2132},
  year={1990},
  organization={IEEE}
}

@article{shreedhar1996efficient,
  title={Efficient fair queuing using deficit round-robin},
  author={Shreedhar, Madhavapeddi and Varghese, George},
  journal={IEEE/ACM Transactions on networking},
  volume={4},
  number={3},
  pages={375--385},
  year={1996},
  publisher={IEEE}
}

@article{dimakis2006sufficient,
  title={Sufficient conditions for stability of longest-queue-first scheduling: Second-order properties using fluid limits},
  author={Dimakis, Antonis and Walrand, Jean},
  journal={Advances in Applied probability},
  volume={38},
  number={2},
  pages={505--521},
  year={2006},
  publisher={Cambridge University Press}
}

@article{Davis2011,
  title = {A survey of hard real-time scheduling for multiprocessor systems},
  volume = {43},
  ISSN = {1557-7341},
  url = {http://dx.doi.org/10.1145/1978802.1978814},
  DOI = {10.1145/1978802.1978814},
  number = {4},
  journal = {ACM Computing Surveys},
  publisher = {Association for Computing Machinery (ACM)},
  author = {Davis,  Robert I. and Burns,  Alan},
  year = {2011},
  month = oct,
  pages = {1–44}
}

@article{liu1973scheduling,
  title={Scheduling algorithms for multiprogramming in a hard-real-time environment},
  author={Liu, Chung Laung and Layland, James W},
  journal={Journal of the ACM (JACM)},
  volume={20},
  number={1},
  pages={46--61},
  year={1973},
  publisher={ACM New York, NY, USA}
}

@inproceedings{Vestal2007,
  title = {Preemptive Scheduling of Multi-criticality Systems with Varying Degrees of Execution Time Assurance},
  url = {http://dx.doi.org/10.1109/RTSS.2007.47},
  DOI = {10.1109/rtss.2007.47},
  booktitle = {28th IEEE International Real-Time Systems Symposium (RTSS 2007)},
  publisher = {IEEE},
  author = {Vestal,  Steve},
  year = {2007},
  month = dec,
  pages = {239–243}
}

@inbook{Masko2006,
  title = {Scheduling Moldable Tasks for Dynamic SMP Clusters in SoC Technology},
  ISBN = {9783540341420},
  ISSN = {1611-3349},
  url = {http://dx.doi.org/10.1007/11752578_106},
  DOI = {10.1007/11752578_106},
  booktitle = {Parallel Processing and Applied Mathematics},
  publisher = {Springer Berlin Heidelberg},
  author = {Masko,  Lukasz and Dutot,  Pierre–Francois and Mounie,  Gregory and Trystram,  Denis and Tudruj,  Marek},
  year = {2006},
  pages = {879–887}
}

@article{Terhal2015,
  title = {Quantum error correction for quantum memories},
  volume = {87},
  ISSN = {1539-0756},
  url = {http://dx.doi.org/10.1103/RevModPhys.87.307},
  DOI = {10.1103/revmodphys.87.307},
  number = {2},
  journal = {Reviews of Modern Physics},
  publisher = {American Physical Society (APS)},
  author = {Terhal,  Barbara M.},
  year = {2015},
  month = apr,
  pages = {307–346}
}

@article{iyer2015hardness,
  title={Hardness of decoding quantum stabilizer codes},
  author={Iyer, Pavithran and Poulin, David},
  journal={IEEE Transactions on Information Theory},
  volume={61},
  number={9},
  pages={5209--5223},
  year={2015},
  publisher={IEEE}
}

@article{Gidney2024Cultivation,
  doi = {10.48550/ARXIV.2409.17595},
  url = {https://arxiv.org/abs/2409.17595},
  author = {Gidney,  Craig and Shutty,  Noah and Jones,  Cody},
  title = {Magic state cultivation: growing T states as cheap as CNOT gates},
  publisher = {arXiv},
  year = {2024},
  copyright = {Creative Commons Attribution 4.0 International}
}

@inproceedings{Wu2025,
  series = {ASPLOS ’25},
  title = {Micro Blossom: Accelerated Minimum-Weight Perfect Matching Decoding for Quantum Error Correction},
  url = {http://dx.doi.org/10.1145/3676641.3716005},
  DOI = {10.1145/3676641.3716005},
  booktitle = {Proceedings of the 30th ACM International Conference on Architectural Support for Programming Languages and Operating Systems,  Volume 2},
  publisher = {ACM},
  author = {Wu,  Yue and Liyanage,  Namitha and Zhong,  Lin},
  year = {2025},
  month = mar,
  pages = {639–654},
  collection = {ASPLOS ’25}
}

@article{Resch2021,
  title = {Benchmarking Quantum Computers and the Impact of Quantum Noise},
  volume = {54},
  ISSN = {1557-7341},
  url = {http://dx.doi.org/10.1145/3464420},
  DOI = {10.1145/3464420},
  number = {7},
  journal = {ACM Computing Surveys},
  publisher = {Association for Computing Machinery (ACM)},
  author = {Resch,  Salonik and Karpuzcu,  Ulya R.},
  year = {2021},
  month = jul,
  pages = {1–35}
}

@article{Gidney2024YState,
  title = {Inplace Access to the Surface Code Y Basis},
  volume = {8},
  ISSN = {2521-327X},
  url = {http://dx.doi.org/10.22331/q-2024-04-08-1310},
  DOI = {10.22331/q-2024-04-08-1310},
  journal = {Quantum},
  publisher = {Verein zur Forderung des Open Access Publizierens in den Quantenwissenschaften},
  author = {Gidney,  Craig},
  year = {2024},
  month = apr,
  pages = {1310}
}

@article{Chamberland2020,
  title = {Very low overhead fault-tolerant magic state preparation using redundant ancilla encoding and flag qubits},
  volume = {6},
  ISSN = {2056-6387},
  url = {http://dx.doi.org/10.1038/s41534-020-00319-5},
  DOI = {10.1038/s41534-020-00319-5},
  number = {1},
  journal = {npj Quantum Information},
  publisher = {Springer Science and Business Media LLC},
  author = {Chamberland,  Christopher and Noh,  Kyungjoo},
  year = {2020},
  month = oct 
}

@misc{Harvard2025MLDecoder,
  doi = {10.48550/ARXIV.2509.11370},
  url = {https://arxiv.org/abs/2509.11370},
  author = {Ataides,  J. Pablo Bonilla and Gu,  Andi and Yelin,  Susanne F. and Lukin,  Mikhail D.},
  title = {Neural Decoders for Universal Quantum Algorithms},
  publisher = {arXiv},
  year = {2025},
  copyright = {arXiv.org perpetual,  non-exclusive license}
}

@article{gidney2025factor,
  title={How to factor 2048 bit RSA integers with less than a million noisy qubits},
  author={Gidney, Craig},
  journal={arXiv preprint arXiv:2505.15917},
  year={2025}
}

@article{wu2024lego,
  title={LEGO: QEC Decoding System Architecture for Dynamic Circuits},
  author={Wu, Yue and Liyanage, Namitha and Zhong, Lin},
  journal={arXiv preprint arXiv:2410.03073},
  year={2024}
}

@article{liyanage2025network,
  title={Network-Integrated Decoding System for Real-Time Quantum Error Correction with Lattice Surgery},
  author={Liyanage, Namitha and Wu, Yue and Houghton, Emmet and Zhong, Lin},
  journal={arXiv preprint arXiv:2504.11805},
  year={2025}
}

@online{nvidia_quera,
  author = "{NVIDIA}",
  title = "{NVIDIA and QuEra Decode Quantum Errors with AI}",
  year={2025},
  organization="{NVIDIA}",
  url = "{https://developer.nvidia.com/blog/nvidia-and-quera-decode-quantum-errors-with-ai/}",
  note = "{Accessed: June 18, 2025}"
}

@article{Fowler2012,
  title = {Surface codes: Towards practical large-scale quantum computation},
  volume = {86},
  ISSN = {1094-1622},
  url = {http://dx.doi.org/10.1103/PhysRevA.86.032324},
  DOI = {10.1103/physreva.86.032324},
  number = {3},
  journal = {Physical Review A},
  publisher = {American Physical Society (APS)},
  author = {Fowler,  Austin G. and Mariantoni,  Matteo and Martinis,  John M. and Cleland,  Andrew N.},
  year = {2012},
  month = sep 
}

@article{Horsman2012,
  title = {Surface code quantum computing by lattice surgery},
  volume = {14},
  ISSN = {1367-2630},
  url = {http://dx.doi.org/10.1088/1367-2630/14/12/123011},
  DOI = {10.1088/1367-2630/14/12/123011},
  number = {12},
  journal = {New Journal of Physics},
  publisher = {IOP Publishing},
  author = {Horsman,  Dominic and Fowler,  Austin G and Devitt,  Simon and Meter,  Rodney Van},
  year = {2012},
  month = dec,
  pages = {123011}
}

@article{Litinski2019,
  title = {A Game of Surface Codes: Large-Scale Quantum Computing with Lattice Surgery},
  volume = {3},
  ISSN = {2521-327X},
  url = {http://dx.doi.org/10.22331/q-2019-03-05-128},
  DOI = {10.22331/q-2019-03-05-128},
  journal = {Quantum},
  publisher = {Verein zur Forderung des Open Access Publizierens in den Quantenwissenschaften},
  author = {Litinski,  Daniel},
  year = {2019},
  month = mar,
  pages = {128}
}

@article{Litinski2019magic,
  title = {Magic State Distillation: Not as Costly as You Think},
  volume = {3},
  ISSN = {2521-327X},
  url = {http://dx.doi.org/10.22331/q-2019-12-02-205},
  DOI = {10.22331/q-2019-12-02-205},
  journal = {Quantum},
  publisher = {Verein zur Forderung des Open Access Publizierens in den Quantenwissenschaften},
  author = {Litinski,  Daniel},
  year = {2019},
  month = dec,
  pages = {205}
}

@misc{Alavisamani2024,
  doi = {10.48550/ARXIV.2404.03136},
  url = {https://arxiv.org/abs/2404.03136},
  author = {Alavisamani,  Narges and Vittal,  Suhas and Ayanzadeh,  Ramin and Das,  Poulami and Qureshi,  Moinuddin},
  title = {Promatch: Extending the Reach of Real-Time Quantum Error Correction with Adaptive Predecoding},
  publisher = {arXiv},
  year = {2024},
  copyright = {Creative Commons Attribution 4.0 International}
}

@inproceedings{Vittal2023,
  series = {ISCA ’23},
  title = {Astrea: Accurate Quantum Error-Decoding via Practical Minimum-Weight Perfect-Matching},
  url = {http://dx.doi.org/10.1145/3579371.3589037},
  DOI = {10.1145/3579371.3589037},
  booktitle = {Proceedings of the 50th Annual International Symposium on Computer Architecture},
  publisher = {ACM},
  author = {Vittal,  Suhas and Das,  Poulami and Qureshi,  Moinuddin},
  year = {2023},
  month = jun,
  collection = {ISCA ’23}
}

@inproceedings{Ravi2023,
  series = {ASPLOS ’23},
  title = {Better Than Worst-Case Decoding for Quantum Error Correction},
  url = {http://dx.doi.org/10.1145/3575693.3575733},
  DOI = {10.1145/3575693.3575733},
  booktitle = {Proceedings of the 28th ACM International Conference on Architectural Support for Programming Languages and Operating Systems,  Volume 2},
  publisher = {ACM},
  author = {Ravi,  Gokul Subramanian and Baker,  Jonathan M. and Fayyazi,  Arash and Lin,  Sophia Fuhui and Javadi-Abhari,  Ali and Pedram,  Massoud and Chong,  Frederic T.},
  year = {2023},
  month = jan,
  collection = {ASPLOS ’23}
}

@article{Smith2023,
  title = {Local Predecoder to Reduce the Bandwidth and Latency of Quantum Error Correction},
  volume = {19},
  ISSN = {2331-7019},
  url = {http://dx.doi.org/10.1103/PhysRevApplied.19.034050},
  DOI = {10.1103/physrevapplied.19.034050},
  number = {3},
  journal = {Physical Review Applied},
  publisher = {American Physical Society (APS)},
  author = {Smith,  Samuel C. and Brown,  Benjamin J. and Bartlett,  Stephen D.},
  year = {2023},
  month = mar 
}

@misc{riverlaneLCD2024,
  doi = {10.48550/ARXIV.2411.10343},
  url = {https://arxiv.org/abs/2411.10343},
  author = {Ziad,  Abbas B. and Zalawadiya,  Ankit and Topal,  Canberk and Camps,  Joan and Gehér,  Gy\"{o}rgy P. and Stafford,  Matthew P. and Turner,  Mark L.},
  title = {Local Clustering Decoder: a fast and adaptive hardware decoder for the surface code},
  publisher = {arXiv},
  year = {2024},
  copyright = {arXiv.org perpetual,  non-exclusive license}
}

@article{FuhuiLin2025,
  title = {Spatially parallel decoding for multi-qubit lattice surgery},
  volume = {10},
  ISSN = {2058-9565},
  url = {http://dx.doi.org/10.1088/2058-9565/adc6b6},
  DOI = {10.1088/2058-9565/adc6b6},
  number = {3},
  journal = {Quantum Science and Technology},
  publisher = {IOP Publishing},
  author = {Fuhui Lin,  Sophia and Peterson,  Eric C and Sankar,  Krishanu and Sivarajah,  Prasahnt},
  year = {2025},
  month = apr,
  pages = {035007}
}

@misc{Google2024QEC,
  doi = {10.48550/ARXIV.2408.13687},
  url = {https://arxiv.org/abs/2408.13687},
  author = {Acharya,  Rajeev and Aghababaie-Beni,  Laleh and Aleiner,  Igor and Andersen,  Trond I. and Ansmann,  Markus and Arute,  Frank and Arya,  Kunal and Asfaw,  Abraham and Astrakhantsev,  Nikita and Atalaya,  Juan and Babbush,  Ryan and Bacon,  Dave and Ballard,  Brian and Bardin,  Joseph C. and Bausch,  Johannes and Bengtsson,  Andreas and Bilmes,  Alexander and Blackwell,  Sam and Boixo,  Sergio and Bortoli,  Gina and Bourassa,  Alexandre and Bovaird,  Jenna and Brill,  Leon and Broughton,  Michael and Browne,  David A. and Buchea,  Brett and Buckley,  Bob B. and Buell,  David A. and Burger,  Tim and Burkett,  Brian and Bushnell,  Nicholas and Cabrera,  Anthony and Campero,  Juan and Chang,  Hung-Shen and Chen,  Yu and Chen,  Zijun and Chiaro,  Ben and Chik,  Desmond and Chou,  Charina and Claes,  Jahan and Cleland,  Agnetta Y. and Cogan,  Josh and Collins,  Roberto and Conner,  Paul and Courtney,  William and Crook,  Alexander L. and Curtin,  Ben and Das,  Sayan and Davies,  Alex and De Lorenzo,  Laura and Debroy,  Dripto M. and Demura,  Sean and Devoret,  Michel and Di Paolo,  Agustin and Donohoe,  Paul and Drozdov,  Ilya and Dunsworth,  Andrew and Earle,  Clint and Edlich,  Thomas and Eickbusch,  Alec and Elbag,  Aviv Moshe and Elzouka,  Mahmoud and Erickson,  Catherine and Faoro,  Lara and Farhi,  Edward and Ferreira,  Vinicius S. and Burgos,  Leslie Flores and Forati,  Ebrahim and Fowler,  Austin G. and Foxen,  Brooks and Ganjam,  Suhas and Garcia,  Gonzalo and Gasca,  Robert and Genois,  Elie and Giang,  William and Gidney,  Craig and Gilboa,  Dar and Gosula,  Raja and Dau,  Alejandro Grajales and Graumann,  Dietrich and Greene,  Alex and Gross,  Jonathan A. and Habegger,  Steve and Hall,  John and Hamilton,  Michael C. and Hansen,  Monica and Harrigan,  Matthew P. and Harrington,  Sean D. and Heras,  Francisco J. H. and Heslin,  Stephen and Heu,  Paula and Higgott,  Oscar and Hill,  Gordon and Hilton,  Jeremy and Holland,  George and Hong,  Sabrina and Huang,  Hsin-Yuan and Huff,  Ashley and Huggins,  William J. and Ioffe,  Lev B. and Isakov,  Sergei V. and Iveland,  Justin and Jeffrey,  Evan and Jiang,  Zhang and Jones,  Cody and Jordan,  Stephen and Joshi,  Chaitali and Juhas,  Pavol and Kafri,  Dvir and Kang,  Hui and Karamlou,  Amir H. and Kechedzhi,  Kostyantyn and Kelly,  Julian and Khaire,  Trupti and Khattar,  Tanuj and Khezri,  Mostafa and Kim,  Seon and Klimov,  Paul V. and Klots,  Andrey R. and Kobrin,  Bryce and Kohli,  Pushmeet and Korotkov,  Alexander N. and Kostritsa,  Fedor and Kothari,  Robin and Kozlovskii,  Borislav and Kreikebaum,  John Mark and Kurilovich,  Vladislav D. and Lacroix,  Nathan and Landhuis,  David and Lange-Dei,  Tiano and Langley,  Brandon W. and Laptev,  Pavel and Lau,  Kim-Ming and Guevel,  Loïck Le and Ledford,  Justin and Lee,  Kenny and Lensky,  Yuri D. and Leon,  Shannon and Lester,  Brian J. and Li,  Wing Yan and Li,  Yin and Lill,  Alexander T. and Liu,  Wayne and Livingston,  William P. and Locharla,  Aditya and Lucero,  Erik and Lundahl,  Daniel and Lunt,  Aaron and Madhuk,  Sid and Malone,  Fionn D. and Maloney,  Ashley and Mandrá,  Salvatore and Martin,  Leigh S. and Martin,  Steven and Martin,  Orion and Maxfield,  Cameron and McClean,  Jarrod R. and McEwen,  Matt and Meeks,  Seneca and Megrant,  Anthony and Mi,  Xiao and Miao,  Kevin C. and Mieszala,  Amanda and Molavi,  Reza and Molina,  Sebastian and Montazeri,  Shirin and Morvan,  Alexis and Movassagh,  Ramis and Mruczkiewicz,  Wojciech and Naaman,  Ofer and Neeley,  Matthew and Neill,  Charles and Nersisyan,  Ani and Neven,  Hartmut and Newman,  Michael and Ng,  Jiun How and Nguyen,  Anthony and Nguyen,  Murray and Ni,  Chia-Hung and O'Brien,  Thomas E. and Oliver,  William D. and Opremcak,  Alex and Ottosson,  Kristoffer and Petukhov,  Andre and Pizzuto,  Alex and Platt,  John and Potter,  Rebecca and Pritchard,  Orion and Pryadko,  Leonid P. and Quintana,  Chris and Ramachandran,  Ganesh and Reagor,  Matthew J. and Rhodes,  David M. and Roberts,  Gabrielle and Rosenberg,  Eliott and Rosenfeld,  Emma and Roushan,  Pedram and Rubin,  Nicholas C. and Saei,  Negar and Sank,  Daniel and Sankaragomathi,  Kannan and Satzinger,  Kevin J. and Schurkus,  Henry F. and Schuster,  Christopher and Senior,  Andrew W. and Shearn,  Michael J. and Shorter,  Aaron and Shutty,  Noah and Shvarts,  Vladimir and Singh,  Shraddha and Sivak,  Volodymyr and Skruzny,  Jindra and Small,  Spencer and Smelyanskiy,  Vadim and Smith,  W. Clarke and Somma,  Rolando D. and Springer,  Sofia and Sterling,  George and Strain,  Doug and Suchard,  Jordan and Szasz,  Aaron and Sztein,  Alex and Thor,  Douglas and Torres,  Alfredo and Torunbalci,  M. Mert and Vaishnav,  Abeer and Vargas,  Justin and Vdovichev,  Sergey and Vidal,  Guifre and Villalonga,  Benjamin and Heidweiller,  Catherine Vollgraff and Waltman,  Steven and Wang,  Shannon X. and Ware,  Brayden and Weber,  Kate and White,  Theodore and Wong,  Kristi and Woo,  Bryan W. K. and Xing,  Cheng and Yao,  Z. Jamie and Yeh,  Ping and Ying,  Bicheng and Yoo,  Juhwan and Yosri,  Noureldin and Young,  Grayson and Zalcman,  Adam and Zhang,  Yaxing and Zhu,  Ningfeng and Zobrist,  Nicholas},
  title = {Quantum error correction below the surface code threshold},
  publisher = {arXiv},
  year = {2024},
  copyright = {Creative Commons Attribution 4.0 International}
}

@online{GoogleQuantumRoadmap,
  author = "{Google Quantum AI}",
  title = "{Google Quantum Roadmap}",
year={2024},
  url = {https://quantumai.google/qecmilestone},
  note = "{Accessed: September 18, 2025}"
}

@online{IBMQuantumRoadmap,
  author = "{IBM Quantum}",
  title = "{IBM Quantum Roadmap}",
  year={2024},
  url = {https://www.ibm.com/quantum/roadmap},
  note = "{Accessed: September 18, 2025}"
}

@online{QuantinuumQuantumRoadmap,
  author = "{Quantinuum}",
  title = "{Quantinuum Unveils Accelerated Roadmap to Achieve Universal, Fully Fault-Tolerant Quantum Computing by 2030}",
  year={2025},
  url = {https://www.quantinuum.com/press-releases/quantinuum-unveils-accelerated-roadmap-to-achieve-universal-fault-tolerant-quantum-computing-by-2030},
  note = "{Accessed: September 18, 2025}"
}

@article{AspuruGuzik2005,
  doi = {10.1126/science.1113479},
  url = {https://doi.org/10.1126/science.1113479},
  year = {2005},
  month = sep,
  publisher = {American Association for the Advancement of Science ({AAAS})},
  volume = {309},
  number = {5741},
  pages = {1704--1707},
  author = {Alan Aspuru-Guzik and Anthony D. Dutoi and Peter J. Love and Martin Head-Gordon},
  title = {Simulated Quantum Computation of Molecular Energies},
  journal = {Science}
}

@inproceedings{Grover1996,
author = {Grover, Lov K.},
title = {A fast quantum mechanical algorithm for database search},
year = {1996},
isbn = {0897917855},
publisher = {Association for Computing Machinery},
address = {New York, NY, USA},
url = {https://doi.org/10.1145/237814.237866},
doi = {10.1145/237814.237866},
booktitle = {Proceedings of the Twenty-Eighth Annual ACM Symposium on Theory of Computing},
pages = {212–219},
numpages = {8},
location = {Philadelphia, Pennsylvania, USA},
series = {STOC '96}
}

@article{Shor1997,
author = {Shor, Peter W.},
title = {Polynomial-Time Algorithms for Prime Factorization and Discrete Logarithms on a Quantum Computer},
year = {1997},
issue_date = {Oct. 1997},
publisher = {Society for Industrial and Applied Mathematics},
address = {USA},
volume = {26},
number = {5},
issn = {0097-5397},
url = {https://doi.org/10.1137/S0097539795293172},
doi = {10.1137/S0097539795293172},
journal = {SIAM J. Comput.},
month = {oct},
pages = {1484–1509},
numpages = {26}
}

@article{Lloyd1996,
  doi = {10.1126/science.273.5278.1073},
  url = {https://doi.org/10.1126/science.273.5278.1073},
  year = {1996},
  month = aug,
  publisher = {American Association for the Advancement of Science ({AAAS})},
  volume = {273},
  number = {5278},
  pages = {1073--1078},
  author = {Seth Lloyd},
  title = {Universal Quantum Simulators},
  journal = {Science}
}

@article{muller2025improved,
  title={Improved belief propagation is sufficient for real-time decoding of quantum memory},
  author={M{\"u}ller, Tristan and Alexander, Thomas and Beverland, Michael E and B{\"u}hler, Markus and Johnson, Blake R and Maurer, Thilo and Vandeth, Drew},
  journal={arXiv preprint arXiv:2506.01779},
  year={2025}
}

@inproceedings{CorriganGibbs2017,
  series = {HotOS ’17},
  title = {Quantum Operating Systems},
  url = {http://dx.doi.org/10.1145/3102980.3102993},
  DOI = {10.1145/3102980.3102993},
  booktitle = {Proceedings of the 16th Workshop on Hot Topics in Operating Systems},
  publisher = {ACM},
  author = {Corrigan-Gibbs,  Henry and Wu,  David J. and Boneh,  Dan},
  year = {2017},
  month = may,
  pages = {76–81},
  collection = {HotOS ’17}
}

@inproceedings{giortamis2025qos,
  title={$\{$QOS$\}$: Quantum Operating System},
  author={Giortamis, Emmanouil and Rom{\~a}o, Francisco and Tornow, Nathaniel and Bhatotia, Pramod},
  booktitle={19th USENIX Symposium on Operating Systems Design and Implementation (OSDI 25)},
  pages={429--447},
  year={2025}
}

@online{vader,
  author="{Satvik Maurya and Abtin Molavi and Aws Albarghouthi and Swamit Tannu}",
  title="{Decoder Resource Estimator}",
  year={2026},
  url = "https://doi.org/10.5281/zenodo.18555904",
  note = "Available at \url{https://doi.org/10.5281/zenodo.18555904}"
}

@misc{Riverlane2023,
  doi = {10.48550/ARXIV.2309.05558},
  url = {https://arxiv.org/abs/2309.05558},
  author = {Barber,  Ben and Barnes,  Kenton M. and Bialas,  Tomasz and Buğdaycı,  Okan and Campbell,  Earl T. and Gillespie,  Neil I. and Johar,  Kauser and Rajan,  Ram and Richardson,  Adam W. and Skoric,  Luka and Topal,  Canberk and Turner,  Mark L. and Ziad,  Abbas B.},
  title = {A real-time,  scalable,  fast and highly resource efficient decoder for a quantum computer},
  publisher = {arXiv},
  year = {2023},
  copyright = {Creative Commons Attribution 4.0 International}
}

@misc{Beverland2022,
  doi = {10.48550/ARXIV.2211.07629},
  url = {https://arxiv.org/abs/2211.07629},
  author = {Beverland,  Michael E. and Murali,  Prakash and Troyer,  Matthias and Svore,  Krysta M. and Hoefler,  Torsten and Kliuchnikov,  Vadym and Low,  Guang Hao and Soeken,  Mathias and Sundaram,  Aarthi and Vaschillo,  Alexander},
  title = {Assessing requirements to scale to practical quantum advantage},
  publisher = {arXiv},
  year = {2022},
  copyright = {Creative Commons Attribution 4.0 International}
}

@article{Beverland2022edpc,
  title = {Surface Code Compilation via Edge-Disjoint Paths},
  volume = {3},
  ISSN = {2691-3399},
  url = {http://dx.doi.org/10.1103/PRXQuantum.3.020342},
  DOI = {10.1103/prxquantum.3.020342},
  number = {2},
  journal = {PRX Quantum},
  publisher = {American Physical Society (APS)},
  author = {Beverland,  Michael and Kliuchnikov,  Vadym and Schoute,  Eddie},
  year = {2022},
  month = may 
}

@article{watkins2023high,
  title = {A High Performance Compiler for Very Large Scale Surface Code Computations},
  volume = {8},
  ISSN = {2521-327X},
  url = {http://dx.doi.org/10.22331/q-2024-05-22-1354},
  DOI = {10.22331/q-2024-05-22-1354},
  journal = {Quantum},
  publisher = {Verein zur Forderung des Open Access Publizierens in den Quantenwissenschaften},
  author = {Watkins,  George and Nguyen,  Hoang Minh and Watkins,  Keelan and Pearce,  Steven and Lau,  Hoi-Kwan and Paler,  Alexandru},
  year = {2024},
  month = may,
  pages = {1354}
}

@article{Skoric2023,
  title = {Parallel window decoding enables scalable fault tolerant quantum computation},
  volume = {14},
  ISSN = {2041-1723},
  url = {http://dx.doi.org/10.1038/s41467-023-42482-1},
  DOI = {10.1038/s41467-023-42482-1},
  number = {1},
  journal = {Nature Communications},
  publisher = {Springer Science and Business Media LLC},
  author = {Skoric,  Luka and Browne,  Dan E. and Barnes,  Kenton M. and Gillespie,  Neil I. and Campbell,  Earl T.},
  year = {2023},
  month = nov 
}

@misc{Delfosse2020,
  doi = {10.48550/ARXIV.2001.11427},
  url = {https://arxiv.org/abs/2001.11427},
  author = {Delfosse,  Nicolas},
  title = {Hierarchical decoding to reduce hardware requirements for quantum computing},
  publisher = {arXiv},
  year = {2020},
  copyright = {arXiv.org perpetual,  non-exclusive license}
}

@misc{Bombin2023,
  doi = {10.48550/ARXIV.2303.04846},
  url = {https://arxiv.org/abs/2303.04846},
  author = {Bombín,  Héctor and Dawson,  Chris and Liu,  Ye-Hua and Nickerson,  Naomi and Pastawski,  Fernando and Roberts,  Sam},
  title = {Modular decoding: parallelizable real-time decoding for quantum computers},
  publisher = {arXiv},
  year = {2023},
  copyright = {arXiv.org perpetual,  non-exclusive license}
}

@misc{Caune2023,
  doi = {10.48550/ARXIV.2306.17142},
  url = {https://arxiv.org/abs/2306.17142},
  author = {Caune,  Laura and Reid,  Brendan and Camps,  Joan and Campbell,  Earl},
  title = {Belief propagation as a partial decoder},
  publisher = {arXiv},
  year = {2023},
  copyright = {Creative Commons Attribution 4.0 International}
}

@misc{Delfosse2023,
  doi = {10.48550/ARXIV.2310.15313},
  url = {https://arxiv.org/abs/2310.15313},
  author = {Delfosse,  Nicolas and Paz,  Andres and Vaschillo,  Alexander and Svore,  Krysta M.},
  title = {How to choose a decoder for a fault-tolerant quantum computer? The speed vs accuracy trade-off},
  publisher = {arXiv},
  year = {2023},
  copyright = {Creative Commons Attribution 4.0 International}
}

@inproceedings{Das2022lilliput,
  series = {ASPLOS ’22},
  title = {LILLIPUT: a lightweight low-latency lookup-table decoder for near-term Quantum error correction},
  url = {http://dx.doi.org/10.1145/3503222.3507707},
  DOI = {10.1145/3503222.3507707},
  booktitle = {Proceedings of the 27th ACM International Conference on Architectural Support for Programming Languages and Operating Systems},
  publisher = {ACM},
  author = {Das,  Poulami and Locharla,  Aditya and Jones,  Cody},
  year = {2022},
  month = feb,
  collection = {ASPLOS ’22}
}

@inproceedings{Das2022afs,
  title = {AFS: Accurate,  Fast,  and Scalable Error-Decoding for Fault-Tolerant Quantum Computers},
  url = {http://dx.doi.org/10.1109/HPCA53966.2022.00027},
  DOI = {10.1109/hpca53966.2022.00027},
  booktitle = {2022 IEEE International Symposium on High-Performance Computer Architecture (HPCA)},
  publisher = {IEEE},
  author = {Das,  Poulami and Pattison,  Christopher A. and Manne,  Srilatha and Carmean,  Douglas M. and Svore,  Krysta M. and Qureshi,  Moinuddin and Delfosse,  Nicolas},
  year = {2022},
  month = apr 
}

@misc{qasm,
  doi = {10.48550/ARXIV.1707.03429},
  url = {https://arxiv.org/abs/1707.03429},
  author = {Cross,  Andrew W. and Bishop,  Lev S. and Smolin,  John A. and Gambetta,  Jay M.},
  title = {Open Quantum Assembly Language},
  publisher = {arXiv},
  year = {2017},
  copyright = {arXiv.org perpetual,  non-exclusive license}
}

@misc{gridsynth,
  doi = {10.48550/ARXIV.1403.2975},
  url = {https://arxiv.org/abs/1403.2975},
  author = {Ross,  Neil J. and Selinger,  Peter},
  title = {Optimal ancilla-free Clifford+T approximation of z-rotations},
  publisher = {arXiv},
  year = {2014},
  copyright = {arXiv.org perpetual,  non-exclusive license}
}

@inproceedings{Byun2022,
  series = {ISCA ’22},
  title = {XQsim: modeling cross-technology control processors for 10+K qubit quantum computers},
  url = {http://dx.doi.org/10.1145/3470496.3527417},
  DOI = {10.1145/3470496.3527417},
  booktitle = {Proceedings of the 49th Annual International Symposium on Computer Architecture},
  publisher = {ACM},
  author = {Byun,  Ilkwon and Kim,  Junpyo and Min,  Dongmoon and Nagaoka,  Ikki and Fukumitsu,  Kosuke and Ishikawa,  Iori and Tanimoto,  Teruo and Tanaka,  Masamitsu and Inoue,  Koji and Kim,  Jangwoo},
  year = {2022},
  month = jun,
  collection = {ISCA ’22}
}

@inproceedings{Stein2023,
  series = {MICRO ’23},
  title = {HetArch: Heterogeneous Microarchitectures for Superconducting Quantum Systems},
  url = {http://dx.doi.org/10.1145/3613424.3614300},
  DOI = {10.1145/3613424.3614300},
  booktitle = {56th Annual IEEE/ACM International Symposium on Microarchitecture},
  publisher = {ACM},
  author = {Stein,  Samuel and Sussman,  Sara and Tomesh,  Teague and Guinn,  Charles and Tureci,  Esin and Lin,  Sophia Fuhui and Tang,  Wei and Ang,  James and Chakram,  Srivatsan and Li,  Ang and Martonosi,  Margaret and Chong,  Fred and Houck,  Andrew A. and Chuang,  Isaac L. and Demarco,  Michael},
  year = {2023},
  month = oct,
  collection = {MICRO ’23}
}

@inproceedings{Ding2018,
  title = {Magic-State Functional Units: Mapping and Scheduling Multi-Level Distillation Circuits for Fault-Tolerant Quantum Architectures},
  url = {http://dx.doi.org/10.1109/MICRO.2018.00072},
  DOI = {10.1109/micro.2018.00072},
  booktitle = {2018 51st Annual IEEE/ACM International Symposium on Microarchitecture (MICRO)},
  publisher = {IEEE},
  author = {Ding,  Yongshan and Holmes,  Adam and Javadi-Abhari,  Ali and Franklin,  Diana and Martonosi,  Margaret and Chong,  Frederic},
  year = {2018},
  month = oct 
}

@misc{Lin2023,
  doi = {10.48550/ARXIV.2305.00138},
  url = {https://arxiv.org/abs/2305.00138},
  author = {Lin,  Sophia Fuhui and Viszlai,  Joshua and Smith,  Kaitlin N. and Ravi,  Gokul Subramanian and Yuan,  Charles and Chong,  Frederic T. and Brown,  Benjamin J.},
  title = {Codesign of quantum error-correcting codes and modular chiplets in the presence of defects},
  publisher = {arXiv},
  year = {2023},
  copyright = {arXiv.org perpetual,  non-exclusive license}
}

@article{Varsamopoulos2017,
  title = {Decoding small surface codes with feedforward neural networks},
  volume = {3},
  ISSN = {2058-9565},
  url = {http://dx.doi.org/10.1088/2058-9565/aa955a},
  DOI = {10.1088/2058-9565/aa955a},
  number = {1},
  journal = {Quantum Science and Technology},
  publisher = {IOP Publishing},
  author = {Varsamopoulos,  Savvas and Criger,  Ben and Bertels,  Koen},
  year = {2017},
  month = nov,
  pages = {015004}
}

@article{Gicev2023,
  title = {A scalable and fast artificial neural network syndrome decoder for surface codes},
  volume = {7},
  ISSN = {2521-327X},
  url = {http://dx.doi.org/10.22331/q-2023-07-12-1058},
  DOI = {10.22331/q-2023-07-12-1058},
  journal = {Quantum},
  publisher = {Verein zur Forderung des Open Access Publizierens in den Quantenwissenschaften},
  author = {Gicev,  Spiro and Hollenberg,  Lloyd C. L. and Usman,  Muhammad},
  year = {2023},
  month = jul,
  pages = {1058}
}

@article{Meinerz2022,
  title = {Scalable Neural Decoder for Topological Surface Codes},
  volume = {128},
  ISSN = {1079-7114},
  url = {http://dx.doi.org/10.1103/PhysRevLett.128.080505},
  DOI = {10.1103/physrevlett.128.080505},
  number = {8},
  journal = {Physical Review Letters},
  publisher = {American Physical Society (APS)},
  author = {Meinerz,  Kai and Park,  Chae-Yeun and Trebst,  Simon},
  year = {2022},
  month = feb 
}

@article{Overwater2022,
  title = {Neural-Network Decoders for Quantum Error Correction Using Surface Codes: A Space Exploration of the Hardware Cost-Performance Tradeoffs},
  volume = {3},
  ISSN = {2689-1808},
  url = {http://dx.doi.org/10.1109/TQE.2022.3174017},
  DOI = {10.1109/tqe.2022.3174017},
  journal = {IEEE Transactions on Quantum Engineering},
  publisher = {Institute of Electrical and Electronics Engineers (IEEE)},
  author = {Overwater,  Ramon W. J. and Babaie,  Masoud and Sebastiano,  Fabio},
  year = {2022},
  pages = {1–19}
}

@misc{ueno2022,
  doi = {10.48550/ARXIV.2208.05758},
  url = {https://arxiv.org/abs/2208.05758},
  author = {Ueno,  Yosuke and Kondo,  Masaaki and Tanaka,  Masamitsu and Suzuki,  Yasunari and Tabuchi,  Yutaka},
  title = {NEO-QEC: Neural Network Enhanced Online Superconducting Decoder for Surface Codes},
  publisher = {arXiv},
  year = {2022},
  copyright = {arXiv.org perpetual,  non-exclusive license}
}

@misc{googleRNN,
  doi = {10.48550/ARXIV.2310.05900},
  url = {https://arxiv.org/abs/2310.05900},
  author = {Bausch,  Johannes and Senior,  Andrew W and Heras,  Francisco J H and Edlich,  Thomas and Davies,  Alex and Newman,  Michael and Jones,  Cody and Satzinger,  Kevin and Niu,  Murphy Yuezhen and Blackwell,  Sam and Holland,  George and Kafri,  Dvir and Atalaya,  Juan and Gidney,  Craig and Hassabis,  Demis and Boixo,  Sergio and Neven,  Hartmut and Kohli,  Pushmeet},
  title = {Learning to Decode the Surface Code with a Recurrent,  Transformer-Based Neural Network},
  publisher = {arXiv},
  year = {2023},
  copyright = {arXiv.org perpetual,  non-exclusive license}
}

@article{Tomita2014,
  title = {Low-distance surface codes under realistic quantum noise},
  volume = {90},
  ISSN = {1094-1622},
  url = {http://dx.doi.org/10.1103/PhysRevA.90.062320},
  DOI = {10.1103/physreva.90.062320},
  number = {6},
  journal = {Physical Review A},
  publisher = {American Physical Society (APS)},
  author = {Tomita,  Yu and Svore,  Krysta M.},
  year = {2014},
  month = dec 
}

@inproceedings{Ueno2021,
  title = {QECOOL: On-Line Quantum Error Correction with a Superconducting Decoder for Surface Code},
  url = {http://dx.doi.org/10.1109/DAC18074.2021.9586326},
  DOI = {10.1109/dac18074.2021.9586326},
  booktitle = {2021 58th ACM/IEEE Design Automation Conference (DAC)},
  publisher = {IEEE},
  author = {Ueno,  Yosuke and Kondo,  Masaaki and Tanaka,  Masamitsu and Suzuki,  Yasunari and Tabuchi,  Yutaka},
  year = {2021},
  month = dec 
}

@inproceedings{Kim2024,
  title = {A Fault-Tolerant Million Qubit-Scale Distributed Quantum Computer},
  url = {https://doi.org/10.1145/3620665.3640388},
  DOI = {10.1145/3620665.3640388},
  booktitle = {Proceedings of the 27th ACM International Conference on Architectural Support for Programming Languages and Operating Systems},
  publisher = {ACM},
  author = {Kim, Junpyo and Min, Dongmoon and Cho, Jungmin and Jeong, Hyeonseong and Byun, Ilkwon and Choi, Junhyuk and Hong, Juwon and Kin, Jangwoo},
  year = {2024},
  month = apr 
}

@article{Gidney2021,
  title = {Stim: a fast stabilizer circuit simulator},
  volume = {5},
  ISSN = {2521-327X},
  url = {http://dx.doi.org/10.22331/q-2021-07-06-497},
  DOI = {10.22331/q-2021-07-06-497},
  journal = {Quantum},
  publisher = {Verein zur Forderung des Open Access Publizierens in den Quantenwissenschaften},
  author = {Gidney,  Craig},
  year = {2021},
  month = jul,
  pages = {497}
}

@misc{Fowler2019,
  doi = {10.48550/ARXIV.1808.06709},
  url = {https://arxiv.org/abs/1808.06709},
  author = {Fowler,  Austin G. and Gidney,  Craig},
  title = {Low overhead quantum computation using lattice surgery},
  publisher = {arXiv},
  year = {2018},
  copyright = {arXiv.org perpetual,  non-exclusive license}
}

@misc{Higott2023,
  doi = {10.48550/ARXIV.2303.15933},
  url = {https://arxiv.org/abs/2303.15933},
  author = {Higgott,  Oscar and Gidney,  Craig},
  title = {Sparse Blossom: correcting a million errors per core second with minimum-weight matching},
  publisher = {arXiv},
  year = {2023},
  copyright = {Creative Commons Attribution 4.0 International}
}

@misc{nvidiaCUDAQX2025,
  author       = {NVIDIA},
  title        = {Streamlining Quantum Error Correction and Application Development with CUDA-QX 0.4},
  howpublished = {\url{https://developer.nvidia.com/blog/streamlining-quantum-error-correction-and-application-development-with-cuda-qx-0.4/}},
  year         = {2025},
  note         = {GPU-accelerated tensor-network decoders, BP+OSD enhancements, automated detector error models}
}

@misc{riverlaneDeltaflow2025,
  author       = {Riverlane},
  title        = {Deltaflow: The Quantum Error Correction Stack},
  howpublished = {\url{https://www.riverlane.com/quantum-error-correction-stack}},
  year         = {2025},
  note         = {Real-time error correction for up to ~250 qubits using FPGA/ASIC decode hardware}
}

\end{document}